%% file: 0_structure.tex
\documentclass{article}
\usepackage[utf8]{inputenc}

\usepackage{graphicx}
\usepackage[english]{babel}
\usepackage{authblk}
\usepackage{numprint}
\usepackage{float}
	
\usepackage[numbers]{natbib}
\usepackage{hyperref}       
\usepackage{url}            
\usepackage{booktabs}       
\usepackage{amsfonts}       
\usepackage{nicefrac}       
\usepackage{microtype}      
\usepackage{wrapfig}
\usepackage{graphicx}
\usepackage[inline]{enumitem}
\usepackage{bm}
\usepackage{amsfonts,amssymb,amsmath}
\usepackage[dvipsnames]{xcolor}
\usepackage{kky}
\usepackage{cleveref}
\usepackage{textcomp}
\usepackage{xcolor}
\usepackage{lscape}
\usepackage{pgfplotstable}
\usepackage{rotating}
\usepackage{fancyhdr}
\usepackage{subcaption}
\usepackage{multirow}
\usepackage{colortbl}
\usepackage{longtable} 
\usepackage{pdflscape} 

\newenvironment{rotatepage}%
    {\clearpage\pagebreak[4]\global\pdfpageattr\expandafter{\the\pdfpageattr/Rotate 90}}%
    {\clearpage\pagebreak[4]\global\pdfpageattr\expandafter{\the\pdfpageattr/Rotate 0}}%

\title{Real-time Prediction of COVID-19 related Mortality using Electronic Health Records} 
\author[1,*]{Patrick Schwab}
\author[2,3]{Arash Mehrjou}
\author[4]{Sonali Parbhoo}
\author[5,6]{Leo Anthony Celi}
\author[7,8]{J\"urgen Hetzel}
\author[8]{Markus Hofer}
\author[2,3]{Bernhard Schölkopf}
\author[2,9]{Stefan Bauer}
\affil[1]{F. Hoffmann-La Roche Ltd, Basel, Switzerland}
\affil[2]{Max Planck Institute for Intelligent Systems, T\"ubingen, Germany}
\affil[3]{ETH Zurich, Switzerland}
\affil[4]{John A. Paulson School of Engineering and Applied Sciences, Harvard University, Cambridge, USA}
\affil[5]{Department of Medicine, Beth Israel Deaconess Medical Center, Harvard Medical School, Boston, USA}
\affil[6]{MIT Critical Data, Laboratory for Computational Physiology, Institute for Medical Engineering and Science, Harvard-MIT Health Sciences \& Technology, Cambridge, USA}
\affil[7]{Department of Medical Oncology and Pneumology, University Hospital of T\"ubingen, Germany}
\affil[8]{Department of Pneumology, Kantonsspital Winterthur, Switzerland}
\affil[9]{CIFAR Azrieli Global Scholar}
\affil[*]{Corresponding author}

\date{} 
\newcommand{\themethod}{CovEWS}
\newcommand{\isep}{\mathrel{{.}\,{.}}\nobreak}
\newcommand{\sig}{\textsuperscript{$\dagger$}\hspace{0.0ex}}

\begin{document}

\maketitle
\thispagestyle{fancy}
\pagestyle{fancy}

\vspace{-4.5em}
\section{Abstract}
Coronavirus Disease 2019 (COVID-19) is an emerging respiratory disease caused by the severe acute respiratory syndrome coronavirus 2 (SARS-CoV-2) with rapid human-to-human transmission and a high case fatality rate particularly in older patients. Due to the exponential growth of infections, many healthcare systems across the world are under pressure to care for increasing amounts of at-risk patients. Given the high number of infected patients, identifying patients with the highest mortality risk early is critical to enable effective intervention and optimal prioritisation of care. Here, we present the COVID-19 Early Warning System (\themethod{}), a clinical risk scoring system for assessing COVID-19 related mortality risk. \themethod{} provides continuous real-time risk scores for individual patients with clinically meaningful predictive performance up to 192 hours (8 days) in advance, and is automatically derived from patients' electronic health records (EHRs) using machine learning. We trained and evaluated \themethod{} using de-identified data from a cohort of \numprint{66430} COVID-19 positive patients seen at over 69 healthcare institutions in the United States (US), Australia, Malaysia and India amounting to an aggregated total of over \numprint{2863} years of patient observation time. On an external test cohort of \numprint{5005} patients, \themethod{} predicts COVID-19 related mortality from $78.8\%$ ($95\%$ confidence interval [CI]: $76.0$, $84.7\%$) to $69.4\%$ ($95\%$ CI: $57.6, 75.2\%$) specificity at a sensitivity greater than $95\%$ between respectively 1 and 192 hours prior to observed mortality events - significantly outperforming existing generic and COVID-19 specific clinical risk scores. \themethod{} could enable clinicians to intervene at an earlier stage, and may therefore help in preventing or mitigating COVID-19 related mortality.

\input{1_main.tex}

\section{Results} 
\input{2_overview.tex}
\input{3_results_survival.tex}

\begin{rotatepage}
\input{4_cohort_table.tex}
\input{4_subcohort_table.tex}
\end{rotatepage}

\section{Discussion} 
\input{5_discussion.tex}

\section{Data availability}
Accredited users may license the TriNetX COVID-19 research and Optum de-identified COVID-19 electronic health record databases used in this study  at TriNetX and Optum, respectively.


\section{Acknowledgements}
SP is supported by the Swiss National Science Foundation under P2BSP2$\_$184359. LAC is funded by the National Institute of Health through NIBIB R01 EB017205. BS is a member of the excellence cluster “Machine Learning in the Sciences” funded by the Deutsche Forschungsgemeinschaft (DFG, German Research Foundation) under Germany’s Excellence Strategy – EXC number 2064/1 – Project number 390727645. We thank Annika Buchholz for helpful discussions. 

\section{Author Contributions}
AM and PS created the new software used in the work. PS, AM, SP, LAC, JH, MH, SB were involved in the conception and design of the work, and the analysis and interpretation of results. JH, LAC, and MH provided the clinical motivation and interpretation. All authors were involved in reviewing, drafting and/or editing of the manuscript. PS and SB supervised the work.

\section{Competing Interests}
PS is an employee and shareholder of F. Hoffmann-La Roche Ltd.

\bibliography{references}
\bibliographystyle{unsrtnat}

\clearpage

\renewcommand{\thesection}{S}
\renewcommand{\thesubsection}{S.\arabic{subsection}}
\renewcommand{\thetable}{S\arabic{table}}
\renewcommand{\thefigure}{S\arabic{figure}}
\section{Methods}
\input{6_app_method.tex}

\input{7_performance_tables.tex}

\end{document}

%% file: 1_main.tex
\section{Introduction}
The coronavirus disease 2019 (COVID-19) pandemic has recently emerged as a major and urgent threat to healthcare systems worldwide. Since early reports of its outbreak in China in December 2019, the number of global cases has risen to over 21 million known infections and resulted in over \numprint{750000} deaths worldwide as of August 16, 2020 \citep{world2020coronavirus}. Despite public health efforts aimed at improving testing \cite{yan2020laboratory}, developing potential vaccines \cite{lurie2020}, and improving prevention strategies \cite{hellewell2020feasibility}, the disease is placing a significant burden on healthcare systems and existing resources in many countries, particularly where its spread has not been mitigated. Efficient early detection of patients likely to develop critical illness is thus crucial to optimise the allocation of limited resources, and monitor overall disease progression \cite{emanuel2020fair,fauci2020}. The use of clinical predictive models from electronic health records (EHRs) can help reduce some of this burden and inform better decisions overall \cite{henry2015targeted, wang2020response, green2018comparison, churpek2017investigating}. For instance, a model able to predict in advance which patients are at higher risk of mortality may help ensure resources are prioritized accordingly for these individuals. In addition, as more observational data are gathered, these models could be used both to discover new risk factors as well as reveal interactions between existing factors, offering better insights and opportunities for appropriate intervention.

Several approaches have been proposed to determine potential risk factors that contribute to COVID-19 mortality. Some of these approaches identify demographics and inflammatory markers associated with increased mortality \cite{Petrillim1966, Ji2020}, but do not account for risk factors potentially changing over time. Moreover, many existing analyses are limited to a single source of data, often from a single hospital, for both learning a model and predicting a patient's risk which may limit the generalisability of these analyses \cite{wynants2020prediction}. Other more traditional measures of patient prognosis such as Sequential Organ Failure Assessment (SOFA) scores \citep{vincent1996sofa} are based on examining a fixed set of risk factors not specifically adapted to COVID-19; such measures fail to account for relevant changes in patient status outside these risk factors, and therefore often do not reach high levels of sensitivity and specificity in identifying high-risk patients. Due to these challenges, to date, there does not yet exist a COVID-19 risk score that (i) makes use of multiple, representative sources of data to account for patient heterogeneity, (ii) includes important short-term and long-term risk factors that have a significant impact on mortality risk, (iii) reacts in real time to potentially rapid changes in patient status, and (iv) is adapted to risk factors relevant to COVID-19. 

To address these issues, we developed the COVID-19 Early Warning System (\themethod{}), a risk assessment system for real-time prediction of COVID-19-related mortality that we trained on a large and representative sample of EHRs collected from more than 69 healthcare institutions using machine learning. In contrast to existing risk scores, \themethod{} provides early warnings with clinically meaningful predictive performance up to 192 hours prior to observed mortality events, hence enabling critical time to intervene to potentially prevent such events from occurring. Since \themethod{} is automatically derived from patient EHRs, it updates in real time without any necessity for manual action to reflect changes in patient status, and accounts for a much larger number of risk factors correlated with COVID-19 mortality than existing risk scores. \themethod{} is based on a time-varying neural Cox model that accounts for risk factors changing over time and potential non-linear interactions between risk factors and COVID-19-related mortality risk\footnote{While these extensions have been pursued in~\cite{zhang2018time} and~\cite{ching2018cox} separately, they have neither been considered in combination nor in the context of COVID-19 risk scoring using EHRs.}, and was derived from the de-identified EHRs of \numprint{66430} diverse COVID-19 positive patients. We demonstrate experimentally that the predictive performance of \themethod{} is superior to existing generic risk scores, such as SOFA \citep{vincent1996sofa}, COVID-19 specific risk scores, such as the machine learning models from \citet{yan2020interpretable} and \citet{liang2020early}, and COVER\_F \citep{williams2020seek}, and a time-varying Cox model with linear interactions \citep{cameron2020lifelines}. We additionally show that the gradient information of our differentiable \themethod{} model can be used to quantify the influence of the input risk factors on the output score in real time. \themethod{} may enable clinicians to identify high-risk patients at an early stage, and may, therefore, help improve patient outcomes through earlier intervention.

%% file: 2_overview.tex
\begin{figure}[ht!]
    \begin{subfigure}{.49\textwidth}
    \includegraphics[width=\linewidth,page=1]{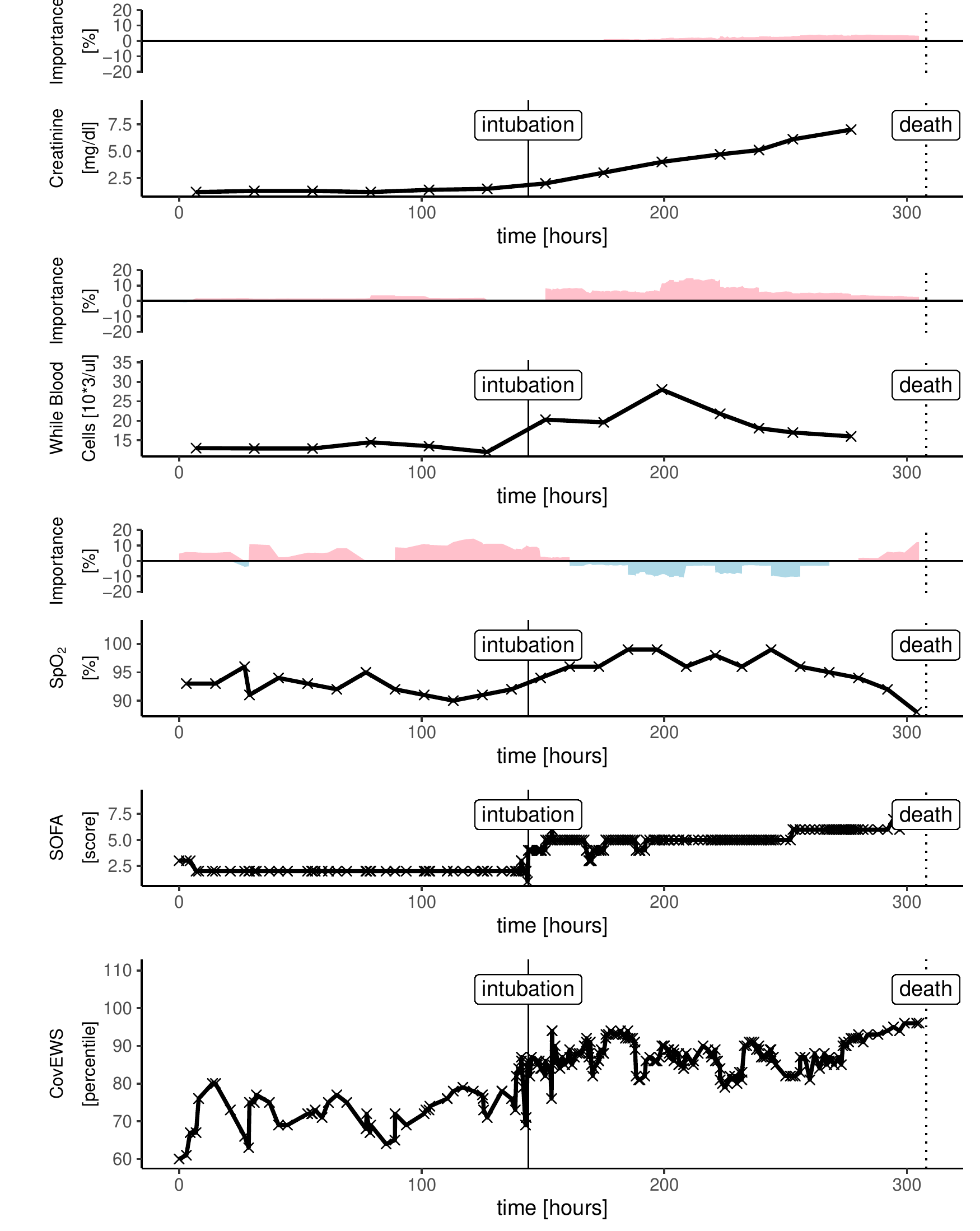}
    \caption{Patient A}
    \end{subfigure}\quad
    \begin{subfigure}{.49\textwidth}
    \includegraphics[width=\linewidth,page=1]{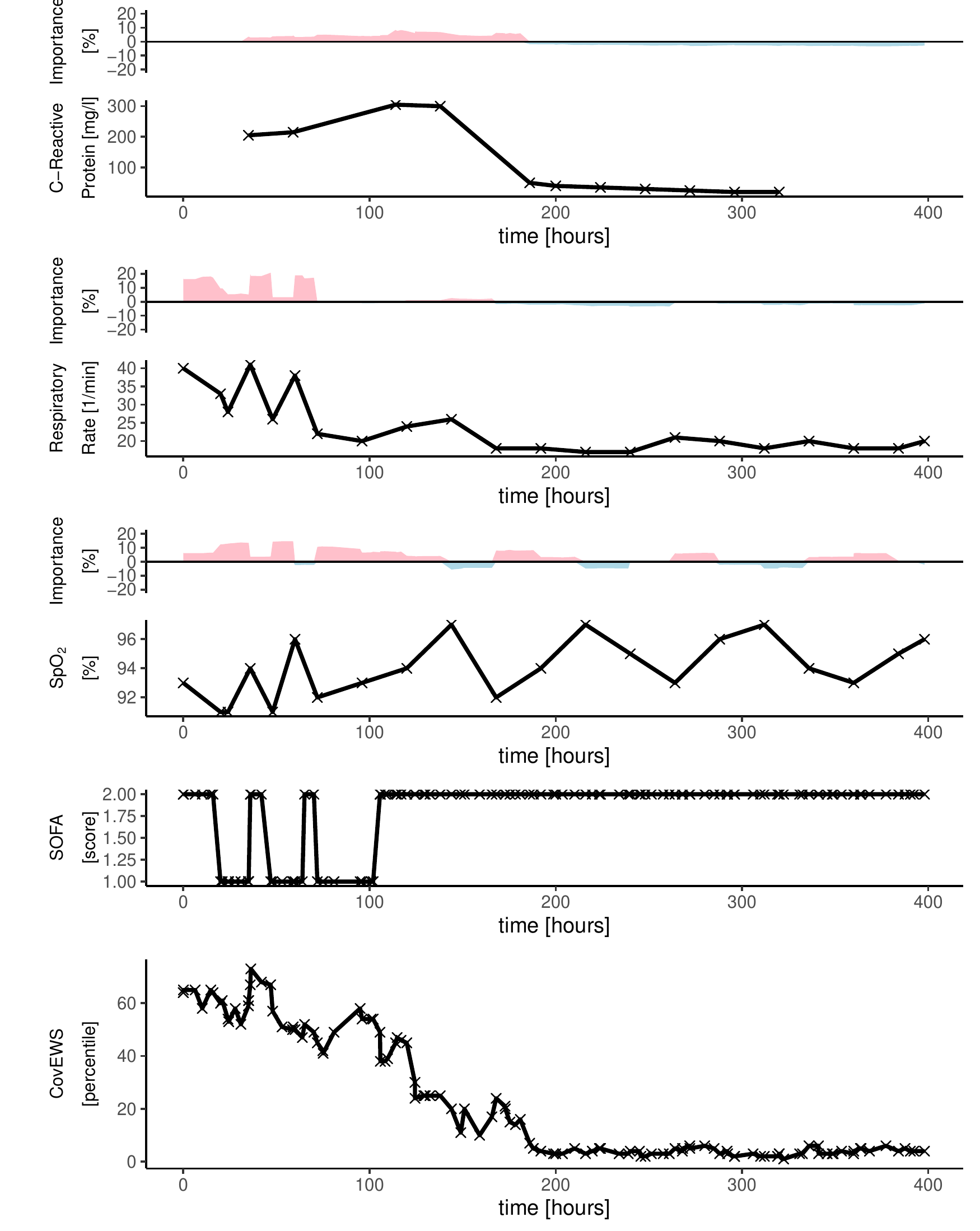}
    \caption{Patient B}
    \end{subfigure}
    \caption{A selected number of clinical risk factors, corresponding Sequential Organ Failure Assessment (SOFA) scores, and COVID-19 Early Warning System (\themethod{}) scores for two contrasting patient timelines. Positive (red) and negative (blue) importance contributions (coloured areas above the clinical time series, see \Cref{sec:feature_importance}) indicate to what degree the risk factor at that time point contributed to increasing or decreasing to the mortality risk predicted by \themethod{}, respectively. \textbf{Patient A's} (left) oxygen saturation (SPO$_2$) fluctuates significantly before dropping below 95\% after around 150 hours since her COVID-19 diagnosis, suggesting respiratory distress. The patient is subsequently intubated. This is followed by a sharp rise in serum creatinine levels, indicating potential acute kidney injury. Both SOFA and \themethod{} reflect these events with an increase in Patient A's risk. Crucially, however, since \themethod{} accounts for early deterioration in SPO$_2$ and white blood cell counts, it identifies the patient as high-risk much sooner than SOFA, triggering re-evaluation of current treatment strategy, including investigation for delayed complication or treatment injury, and/or the initiation of goals of care discussion. In \textbf{Patient B} (right), different risk factors, including c-reactive protein (CRP), respiratory rate (RR) and SPO$_2$, weigh heavily in risk assessment. Initially, Patient B's RR increases significantly to over 30 breaths per minute while her SPO$_2$ drops below 95\%, reflected by a corresponding increase in both SOFA and \themethod{}. Patient B's RR and CRP levels however stabilise, which is correctly reflected in a lowering of the mortality risk by \themethod{}. Intubation is averted for this patient. In contrast, SOFA does not account for the improvements in SPO$_2$, RR and does not reflect Patient B's now improved state.}
    \label{fig:timeline_1} 
    \end{figure}
    
    \paragraph{COVID-19 Early Warning System (\themethod{}).} \themethod{} is a clinical mortality risk prediction system for COVID-19 positive patients to be used in a continuous manner in both inpatient and outpatient settings. \themethod{} uses clinical risk factors from a patient's EHR to automatically calculate a mortality risk score between 0 and 100 that indicates the current risk percentile that this patient is in relative to the reference cohort\footnote{See \Cref{sec:method} for a mathematical definition of \themethod{}.}. A \themethod{} score of 90 indicates, for example, that the patient has a higher COVID-19 related mortality risk than 90\% of COVID-19 positive patients in the reference cohort. An important property of \themethod{} scores is that they always reflect the momentary risk of patients in their current states, and that they update instantaneously to reflect relevant, EHR-derived changes, which is a key differentiator of \themethod{} compared to existing COVID-19 related mortality risk prediction systems that are not designed to take into account new, incoming clinical evidence. \Cref{fig:timeline_1} demonstrates the application of \themethod{} to two contrasting patient timelines (a deteriorating patient that ultimately died and a patient that initially deteriorates but then recovers) by visualising a selected number of clinical risk factors, such as respiratory rate, oxygen saturation, and creatinine levels, alongside the corresponding momentary risk assessment output by \themethod{}. As shown in \Cref{fig:timeline_1}, \themethod{} additionally maintains a high degree of interpretability for clinicians by indicating the relative positive and negative influences of each clinical risk factor over time on the predicted risk score (see \Cref{sec:feature_importance}). The information conveyed by \themethod{} can be used to quickly and objectively assess individual COVID-19 related mortality risk in order to prevent or mitigate mortality, and optimise prioritisation of scarce healthcare resources.
    
    To develop \themethod{}, we used EHR data from two federated networks of US and international healthcare organisations (HCOs), Optum (US) and TriNetX (US + international), that include de-identified EHRs containing data on demographics, clinical measurements, vital signs, lab tests and diagnoses of \numprint{47384} and \numprint{5005} patients seen between March 21$^\text{st}$ and June 5$^\text{th}$ 2020 (11 weeks) and March 21$^\text{st}$ and June 25$^\text{st}$ 2020 (13 weeks), respectively. To demonstrate the generalisability of predictions made by \themethod{}, we limited the training of \themethod{} to a training cohort of \numprint{14215} (30\%) patients from the Optum cohort, used \numprint{9477} (20\%) Optum patients for model selection, and evaluated \themethod{} against both a held-out test cohort of \numprint{14215} (30\%) patients from the Optum cohort and a separate external test cohort consisting of the entire TriNetX cohort of  \numprint{5005} (100\%) patients (\Cref{tbl:cohort_description}, stratification details in \Cref{sec:stratification}). In addition, we collected supplementary EHR data on new patients diagnosed with COVID-19 between June 6$^\text{th}$ to July 13$^\text{th}$ 2020 (5 weeks) from Optum - the Optum future cohort (\numprint{14041} patients) - after \themethod{} had been trained to demonstrate the robustness of \themethod{} under rapidly changing treatment regimes\footnote{During this period, the RECOVERY Collaborative Group reported results of randomised clinical trials demonstrating the lack of efficacy of hydroxychloroquine \cite{horby2020hydroxychloroquine} and the efficacy of dexamethasone \cite{recovery2020dexamethasone} in COVID-19 patients on June 5th 2020 and June 16th, respectively - which significantly impacted clinical treatment practice of COVID-19 patients.} and other temporal effects. The data formats were normalised across the two federated networks of HCOs (\Cref{sec:data_normalisation}), and all data were preprocessed to address the missingness that is characteristic for real-world clinical data (\Cref{sec:preprocessing}).

%% file: 3_results_survival.tex
\begin{figure}
    \includegraphics[width=0.419\linewidth,page=1]{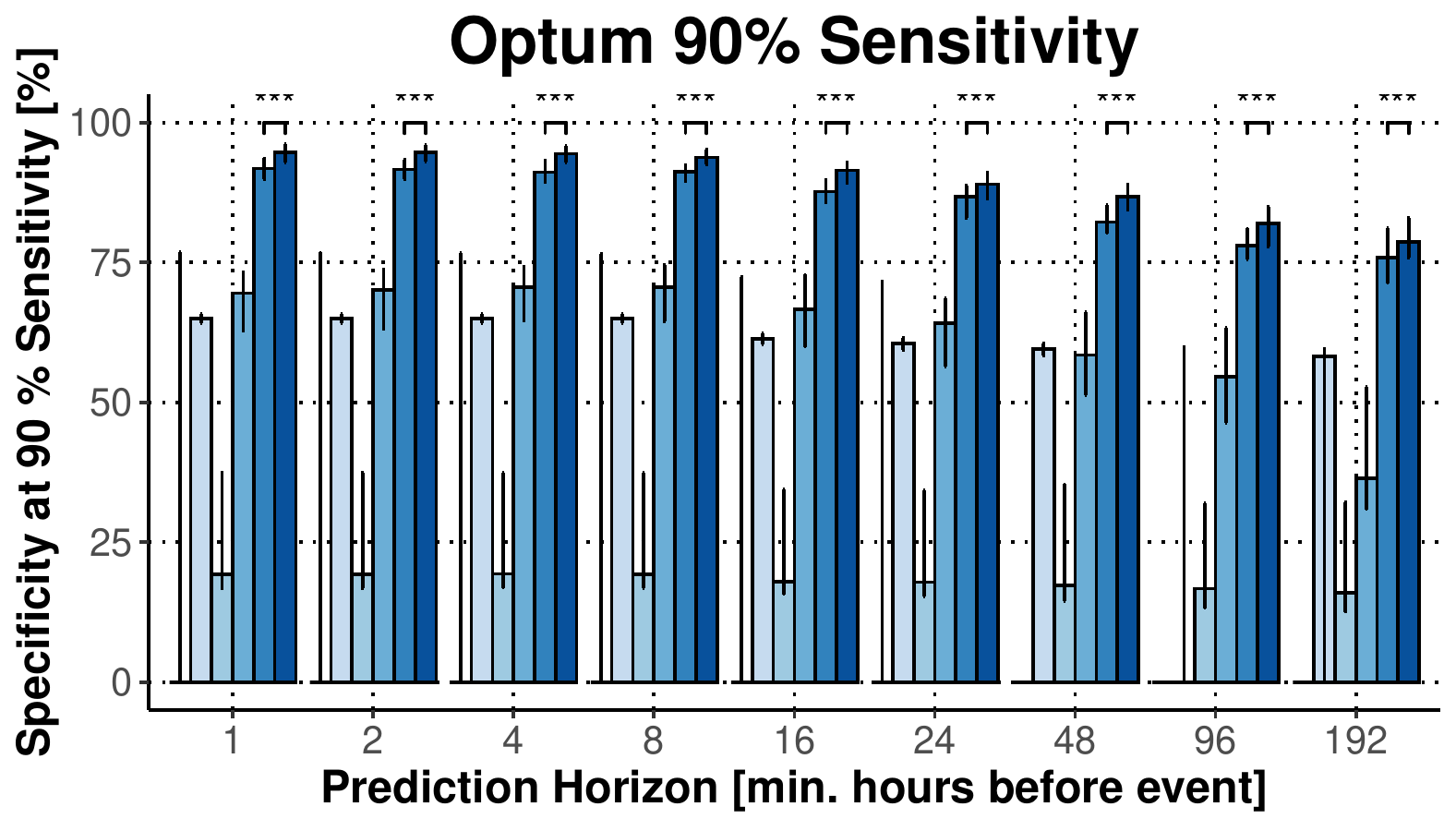}\quad
    \includegraphics[width=0.525\linewidth,page=1]{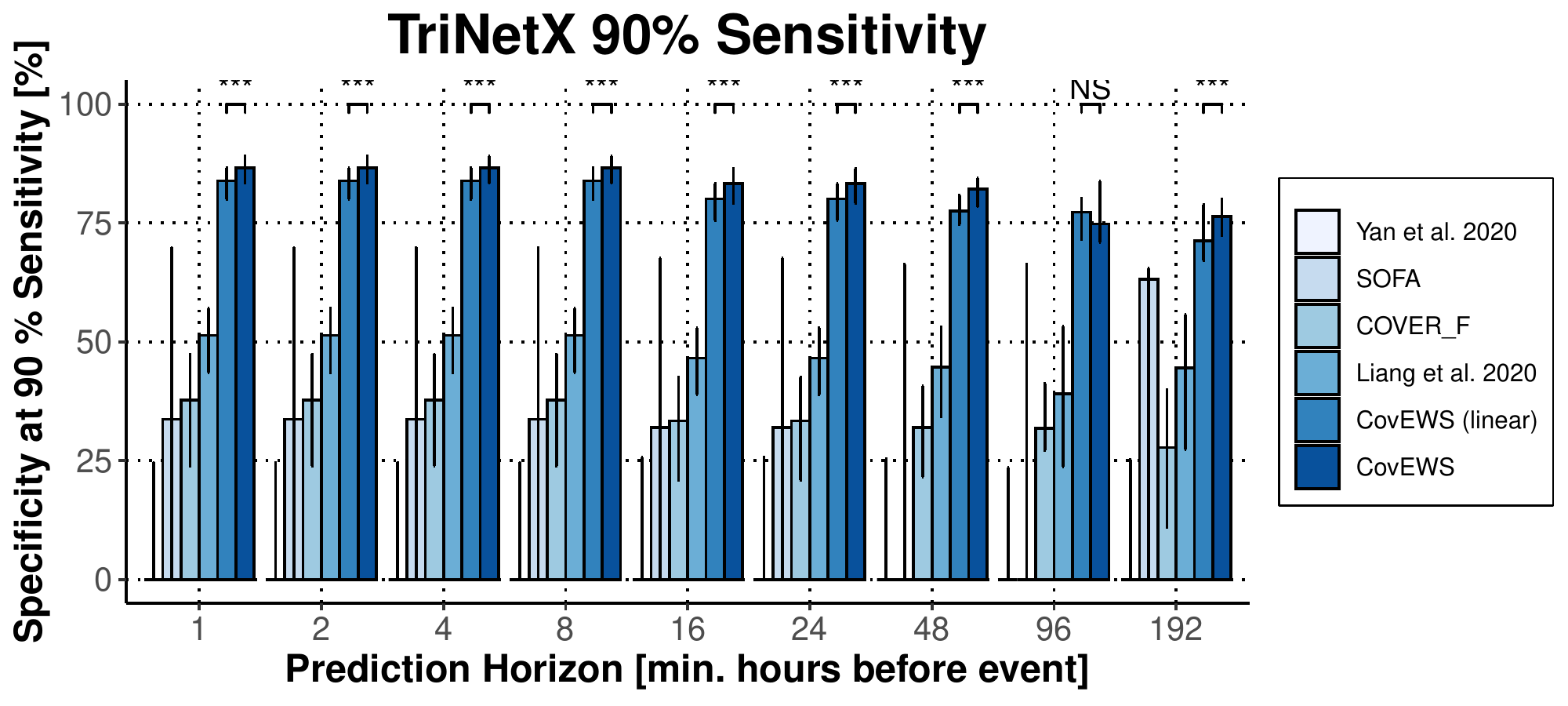}\quad
    \includegraphics[width=0.419\linewidth,page=1]{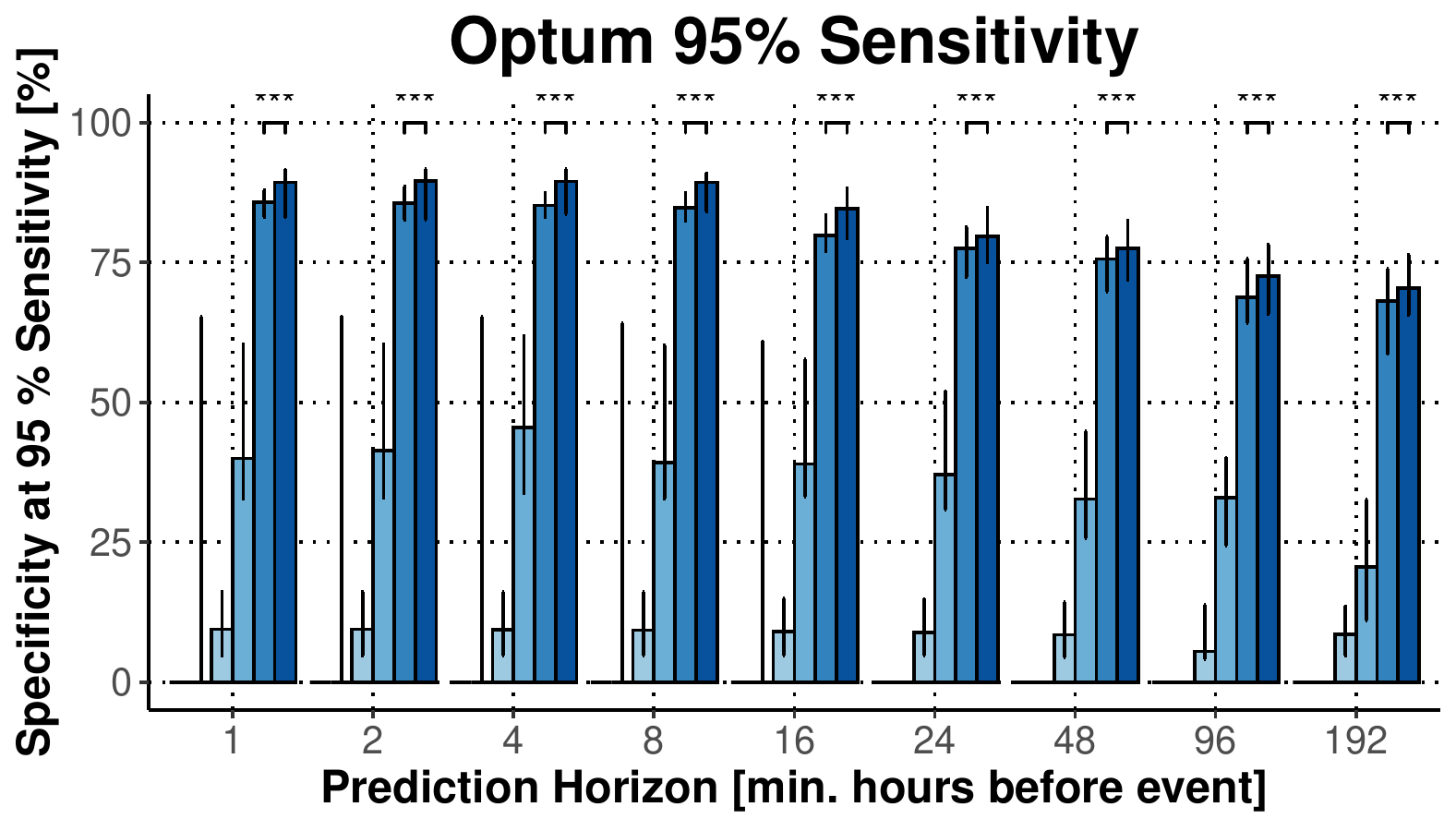}\quad
    \includegraphics[width=0.419\linewidth,page=1]{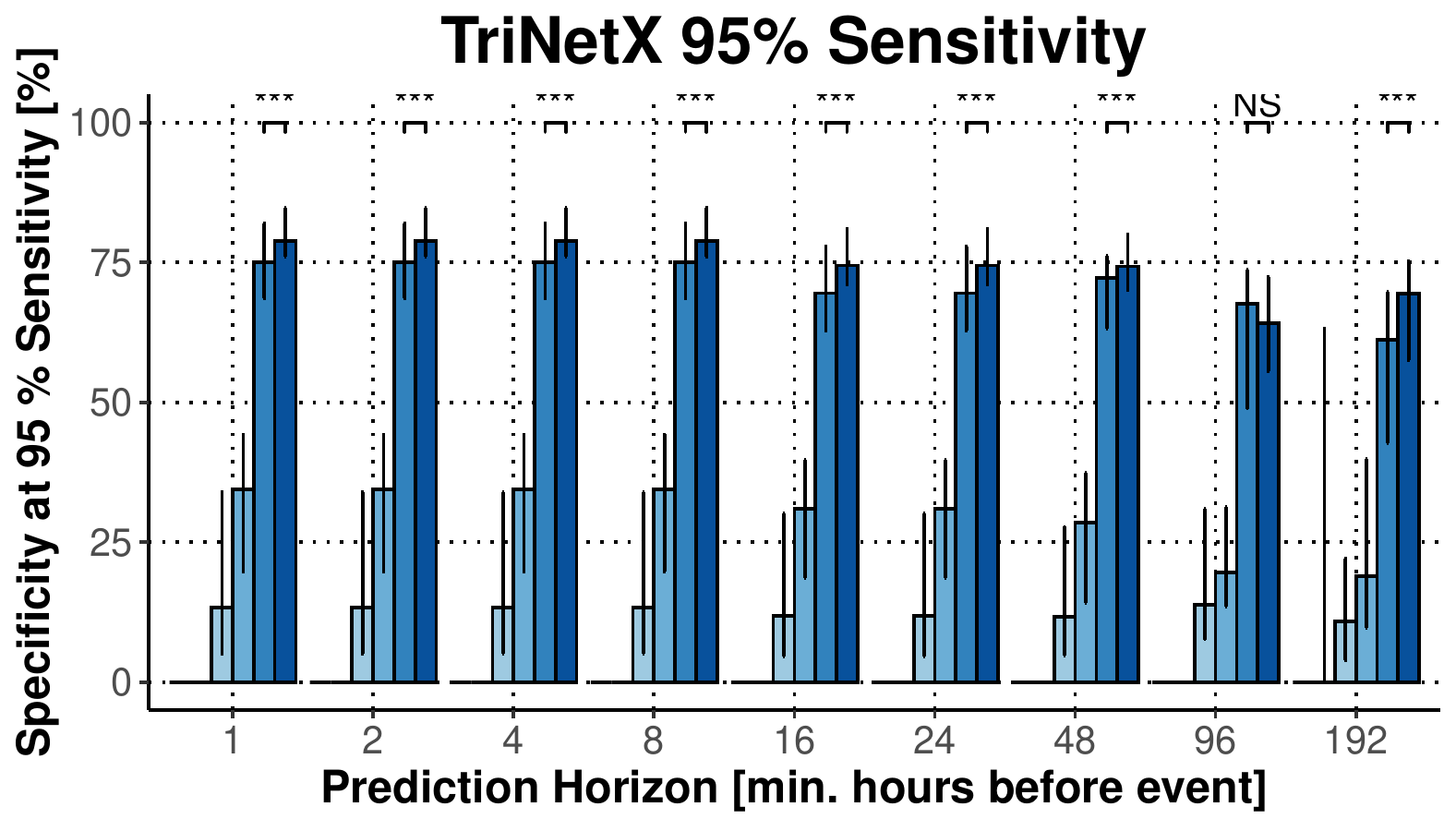}\quad
    \includegraphics[width=0.419\linewidth,page=1]{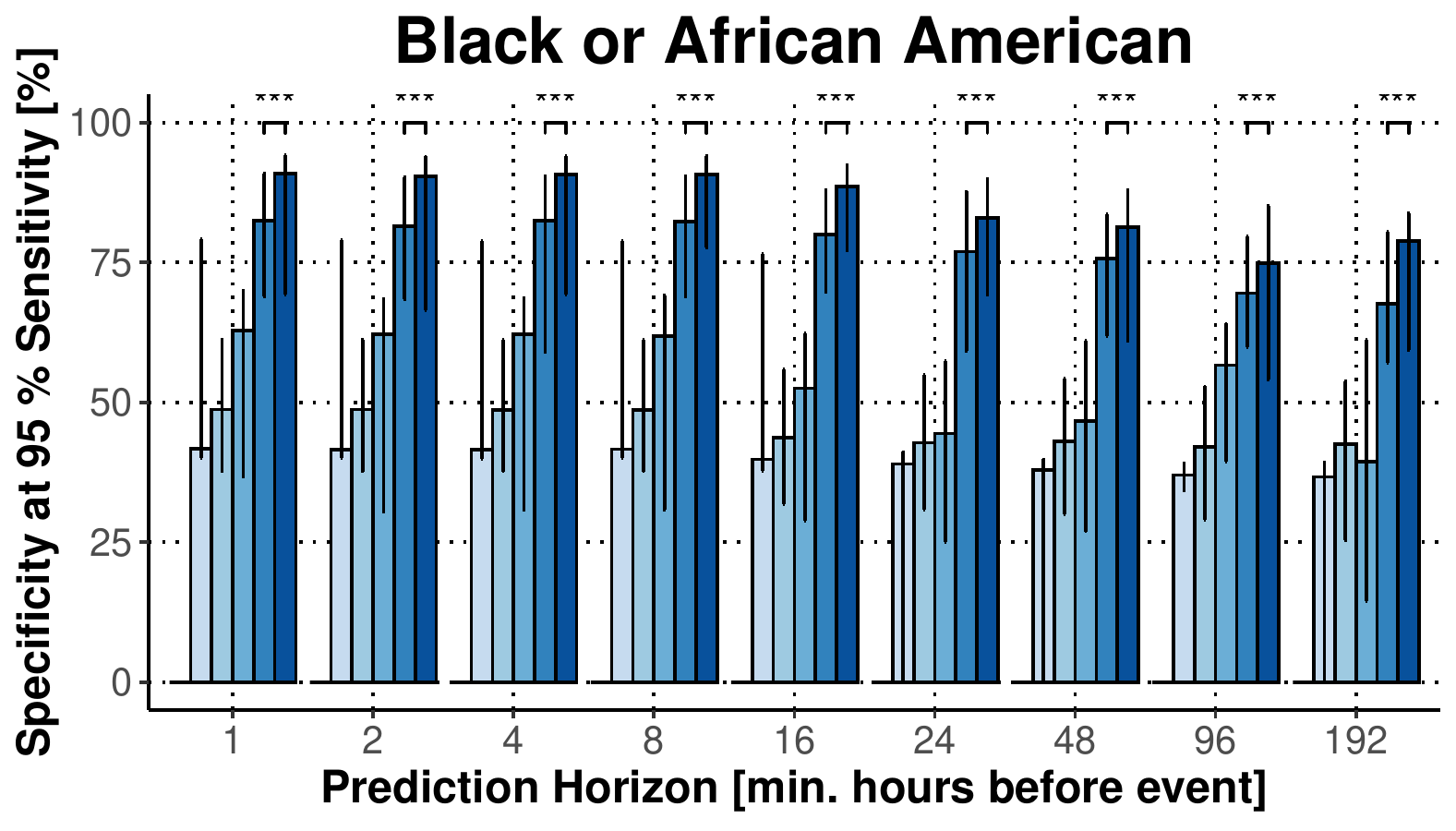}\quad
    \includegraphics[width=0.419\linewidth,page=1]{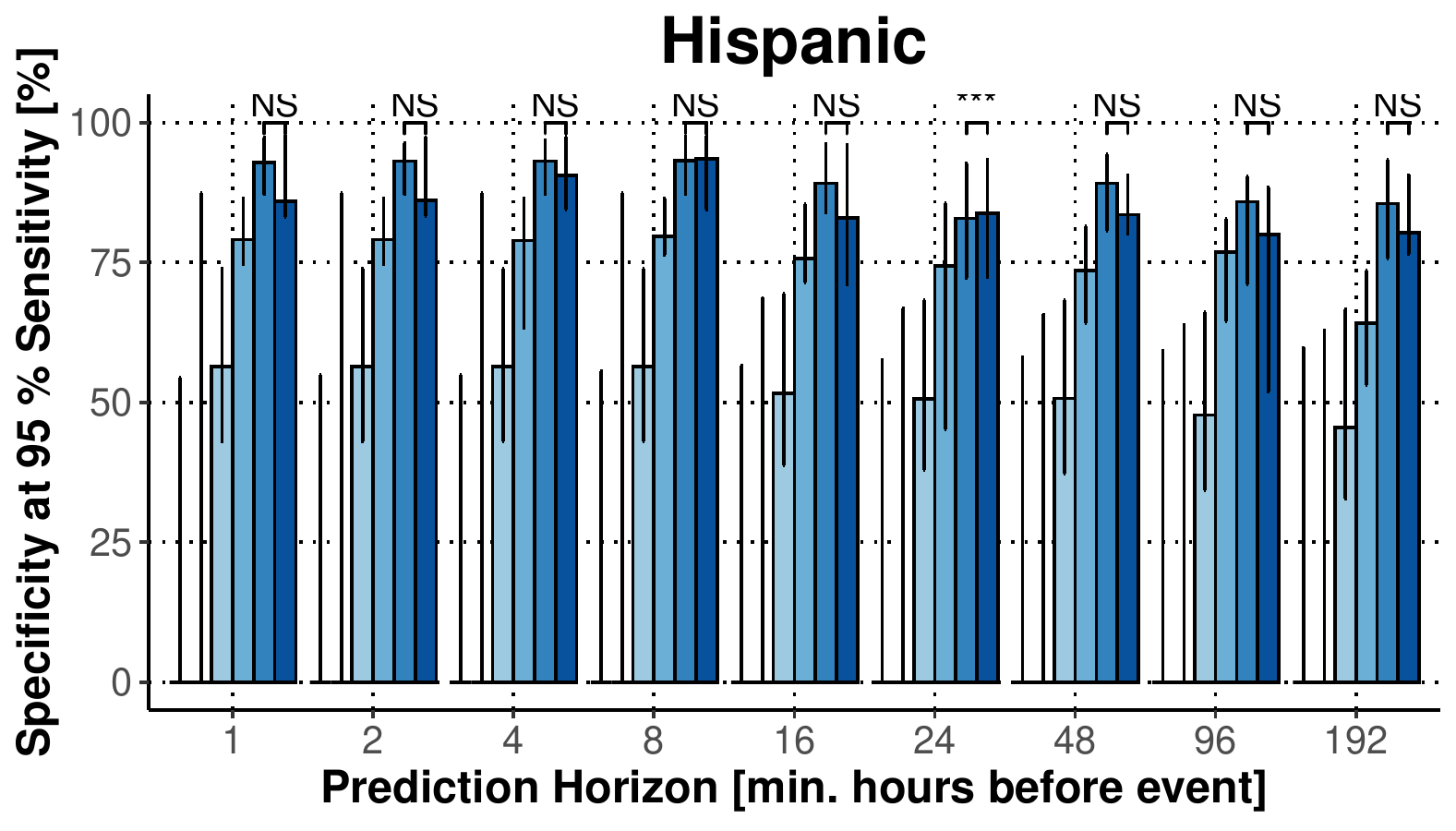}\quad 
    \includegraphics[width=0.419\linewidth,page=1]{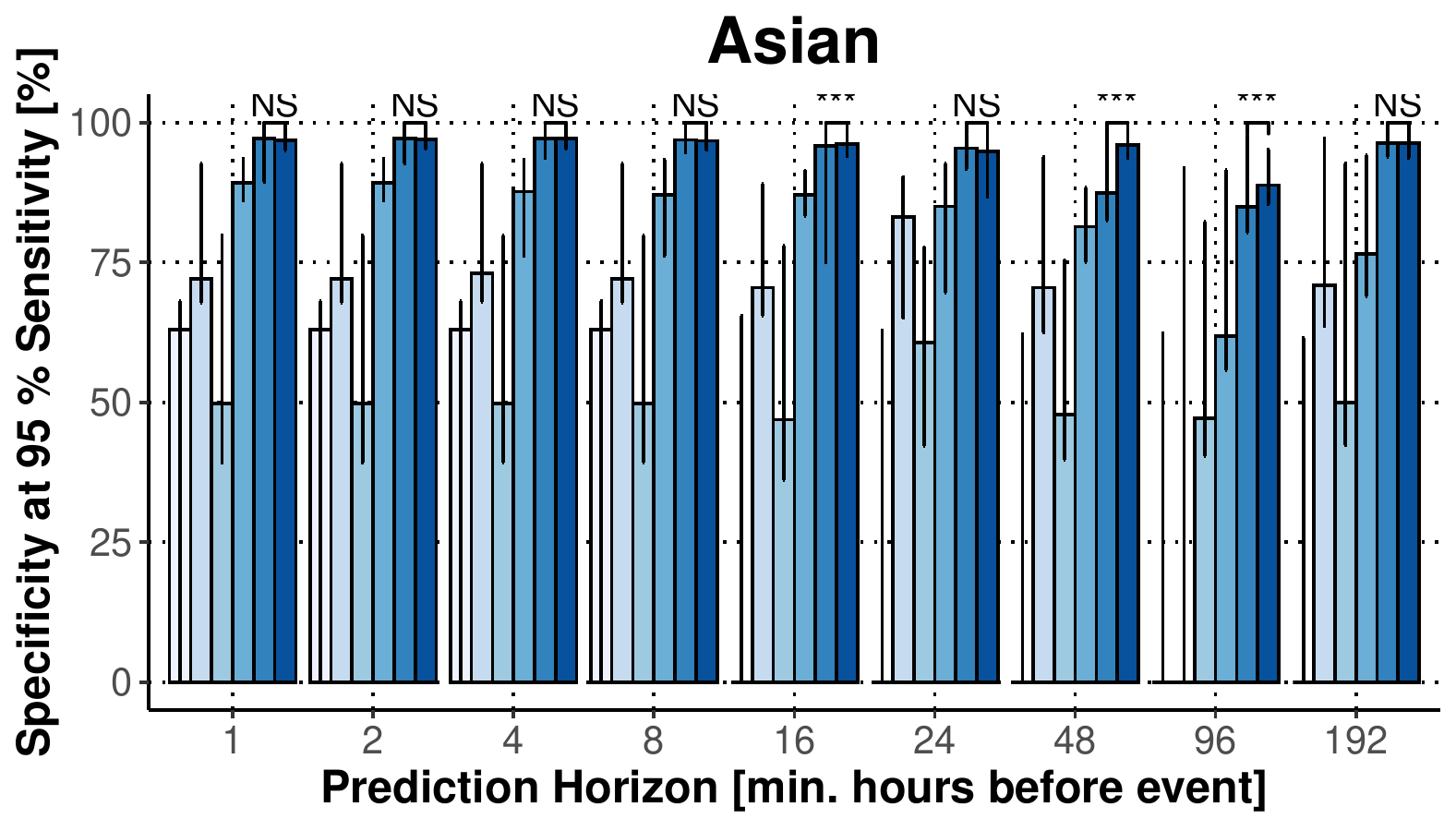}\quad
    \includegraphics[width=0.419\linewidth,page=1]{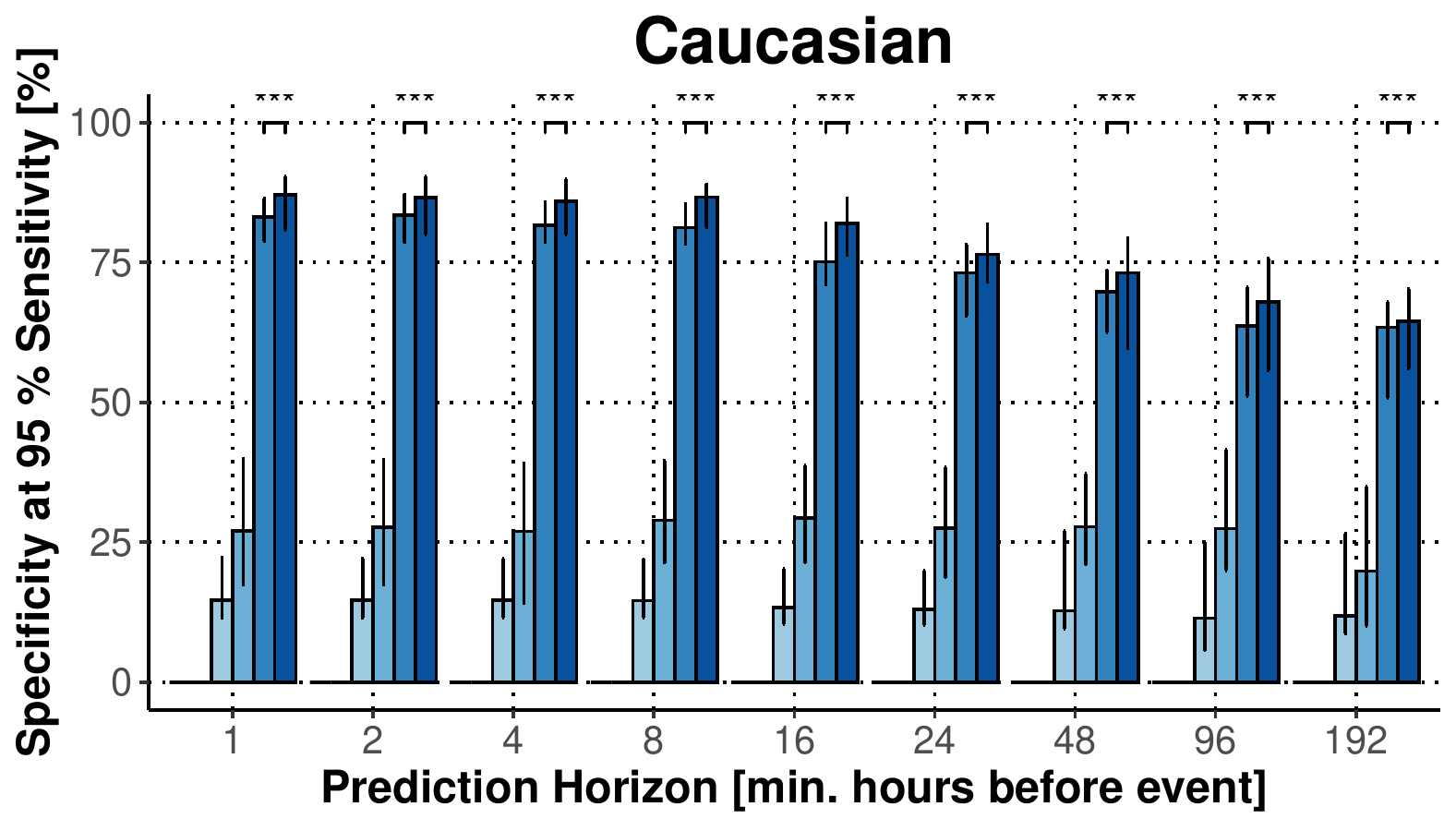}\quad
    \includegraphics[width=0.419\linewidth,page=1]{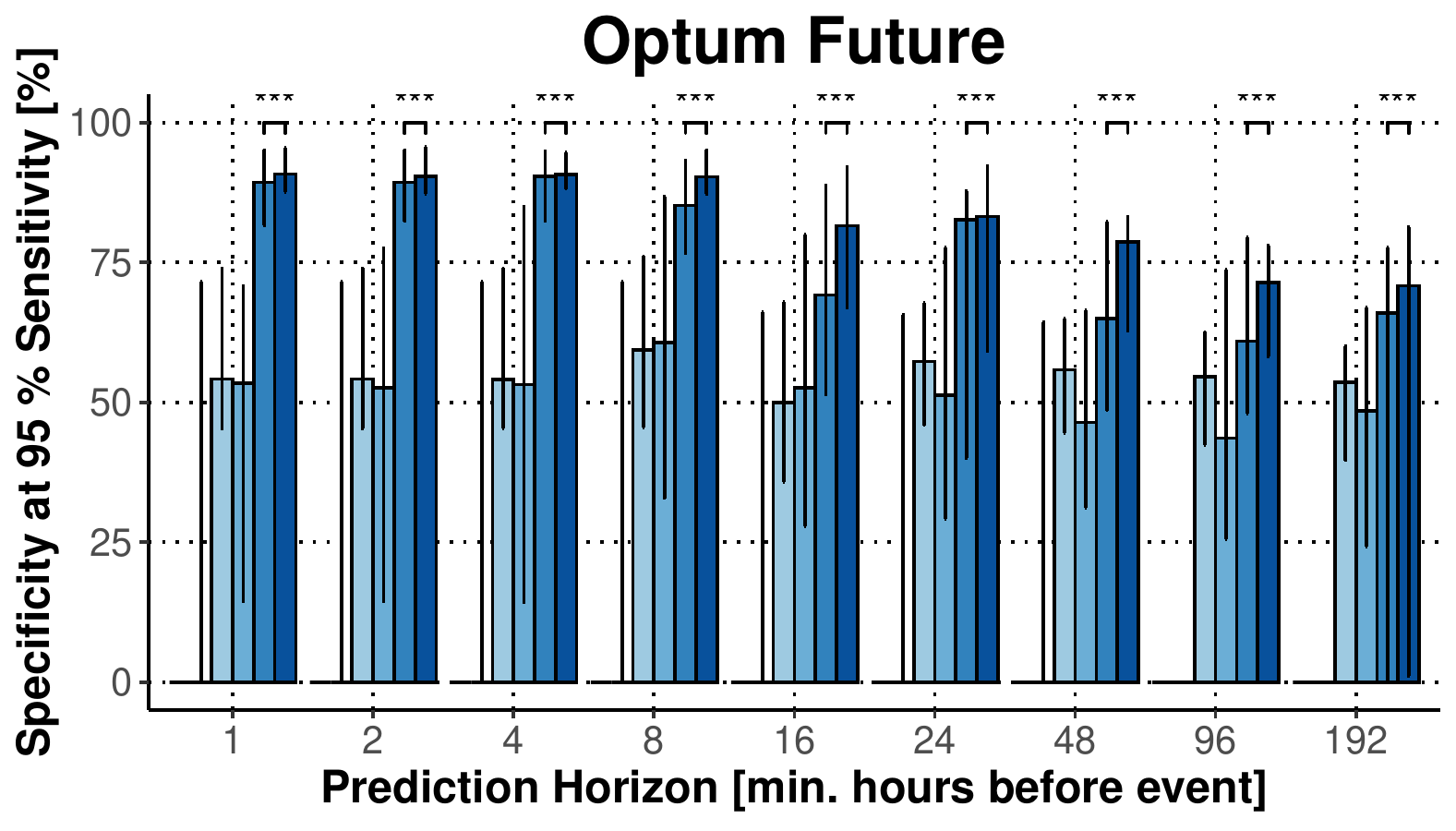}\quad
    \includegraphics[width=0.419\linewidth,page=1]{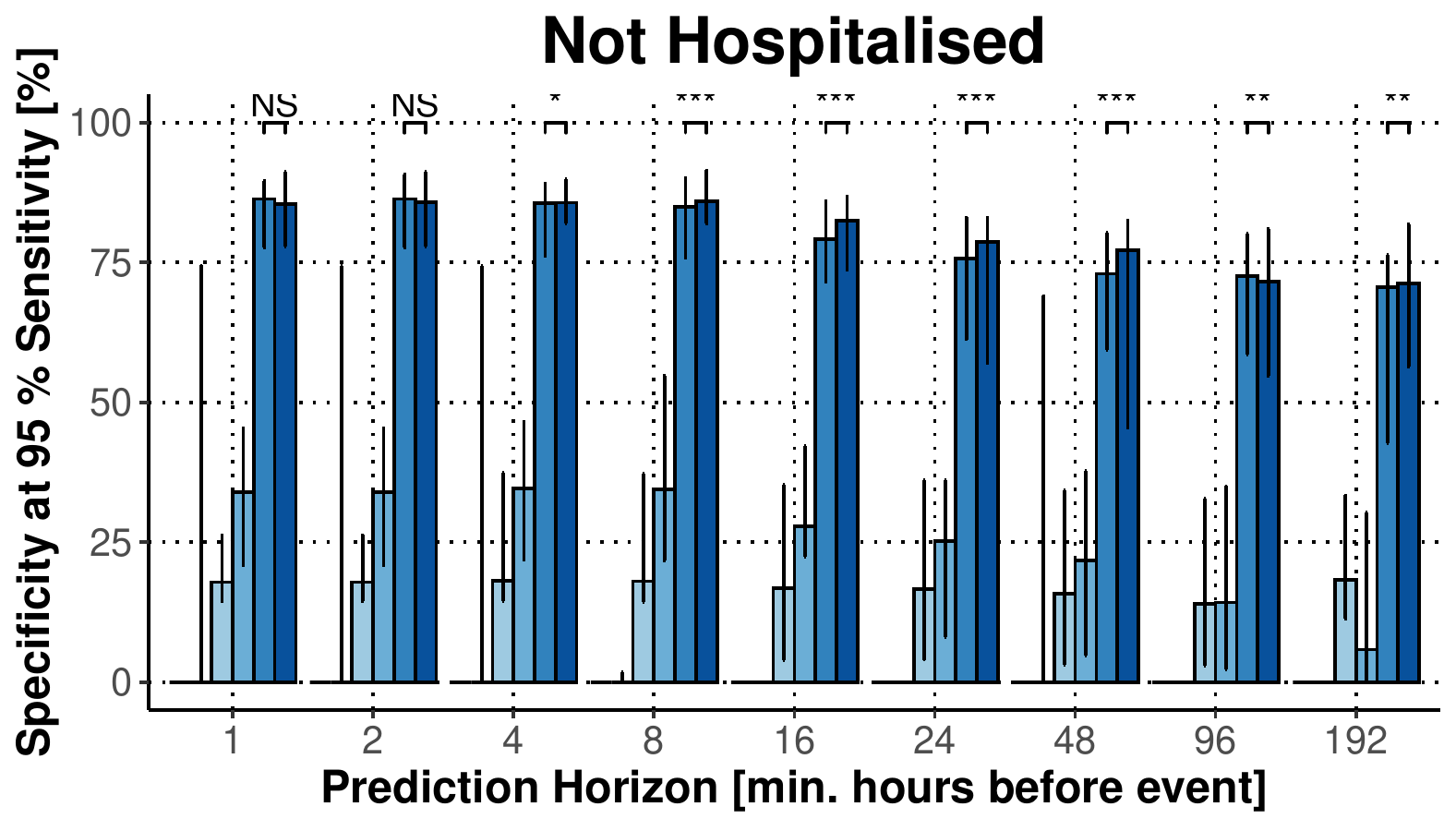}\quad
    \caption{Performance comparison in terms of Specificity at greater than either $95\%$ or $90\%$ Sensitivity (y-axis) for different prediction horizons ahead of observed mortality events (in hours, x-axis) for \themethod{}, \themethod{} (linear), COVID-19 Estimated Risk for Fatality (COVER\_F) \citep{williams2020seek}, Sequential Organ Failure Assessment (SOFA) \citep{vincent1996sofa}, \citet{liang2020early}, and \citet{yan2020interpretable} on the held-out Optum test set, the external TriNetX test set, and selected patient subgroups from the Optum test set. Some methods do not reach $90\%$ and $95\%$ sensitivity for some horizons, and may therefore not be visible in all plots. Bars indicate median and error bars indicate 95\% confidence intervals (CIs) obtained via bootstrapping with 200 samples. Detailled results are available in \Cref{sec:performance}. (* = $p < 0.10$, ** = $p < 0.05$, *** = $p < 0.01$, NS = not significant, one-sided Mann-Whitney-Wilcoxon for superiority of \themethod{} over \themethod{} [linear]).}
    \label{fig:baselines}
    \end{figure}
    
    \paragraph{Predictive Performance for Different Prediction Horizons.} We compared the predictive performance of \themethod{}, several baselines and existing risk prediction scores (\Cref{sec:baselines}), including a version of \themethod{} based on a linear time-varying Cox model \citep{cameron2020lifelines} (\themethod{} [linear], \Cref{sec:cox_proportional_hazard_with_time_varying_covariates}), COVID-19 Estimated Risk for Fatality (COVER\_F) \citep{williams2020seek}, Sequential Organ Failure Assessment (SOFA) \citep{vincent1996sofa}, the decision tree developed by \citet{yan2020interpretable} and the deep learning model developed by \citet{liang2020early}, in terms of their respective specificity for identifying COVID-19 related mortality with a conservative fixed sensitivity of at least $95\%$ and a slightly more relaxed level of $90\%$ at a minimum of 1, 2, 4, 8, 16, 24, 48, 96 and 192 hours (8 days) prior to observed mortality events\footnote{The last observed EHR entry's date was taken as a reference time for those patients that did not have an observed mortality event during the data collection period.} on both the hold-out test data of Optum cohort and the external test cohort from the TriNetX network (\Cref{fig:baselines}). In terms of specificity at a sensitivity greater than $95\%$, we found that \themethod{} significantly ($p<0.05$, one-sided Mann-Whitney-Wilcoxon with Bonferroni correction, see \Cref{tbl:results_all_optum} and \Cref{tbl:results_all_trinetx}) outperformed other baselines and existing risk prediction scores at each prediction horizon and on both the Optum and TriNetX cohorts with few exceptions. By comparing the predictive performances of the mortality prediction scores at different time horizons, we additionally quantified the degree to which risk prediction methods give more accurate predictions when the mortality event is closer to the prediction date. For example, the predictive performance of \themethod{} in terms of specificity at a sensitivity greater than $95\%$ dropped from $89.3\%$ ($95\%$ confidence interval [CI]: $83.0$, $91.6\%$) to $70.5\%$ ($95\%$ CI: $65.6, 76.4\%$) and from $78.8\%$ ($95\%$ CI: $76.0$, $84.7\%$) to $69.4\%$ ($95\%$ CI: $57.6, 75.2\%$) from 1 hour to 192 hours prior to an observed mortality event on the held-out Optum test cohort and the external TriNetX test cohort, respectively. When comparing the predictive performance across the held-out Optum test cohort and the external TriNetX test cohort, we saw the same trends in performance. However, all methods were roughly $10\%$ less specific at greater than $95\%$ sensitivity. This difference persisted even in those risk assessment systems that were not originally trained on the Optum training cohort, such as COVER\_F. We thus attributed this apparent difference in performance not to overfitting to the Optum training cohort, but to (i) the difference of $5.38\%$ against $6.91\%$ in baseline mortality between the held-out Optum test cohort and the external TriNetX test cohort, respectively, and (ii) the higher degree of missingness in short-term mortality risk factors, such as, e.g., respiratory rate, SpO$_2$ and blood pressure, in TriNetX (\Cref{tbl:cohort_description}). In addition to assessing predictive performance, we also evaluated the calibration \citep{van2019calibration} of the risk scores predicted by \themethod{}. We found that \themethod{} overestimates mortality risk when interpreted as the probability of a mortality event occurring within the next 24 hours because patients' states may change between the prediction time and the end of the prediction horizon (\Cref{fig:calibration_curves}).
    
    \paragraph{Predictive Performance for Different Subgroups.} We also compared the predictive performance of \themethod{} against the baselines and existing scores across various ethnic subgroups, on patients that were not hospitalised, and on the Optum future cohort (\Cref{fig:baselines}; cohort statistics in \Cref{tbl:subcohort_description}). Overall, across each of these cohorts, we found that \themethod{} significantly ($p<0.05$, one-sided Mann-Whitney-Wilcoxon with Bonferroni correction) outperformed all of the baselines at each prediction horizon with the sole exception being the 96 and 192 hours prediction horizons on the Optum future cohort - where the performance difference was not in all cases significant. The performance difference was more pronounced across Caucasian and African American populations which is likely reflective of the fact that several baselines have been developed using data from predominantly Asian populations. On the subgroup of patients that was not hospitalised, we found that, although lower than the overall performance on the entire Optum test set, \themethod{} maintained a high level of performance. We attributed the lower performance on the non-hospitalised group compared to the overall Optum test set to (i) the considerably higher missingness in this patient group caused by non-hospitalised patients not being monitored as closely as hospitalised patients (\Cref{tbl:missingness_hospitalised}), and (ii) the overall considerably lower mortality rate in this patient group. Respectable performance on the non-hospitalised patient group is particularly important since the majority of COVID-19 patients are treated in an outpatient setting. In addition, when evaluating the various risk assessment methods on the Optum future cohort, we found that \themethod{} was largely robust to changes in treatment policies and other temporal effects. A notable anomaly was the 96 and 192 hours prediction horizons where the variance in our performance estimates was relatively high since fewer patients with recorded mortality outcomes and long-term monitoring data were available due to the shorter data collection time (5 weeks) of the Optum future cohort compared to the Optum test set (11 weeks) and the TriNetX test set (13 weeks).
    
    \begin{figure}
    \begin{subfigure}{.49\textwidth}
    \centerline{\includegraphics[width=1.00\linewidth,page=1]{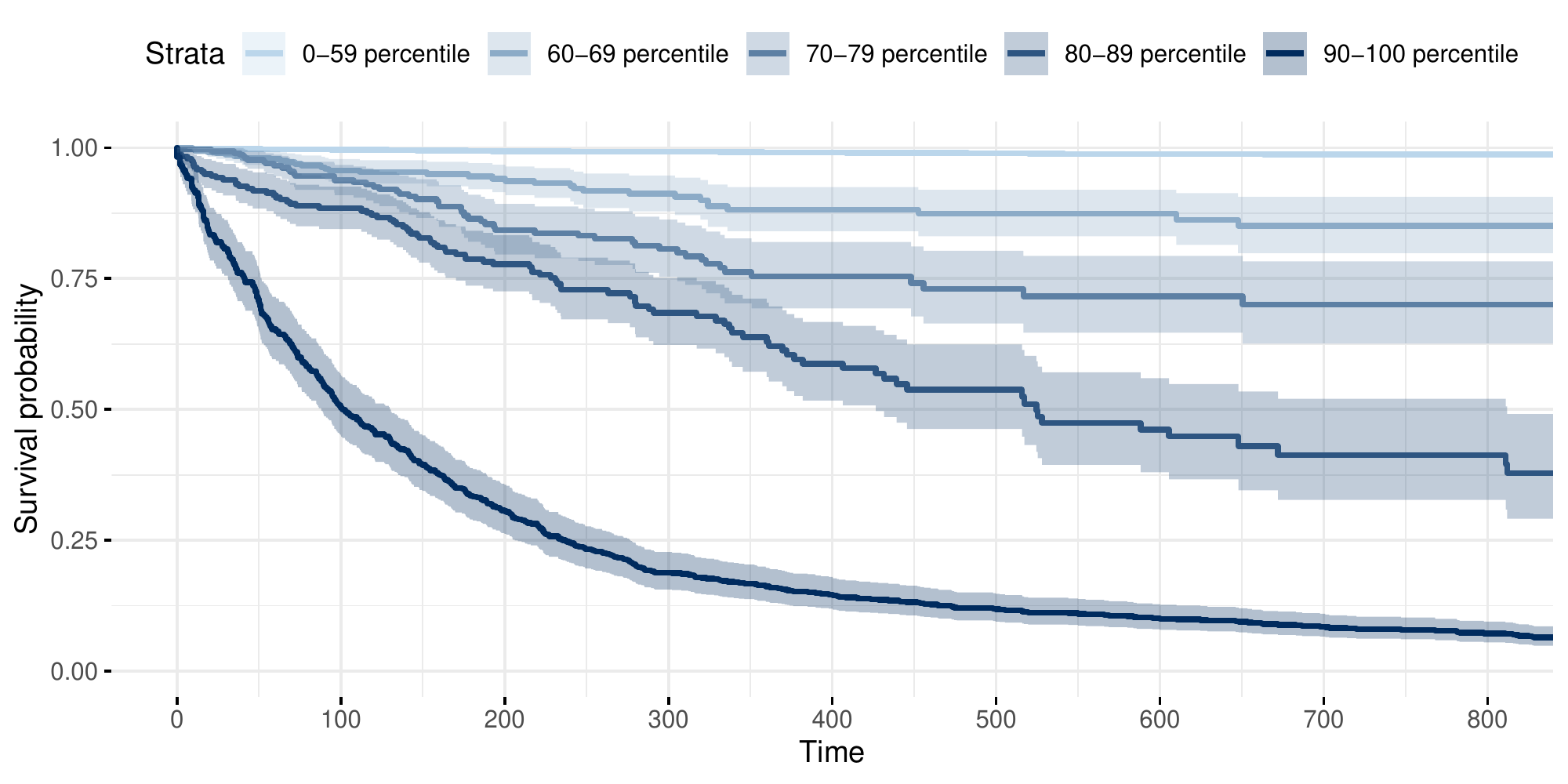}}
    \centerline{\includegraphics[width=1.00\linewidth,page=2]{figs/test_survival_strata.pdf}}
    \centerline{\includegraphics[width=1.00\linewidth,page=3]{figs/test_survival_strata.pdf}}
    \caption{Held-out Optum Test Set} 
    \end{subfigure}\quad
    \begin{subfigure}{.49\textwidth}
    \centerline{\includegraphics[width=1.00\linewidth,page=1]{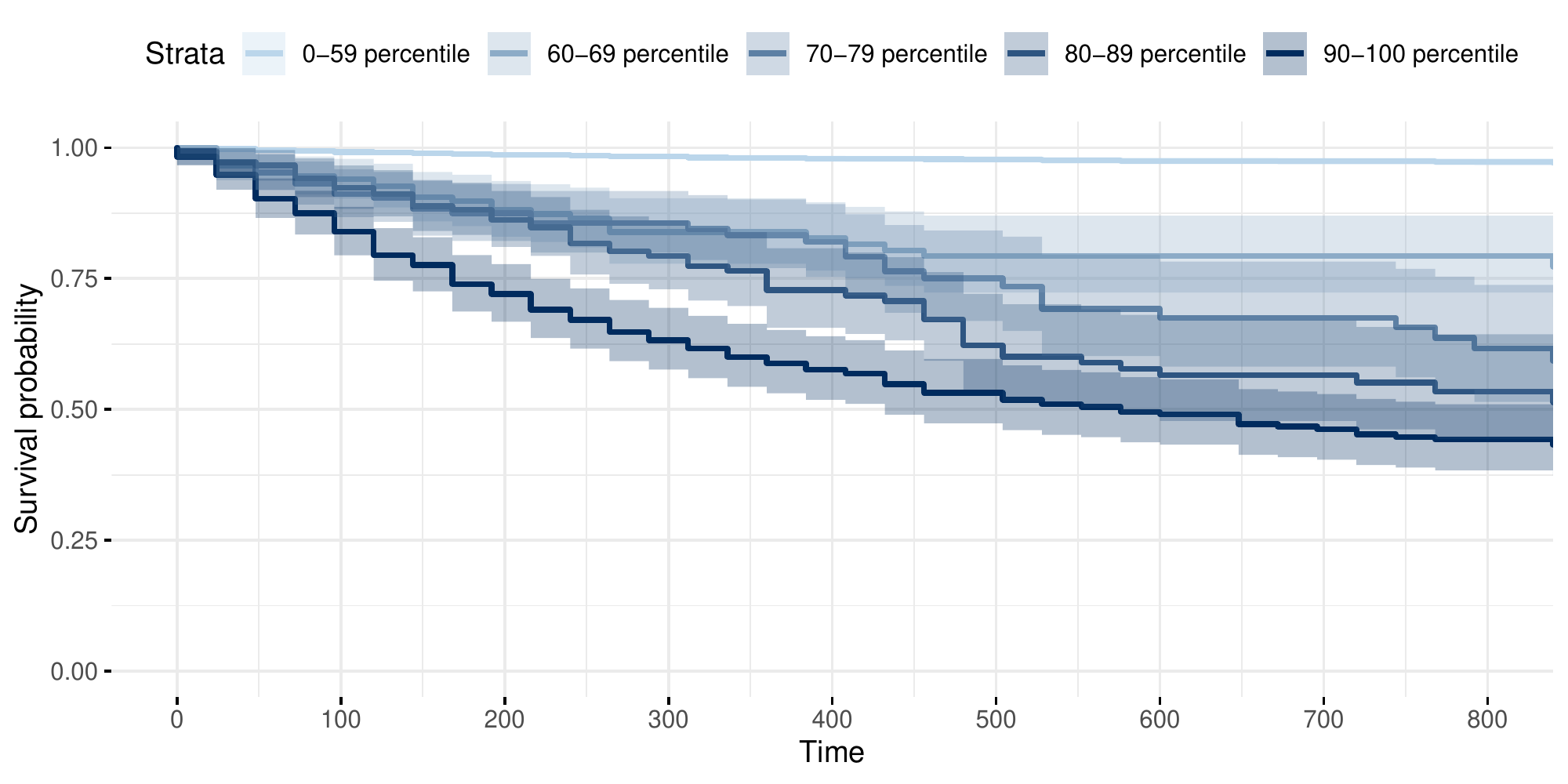}}
    \centerline{\includegraphics[width=1.00\linewidth,page=2]{figs/test_survival_strata_ext.pdf}}
    \centerline{\includegraphics[width=1.00\linewidth,page=3]{figs/test_survival_strata_ext.pdf}}
    \caption{External TriNetX Test Set} 
    \end{subfigure}
    \label{fig:survival_curves}
    \caption{Stratification of patients in the held-out Optum test cohort (left) and the external TriNetX test cohort according to their assigned \themethod{} score over time (in hours since COVID-19 diagnosis) into those patients that were assigned a \themethod{} score below 60, from 60 to 69, 70 to 79, 80 to 89, and 90 to 100 on the test fold of the Optum dataset. Note that the five strata and their respective limits were chosen for clarity of visualisation - other strata are possible, and may, depending on context, have better clinical utility. Rows show time-varying survival probabilities (top row), the number of patients (centre row), and the cumulative number of mortality events observed (bottom row) for patients in each stratum of assigned  \themethod{} scores. Steeper curves indicate that more patients died while assigned a \themethod{} score in the respective stratum. In contrast to traditional survival curves, cohorts as defined by strata of \themethod{} scores are not static over time, and patients move between the stratified groups as they are assigned lower or higher \themethod{} scores in response to their status improving or deteriorating, respectively. The results showed that \themethod{} enables effective stratification of patients into risk groups over the course of their disease, as patients that were assigned a higher \themethod{} score were more likely to die over time on both test cohorts while maintaining separation between the stratified cohorts. } 
    \end{figure}
    
    \paragraph{Stratified Time-varying Survival Analysis.} As illustrated in the examples in \Cref{fig:timeline_1}, \themethod{} continuously varies over time since it accounts for the status of patients deteriorating or improving. To add to the analysis of the predictive performance of \themethod{} in identifying the mortality of individual patients at fixed prediction horizons prior to observed mortality events presented in the previous paragraph, we therefore additionally evaluated whether \themethod{} enables stratification of high-risk patients continuously over time. To do so, we stratified the held-out Optum test cohort and the external TriNetX cohort into five strata of the \themethod{} score respectively assigned to each patient (\Cref{fig:survival_curves}). We found that \themethod{} effectively separated patients into risk groups with distinct COVID-19 related mortality risk profiles, as patients that were assigned to higher strata of \themethod{} scores were more likely to die across all strata over the course of their disease. When comparing stratification results between the held-out Optum test cohort and the external TriNetX cohort, we observed that the ability to stratify patients into risk groups generalised across the two datasets - indicating that the predictive performance of \themethod{} can transfer to other sources of data collected with different protocols, from different locations, and under different treatment policies. We also observed that the highest risk stratum of patients assigned \themethod{} scores between 90 and 100 was considerably steeper than other strata in the held-out Optum test cohort and this anomaly did not persist to the same degree in the external TriNetX cohort. Qualitatively, we reasoned that this difference between the two datasets was due to the considerably higher missingness of short-term risk factors associated with mortality, such as, e.g., respiratory rate, SpO$_2$ and blood pressure, in the TriNetX cohort (\Cref{tbl:cohort_description}). Rapid changes in these short-term risk factors often result in substantially increased near-term mortality risk and \themethod{} scores reflected this increased risk immediately (\Cref{fig:timeline_1}), moving patients with extreme short-term risk indicators into the highest risk stratum. Since these short-term risk factors were not included as frequently in the TriNetX cohort, \themethod{} was considerably less able to react to short-term deteriorations in the status of the patients, which was reflected in a relatively flatter time-varying survival curve of the highest risk stratum in the TriNetX cohort.

%% file: 4_cohort_table.tex
\definecolor{shadecolor}{RGB}{150,150,150}
\pgfplotstableread{data/cohort.train.txt}{\loadedtabletrain}
\pgfplotstableread{data/cohort.val.txt}{\loadedtableval}
\pgfplotstableread{data/cohort.test.txt}{\loadedtabletest}
\pgfplotstableread{data/cohort.ext.txt}{\loadedtableext}
\pgfplotstablecreatecol[
  copy column from table={\loadedtableval}{[index] 1},
  ]{val}{\loadedtabletrain}
\pgfplotstablecreatecol[
  copy column from table={\loadedtableval}{[index] 2},
  ]{val2}{\loadedtabletrain}
\pgfplotstablecreatecol[
  copy column from table={\loadedtabletest}{[index] 1},
  ]{test}{\loadedtabletrain}
\pgfplotstablecreatecol[
  copy column from table={\loadedtabletest}{[index] 2},
  ]{test2}{\loadedtabletrain}
\pgfplotstablecreatecol[
  copy column from table={\loadedtableext}{[index] 1},
  ]{ext}{\loadedtabletrain}
\pgfplotstablecreatecol[
  copy column from table={\loadedtableext}{[index] 2},
  ]{ext2}{\loadedtabletrain}
\pgfplotstablegetrowsof{\loadedtabletrain}
\pgfmathsetmacro{\N}{\pgfplotsretval-8}  

\begin{sidewaystable}
\centering
\caption{Descriptive statistics and percentage of patients with missing entries (Miss. \%) for the training, validation and test sets of the Optum cohort and the external TriNetX test cohort. Input covariates of \themethod{} are placed towards the bottom of the table and separated from covariates that are not inputs by a horizontal line. For binary covariates, the Value columns indicate the percentage of patients presenting with the condition at the end of their respective observation periods. For continuous measurements, the Value columns indicate the median and 10$^\text{th}$ and 90$^\text{th}$ percentiles in parentheses of the observed value for measurements that are collected once per patient, such as age, and the median of observed values for measurements that are collected multiple times per patient, such as heart rate. \Cref{tbl:icd_codes} presents the ICD codes corresponding to the shown diagnoses. BMI = Body Mass Index, HIV = Human Immunodeficiency Virus, COPD = Chronic Obstructive Pulmonary Disease, GGT = Gamma Glutamyl Transferase, AAT = Aspartate Aminotransferase, IL6 = Interleukin 6, n/a = not available. }
 \label{tbl:cohort_description}
\begin{tiny}
\setlength{\tabcolsep}{0.65em}
\begin{tabular}{p{0.25ex}}
\toprule
\\  \\  \\ \\
\midrule
{\parbox[t]{0.25ex}{\multirow{8}{*}{\rotatebox[origin=c]{90}{}}}}  \\ \\  \\ \\ \\  \\ \\ \\  \\ \midrule
{\parbox[t]{0.25ex}{\multirow{\N}{*}{\rotatebox[origin=c]{90}{Model Inputs}}}}  \\ \\ \\ \\ \\  \\ \\ \\ \\  \\ \\ \\ \\  \\  \\ \\ \\  \\ \\ \\ \\  \\ \\ \\
\\ \\  \\ \\ \\  \\ \\ \\ \\  \\ \\ \\ \\  \\ \\ \\ \\  \\ \\ \\ \\  \\  \\ \\ \\  \\ \\ \\  \\ \\ \\  \\ \\ \\ [1pt] 
\bottomrule
\end{tabular}%
\pgfplotstabletypeset[
font={\tiny},
display columns/0/.style={column name = {}, column type = {l}},
display columns/1/.style={column name = {Value}, column type = {r}},
display columns/2/.style={column name = {Miss.\%}, column type = {r}},
display columns/3/.style={column name = {Value}, column type = {r}},
display columns/4/.style={column name = {Miss.\%}, column type = {r}},
display columns/5/.style={column name = {Value}, column type = {r}},
display columns/6/.style={column name = {Miss.\%}, column type = {r}},
display columns/7/.style={column name = {Value}, column type = {r}},
display columns/8/.style={column name = {Miss.\%}, column type = {r}},
  string type,
  every row no 9/.style={before row=\midrule},
  every head row/.style={before row=\toprule,after row=\midrule},
  every last row/.style={after row=\bottomrule},
  every head row/.append style={before row={%
   \toprule& \multicolumn{6}{c}{Optum} & \multicolumn{2}{c}{TriNetX}\\ 
    & \multicolumn{6}{c}{March 21 - June 5 2020} & \multicolumn{2}{c}{March 21 - June 25 2020}\\ 
    & \multicolumn{2}{c}{Training Set} & \multicolumn{2}{c}{Validation Set}  & \multicolumn{2}{c}{Test Set} & \multicolumn{2}{c}{External Test Set}\\ 
    }},%
]{\loadedtabletrain}
\end{tiny}
\end{sidewaystable}

%% file: 4_subcohort_table.tex
\definecolor{shadecolor}{RGB}{150,150,150}
\pgfplotstableread{data/cohort.black.txt}{\loadedtabletrain}
\pgfplotstableread{data/cohort.hispanic.txt}{\loadedtableval}
\pgfplotstableread{data/cohort.asian.txt}{\loadedtabletest}
\pgfplotstableread{data/cohort.caucasian.txt}{\loadedtableext}
\pgfplotstableread{data/cohort.hospitalised.txt}{\loadedtablehosp}
\pgfplotstableread{data/cohort.next.txt}{\loadedtablenext}
\pgfplotstablecreatecol[
  copy column from table={\loadedtableval}{[index] 1},
  ]{val}{\loadedtabletrain}
\pgfplotstablecreatecol[
  copy column from table={\loadedtabletest}{[index] 1},
  ]{test}{\loadedtabletrain}
\pgfplotstablecreatecol[
  copy column from table={\loadedtableext}{[index] 1},
  ]{ext}{\loadedtabletrain}
\pgfplotstablecreatecol[
  copy column from table={\loadedtablehosp}{[index] 1},
  ]{next}{\loadedtabletrain}
\pgfplotstablecreatecol[
  copy column from table={\loadedtablenext}{[index] 1},
  ]{hosp}{\loadedtabletrain}
\pgfplotstablegetrowsof{\loadedtabletrain}
\pgfmathsetmacro{\N}{\pgfplotsretval-8}  
\begin{sidewaystable}
\centering
\caption{Descriptive statistics for selected subgroups (Caucasian, Asian, Black or African American, Hispanic, Not Hospitalised) of the held-out Optum test cohort, and for the Optum future cohort. Input covariates of \themethod{} are placed towards the bottom of the table and separated from covariates that are not inputs by a horizontal line. For binary covariates, the Value columns indicate the percentage of patients presenting with the condition at the end of their respective observation periods. For continuous measurements, the Value columns indicate the median and 10$^\text{th}$ and 90$^\text{th}$ percentiles in parentheses of the observed value for measurements that are collected once per patient, such as age, and the median of observed values for measurements that are collected multiple times per patient, such as heart rate. BMI = Body Mass Index, HIV = Human Immunodeficiency Virus, COPD = Chronic Obstructive Pulmonary Disease, GGT = Gamma Glutamyl Transferase, AAT = Aspartate Aminotransferase, IL6 = Interleukin 6, n/a = not available, \sig{} = see \Cref{sec:data} for an explanation of the non-zero intubation rate.}
 \label{tbl:subcohort_description}
\begin{tiny}
\setlength{\tabcolsep}{0.36em}
\begin{tabular}{p{0.25ex}}
\toprule
\\  \\  \\
\midrule
{\parbox[t]{0.25ex}{\multirow{8}{*}{\rotatebox[origin=c]{90}{}}}}  \\ \\  \\ \\ \\  \\ \\ \\  \\ \midrule
{\parbox[t]{0.25ex}{\multirow{\N}{*}{\rotatebox[origin=c]{90}{Model Inputs}}}}  \\ \\ \\ \\  \\ \\ \\ \\  \\ \\ \\ \\  \\  \\ \\ \\  \\ \\ \\ \\  \\ \\ \\ \\
\\ \\  \\ \\ \\  \\ \\ \\ \\  \\ \\ \\ \\  \\ \\ \\ \\  \\ \\ \\ \\  \\  \\ \\ \\  \\ \\ \\  \\ \\ \\  \\ \\ \\[2.5pt] 
\bottomrule
\end{tabular}%
\pgfplotstabletypeset[
font={\tiny},
display columns/0/.style={column name = {}, column type = {l}},
display columns/1/.style={column name = {Value}, column type = {r}},
display columns/2/.style={column name = {Value}, column type = {r}},
display columns/3/.style={column name = {Value}, column type = {r}},
display columns/4/.style={column name = {Value}, column type = {r}},
display columns/5/.style={column name = {Value}, column type = {r}},
display columns/6/.style={column name = {Value}, column type = {r}},
  string type,
  every row no 9/.style={before row=\midrule},
  every head row/.style={before row=\toprule,after row=\midrule},
  every last row/.style={after row=\bottomrule},
  every head row/.append style={before row={%
   \toprule
   & \multicolumn{5}{c}{Held-out Optum Test Set} & Future Cohort \\ 
    & \multicolumn{1}{r}{Black} & \multicolumn{1}{r}{Hispanic}  & \multicolumn{1}{r}{Asian} & \multicolumn{1}{r}{Caucasian} & \multicolumn{1}{r}{Not Hospitalised} & June 6 - July 13 2020\\ 
    }},%
]{\loadedtabletrain}
\end{tiny}
\end{sidewaystable}

%% file: 5_discussion.tex
We developed and validated \themethod{}, a real-time early warning system for predicting mortality of COVID-19 positive patients, using routinely collected clinical measurements and laboratory results from EHRs. When compared to competitive baselines, our method not only provides accurate mortality predictions for each patient, but also provides a real-time early warning system of up to 192 hours (8 days) prior to an observed mortality event for individuals, while identifying clinically-relevant factors for predictive performance. These results are sustained across various ethnic groups and cohorts. Notably, in comparison to existing mortality risk scoring systems, our method achieves significantly higher performance in terms of specificity at greater than 95\% sensitivity across all evaluated prediction time frames, and generalises well to data collected under different treatment and data collection policies and environmental conditions. The implications of providing such an early warning system are significant. The provided risk assessment could potentially broadly aid in clinical decision-making as well as in the prioritisation of care and resource allocation. More specifically, \themethod{} could enable clinicians to intensify monitoring and therefore initiate treatments earlier in patients with a higher risk of mortality. Moreover, as an additional information source, \themethod{} could also help clinicians to decide when to initiate palliative care to improve the quality of remaining life for patients with this need. Additional studies investigating if and how \themethod{} can influence clinical decision-making would be necessary to improve both treatment outcomes, the involvement of palliative care, or resource allocation to reduce COVID-related mortality.

Before applying \themethod{} in clinical practice, it is important to decide and calibrate appropriate warning thresholds, e.g. at the $85\%$, $90\%$ or $95\%$ sensitivity level (\Cref{sec:thresholds}). Especially when hospitals are overwhelmed and need to strictly allocate resources, alarm fatigue due to ill-calibrated thresholds ought to be minimised.  In addition, while the data used in this study already comprises multiple hospitals, a further analysis including hospitals from other countries would be useful to investigate the impact of geographic and cultural differences - particularly in those geographic contexts that are not well covered by this study. Due to differences in data collection methodology and expected data formats, another limitation of this study is that the implementation of some existing risk scoring systems is based on certain assumptions that may adversely influence their comparative performance (\Cref{sec:baselines}). Moreover, this work only concerns risk scores from routinely collected clinical data and patients who are already seeking care at healthcare providers. For efficient mitigation of COVID-19, additional, potentially preventative efforts like tracking apps, risk scores of infection prior to admission, masks and social distancing are necessary.

It is also important to acknowledge upfront the pitfalls of mortality prediction of hospitalised patients. A significant proportion of patients who die in the hospital, do so after cessation of treatment. One may argue that models that predict mortality thus actually predict the likelihood of treatment discontinuation. Numerous factors go into the decision with regard to continuing or stopping interventions, including whether the outcome, if the patient were to survive, is aligned with the patient’s preferences. It will only be accurate in a clinical context where clinicians make predictions in a similar manner, where patients share the same values and preferences around the quality of life, and where the decision-making process resembles that of the training cohort.

From the perspective of medical staff, prognostication as well as the perception of the quality of life if the patient were to survive, determine the framing of patient status to the family and friends; these are vulnerable to bias, both conscious or unconscious and influence the decision to admit the patient to the intensive care unit, as well as the decision to discontinue treatment (which almost certainly lead to death among those who are most severely ill). In a perfect world without bias and health disparities, only patient and disease factors determine hospital mortality, but studies have repeatedly demonstrated that this is far from the case. Recently, mortality from critical illness has been shown to be higher in disproportionately minority-serving hospitals after adjustment for illness severity and other biological factors that pertain to the patient and to the disease \cite{danziger2020temporal,rush2020treatment}. It is nearly impossible to incorporate these factors precisely in a model that is trained on mortality as an outcome. As a decision support tool to inform discussion around goals of care, \themethod{} is subject to the same limitations that mortality prediction models have – it may permeate or even magnify existing health disparities and provider bias. As an early warning system, however, we speculate that the impact of the exclusion of social determinants on model performance is acceptable.

In summary, we presented, developed, and experimentally validated \themethod{}, a real-time early warning system that provides clinically meaningful predictions of COVID-19 related mortality up to 192 hours (8 days) in advance for individual patients using routinely collected EHR data. In contrast to existing risk scoring systems, \themethod{} provides real-time continuous risk assessment that accounts for a large set of short-term and long-term risk factors associated with COVID-19 related mortality, is automatically derived from readily available EHR data, and was externally validated using data from multiple hospitals, diverse patient groups, and across time frames. Accessible risk assessment from readily available EHRs is especially important in the ongoing COVID-19 pandemic since access to advanced clinical lab testing and imaging techniques may be limited in many hospitals. \themethod{} allows for critical time in clinical decision making, even without access to specialised lab tests or advanced diagnostic equipment. Prospective studies are needed to conclusively establish if the availability of early warnings for COVID-19 related mortality through \themethod{} improves patient outcomes compared to the standard of care.

%% file: 6_app_method.tex
\begin{figure}[t!]
    \centering
  \includegraphics[width=1.0\textwidth]{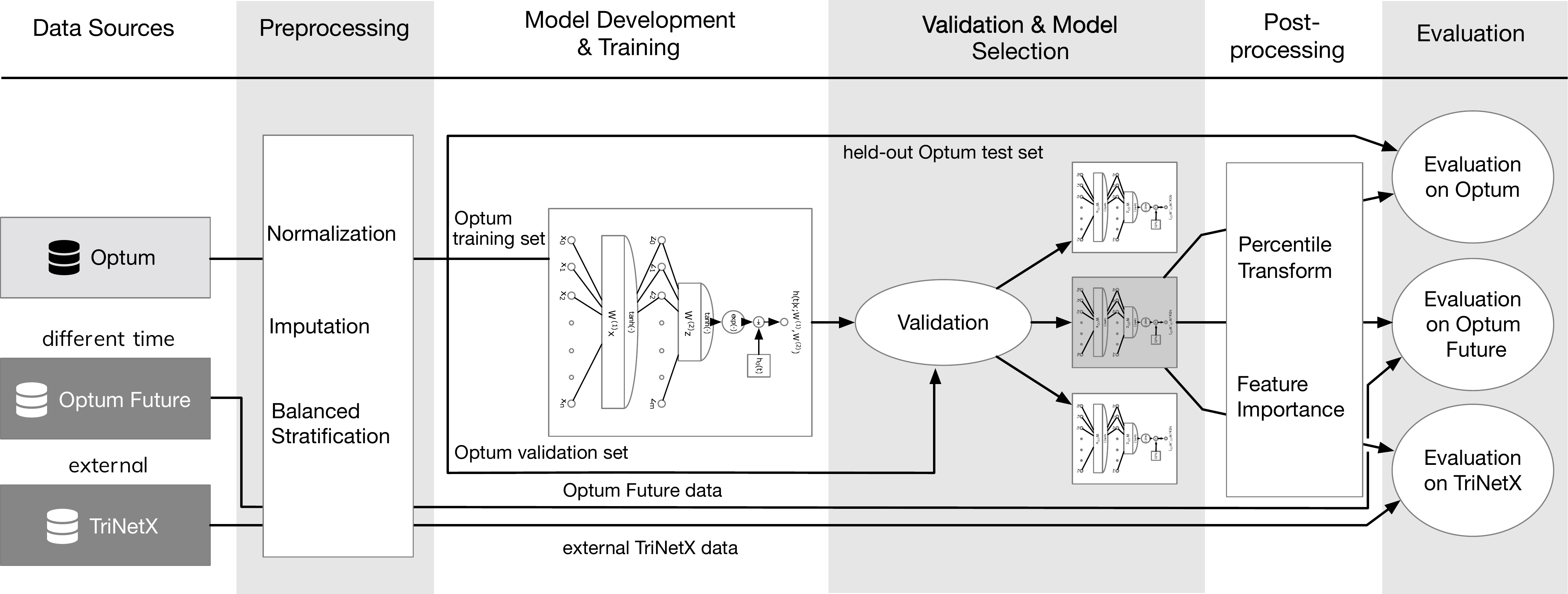}
    \caption{During training the model is provided with EHR data, including lab tests, clinical measurements, and information about pre-existing conditions. Before the data is fed to the model, the preprocessing phase handles missing values and standardises scaling of each covariate (\Cref{sec:preprocessing}). \themethod{} was trained on the training fold of the Optum cohort, and evaluated on the held-out Optum test cohort, the Optum Future cohort and the external TriNetX cohort (\Cref{sec:data}). The proposed model accommodates the effect of time-varying and nonlinearly interacting covariates and is trained using partial likelihood as detailed in \Cref{sec:algorithm_details}. \Cref{sec:preprocessing} presents further details on preprocessing.}
    \label{fig:overall_pipeline}
\end{figure}

\subsection{Overview}
\label{sec:overview}
The overall pipeline of the method is shown in~\Cref{fig:overall_pipeline}. We refer to~\Cref{sec:nonlinear_time_varying} for a detailed presentation of the predictive model used by \themethod{} and \Cref{fig:neuralnet_schematic} for a detailed diagram of the model architecture.

\subsection{Data Collection}
\label{sec:data}

We used data collected by two federated networks of healthcare organisations:

\paragraph{Optum.} The Optum de-identified COVID-19 electronic health records database includes de-identified electronic medical records and clinical administrative data including bedside observations and laboratory data from a geographically diverse set of healthcare institutions in the United States (US). The EHR data was sourced from more than 45 provider groups and integrated delivery networks. We used Optum cohort data collected between 21st March and 5th June 2020, and another cohort separated in time from 6th June to 13th July 2020 for our analysis.

\paragraph{TriNetX.} TriNetX is a global health research network providing a de-identified dataset of electronic medical records (diagnoses, procedures, medications, laboratory values, genomic information) including patients diagnosed with COVID-19. The data is de-identified based on standard defined in Section $\S$164.514(a) of the Health Insurance Portability and Accountability Act (HIPAA) Privacy Rule. The process by which Data Sets are de-identified is attested to through a formal determination by a qualified expert as defined in Section $\S$164.514(b)(1) of the HIPAA Privacy Rule. We used TriNetX cohort data collected between 21st March and 25th June 2020 from 24 healthcare organisations in the US, Australia, Malaysia and India for our analysis.

\paragraph{Data Quality.} Both Optum and TriNetX as well as the data providing healthcare institutions applied quality control steps to their data, but these procedures are not standardised neither across the federated networks nor across healthcare institutions. Varying levels of data quality across EHRs collected at different healthcare institutions and networks are therefore expected. However, heterogeneous data quality standards are characteristic for real-world data collected at different healthcare institutions. By evaluating \themethod{} against an external test cohort from healthcare institutions with data collection policies different from our training cohort, we are able to give a fair assessment as to how robust and transferable \themethod{} is in presence of realistic variations in data quality.

\begin{table}[t!] 
\setlength{\tabcolsep}{0.91em}
\caption{Comparison of the percentage of COVID-19 diagnosed patients without (w/o) a recorded SARS-CoV-2 test result across the analysed datasets (Optum, Optum Future, TriNetX). The higher fraction of COVID-19 diagnoses without a recorded SARS-CoV-2 test result in the Optum cohort compared to the Optum Future cohort is likely a result of the available testing resources having been more scarce early on in the pandemic. Since the Optum database indicates a generally high testing ratio substantiating COVID-19 diagnoses, the relatively higher percentage of COVID-19 diagnoses without corresponding tests in TriNetX is likely a result of inconsistent coding of SARS-CoV-2 tests early in the pandemic.}
\label{tbl:diagnoses_without_test_results}
\centering
\begin{small}
\begin{tabular}{l@{\hskip 0.85ex}rrr}
\toprule
 & Optum & Optum Future & TriNetX \\
\midrule
COVID-19 diagnoses w/o test results [\%] & 6.31 & 0.00 & 37.92 \\
\bottomrule
\end{tabular}
\end{small}
\end{table}

\paragraph{Inclusion Criteria.} We only included patients that were COVID-19 positive in our analysis. In both datasets, we considered patients COVID-19 positive if they either (i) were diagnosed with any of the International Statistical Classification of Diseases and Related Health Problems 10$^\text{th}$ revision (ICD-10) codes J12.89, J20.8, J40, J22, J98.8, and J80 together with B97.29\footnote{The listed criteria correspond to the Centers for Disease Control and Prevention (CDC) COVID-19 coding guidelines effective February 20, 2020 (see \url{https://www.cdc.gov/nchs/data/icd/ICD-10-CM-Official-Coding-Gudance-Interim-Advice-coronavirus-feb-20-2020.pdf}.}, or (ii) had a positive COVID-19 lab test result (\Cref{tbl:diagnoses_without_test_results}). For patients identified as COVID-19 positive via ICD diagnosis codes, we used the date of diagnosis as the reference diagnosis date for our analyses. For those patients identified as COVID-19 positive via a positive lab test, we used the date of the test sample collection as the diagnosis date. For patients with both a positive COVID-19 lab test and diagnosis, the available diagnosis date took precedence. For the subgroup of patients that were not hospitalised, we included all patients that were neither admitted to a hospital as inpatients nor an intensive care unit (ICU) at any point according to their EHRs. We note that it is possible that hospitals did not in all cases record inpatient hospital admissions and ICU admissions in their respective EHRs - which may  explain the observed non-zero rate of intubations in the non-hospitalised group. Membership in the Asian, Caucasian and Black or African American subgroups was mutually exclusive in the underlying EHR data model, and a patient could therefore only be assigned to one of the subgroups. In contrast, hispanic ethnicity was assigned in conjunction with any of the previous race categorisations.

\paragraph{Feature Selection.} We selected EHR-derived covariates for inclusion as input variables for \themethod{} based on (i) previously published research on clinical risk factors for COVID-19 \cite{zheng2020risk,zhou2020clinical,richardson2020presenting}, and (ii) expert input from several medical professionals involved in the treatment of COVID-19 patients. In addition, we aimed to include both short-term and long-term risk factors for COVID-19 related mortality due to the continuous real-time evaluation of \themethod{}. We present the list of all included model input covariates including their p-values in \Cref{tbl:feature_pvalues}, and their distributions across the datasets in \Cref{tbl:cohort_description}.

\begin{longtable}{l@{\hskip 4ex}p{33.5ex}r}
\caption{Descriptions, Logical Observation Identifiers Names and Codes (LOINC) codes (if available), and p-values for each input covariate used by \themethod{}. The p-values were calculated using Wald's $\chi^2$ tests for effect (H$_0$ = no effect) of the coefficients in CovEWS (linear) corresponding to the respective input covariates on the Optum training cohort.}
\label{tbl:feature_pvalues}
\centering
\endfirsthead
\caption*{\Cref{tbl:feature_pvalues} cont.}
\endhead
\toprule
Covariate & Description & p-value \\
\midrule
Sex & - & $0.41$ \\
Age & - & $<0.005$\\
Weight & - & $0.92$ \\
Height & - & $0.80$\\
Body Mass Index (BMI) & - & $0.85$\\
Intubation & Whether or not the patient is intubated & $0.04$ \\
Temperature & Body temperature (LOINC: 8310-5 on TriNetX) & $0.89$ \\
SpO$_2$ & Oxygen saturation by pulse oximetry (LOINC: 59408-5) & $<0.005$\\
Heart rate & - & $0.75$\\
Respiratory rate & - & $0.32$ \\
Systolic blood pressure & - & $0.01$\\
Diastolic blood pressure & - & $<0.005$\\
Kidney disease & see \Cref{tbl:icd_codes} & $0.05$ \\
Ischemic heart disease & see \Cref{tbl:icd_codes} & $0.04$ \\
Other heart disease &see \Cref{tbl:icd_codes} & $0.01$ \\
Cerebrovascular disease &see \Cref{tbl:icd_codes} & $0.20$\\
Hypertension &see \Cref{tbl:icd_codes} &  $0.03$ \\
Diabetes &see \Cref{tbl:icd_codes} &  $0.15$\\
Hyperlipidemia &see \Cref{tbl:icd_codes} &  $0.17$ \\
Cancer & see \Cref{tbl:icd_codes} &  $0.65$ \\
Dyspnea & see \Cref{tbl:icd_codes} & $0.42$ \\
COPD & see \Cref{tbl:icd_codes} &  $0.18$ \\
Asthma & see \Cref{tbl:icd_codes} &  $0.80$ \\
Pulmonary embolism & see \Cref{tbl:icd_codes} & $0.58$ \\
Connective tissue disease & see \Cref{tbl:icd_codes} &  $0.84$ \\
Inflammatory bowel disease & see \Cref{tbl:icd_codes} & $0.99$ \\
Osteoarthritis & see \Cref{tbl:icd_codes} & $0.37$ \\
Rheumatoid arthritis & see \Cref{tbl:icd_codes} & $0.79$ \\
HIV & see \Cref{tbl:icd_codes} & $0.97$ \\
Smoking (never) & - & $0.81$ \\
Smoking (previous) & - & $0.61$ \\
Smoking (current) & - & $0.84$ \\
Smoking (unknown) & - & $0.85$ \\
White blood cells & White blood cell count (LOINC: 26464-8) & $<0.005$ \\
Neutrophil & Neutrophils per 100 leukocytes in blood (LOINC: 26511-6) & $<0.005$\\
Lymphocytes & Lymphocytes per 100 leukocytes in blood (LOINC: 26478-8) & $<0.005$ \\
Eosinophil & Eosinophils per 100 leukocytes in blood (LOINC: 26450-7) & $0.38$ \\
Basophil & Basophils per 100 leukocytes in blood (LOINC: 30180-4) & $0.33$\\
Platelets & Platelets [\#/volume] in blood (LOINC: 26515-7) & $0.28$ \\
C-reactive protein & C-reactive protein [mass/volume] in serum or plasma (LOINC: 1988-5) & $0.52$\\
hs. C-reactive protein & C-reactive protein [mass/volume] in serum or plasma by high sensitivity method (LOINC: 30522-7) & $<0.005$ \\
Procalcitonin & Procalcitonin [mass/volume] in serum or plasma (LOINC: 33959-8)& $0.46$ \\
Fibrin D-dimer & Fibrin D-dimer Fibrinogen Equivalent Units (FEU) [mass/volume] in platelet poor plasma & $0.13$ \\
Ferritin & Ferritin [mass/volume] in serum or plasma (LOINC: 2276-4) & $0.12$ \\
Cardiac troponin T & Cardiac troponin T [mass/volume] in serum or plasma (LOINC: 6598-7) & $0.42$ \\
Creatinine & Creatinine [mass/volume] in Serum or Plasma (LOINC: 2160-0) & $0.03$ \\
Lactate dehydrogenase & Lactate dehydrogenase [enzymatic activity/volume] in serum or plasma (LOINC: 2532-0, 14804-9) & $0.01$ \\
Gamma glutamyl transferase & Gamma glutamyl transferase [enzymatic activity/volume] in serum or plasma (LOINC: 2324-2) & $0.69$ \\
Aspartate aminotransferase & Aspartate aminotransferase [enzymatic activity/volume] in serum or plasma (LOINC: 1920-8) & $0.10$ \\
Creatine kinase & Creatine kinase enzymatic activity/volume] in serum or plasma (LOINC: 2157-6) & $0.60$ \\
Bilirubin & Bilirubin [mass/volume] in serum or plasma (LOINC: 1975-2) & $0.28$\\
Albumin & Albumin [mass/volume] in serum or plasma (LOINC: 1751-7) & $<0.005$ \\
Interleukin 6 (IL-6) & Interleukin 6 [mass/volume] in serum or plasma (LOINC: 26881-3) & $0.68$ \\
pH & pH of blood (LOINC: 2744-1, 2746-6) & $<0.005$ \\
PCO$_2$ & Carbon dioxide [partial pressure] in arterial blood (LOINC: 2019-8) & $0.02$ \\
PaO$_2$ & Oxygen [partial pressure] in arterial blood (LOINC: 2703-7) & $0.63$ \\
HCO$_3$ & Bicarbonate [moles/volume] in blood (LOINC: 1959-6, 1960-4) & $0.17$ \\
CO$_2$ & Carbon dioxide, total [moles/volume] in serum or plasma (LOINC: 2028-9) & $0.21$ \\
\bottomrule
\end{longtable}
\clearpage

\begin{figure}[t!]
\centering
\includegraphics[width=1.0\textwidth]{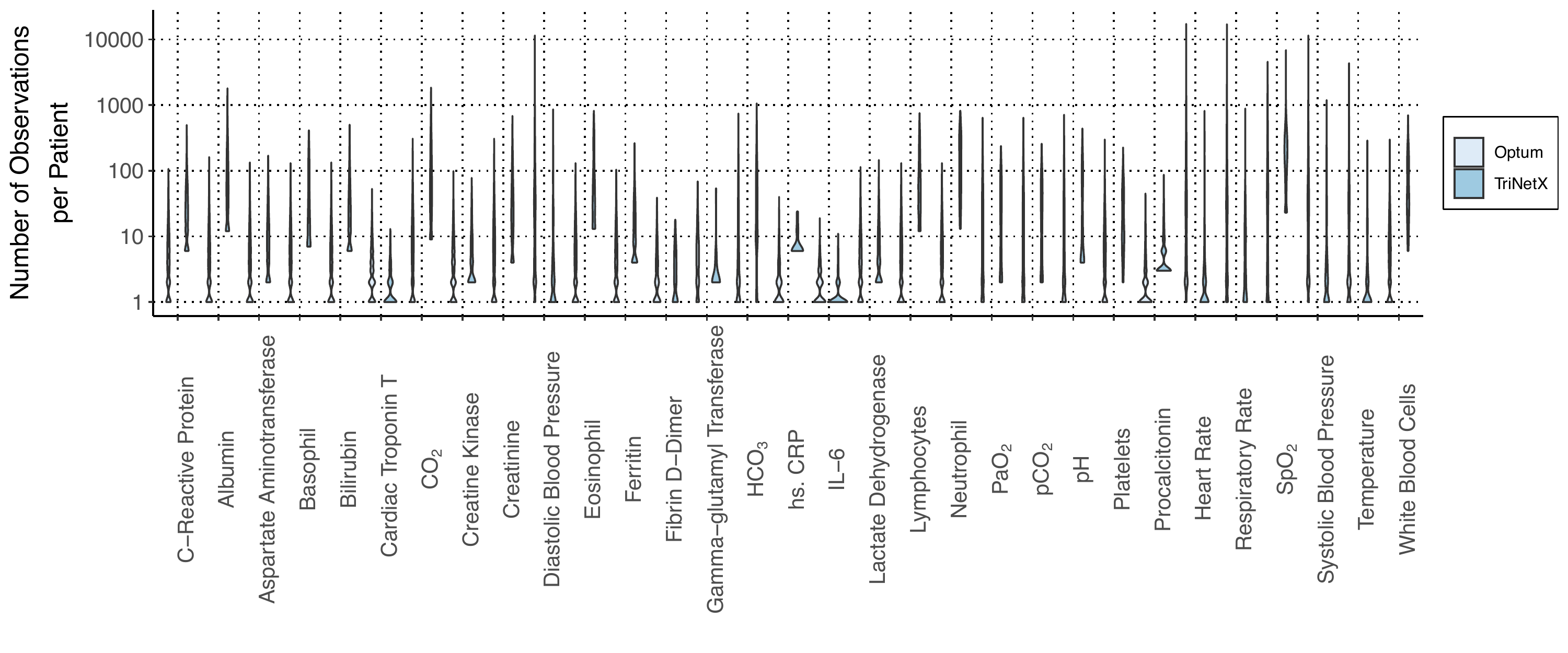}
\caption{Distributions of the number of observations per patient (y-axis, log scale, violin plots) for time-varying covariates (x-axis) in the Optum and TriNetX datasets for patients that have at least one observation of a given covariate. The percentage of patients with no observation for each covariate is shown in \Cref{tbl:cohort_description}.}
\label{fig:test_frequency}
\end{figure}

\paragraph{Data Characteristics.} The cohort statistics of the two datasets are presented in \Cref{tbl:cohort_description}, the ICD-9 and ICD-10 codes corresponding to the diagnoses shown in \Cref{tbl:cohort_description} are given in \Cref{tbl:icd_codes}, and the number of observations per patient for time-varying covariates for the two datasets is visualised in \Cref{fig:test_frequency}. As is characteristic for clinical data collected in real-world contexts, missing covariates are common in both datasets. Missingness in real-world EHR data is caused primarily by differences in laboratory testing guidelines, data collection practices, available testing resources and measurement devices between hospitals, and may in some cases depend on patient status and preferences of clinical staff. For example, \Cref{tbl:missingness_hospitalised} compares the missingness between the Optum test set and the non-hospitalised patient subgroup of the Optum test set. In contrast to traditional clinical studies, realistic missingness patterns in both the training and evaluation datasets are a desirable feature in the context of our study as \themethod{} is designed to be deployed in clinical contexts with similar missingness, and therefore has to be trained and evaluated in the presence of missingness patterns seen across a representative range of heterogeneous hospitals. Covariates were mostly balanced across the Optum and TriNetX datasets. The primary differences were a higher observed mortality rate, and higher ratios of intubations, connective tissue disease, and rheumatoid arthritis in the TriNetX data compared to the Optum data. In addition, we note that the majority of admissions were recorded as being of the "Unknown" type in the TriNetX database. Since the large fraction of unknown admission entries limited potential admission outcome analyses, we reported hospital and ICU admission outcomes as not available for TriNetX (\Cref{tbl:cohort_description}). In compliance with the HIIPA Privacy Rule Section $\S$164.514(a), patients' exact dates of death were not available to protect patient privacy. In our analysis, we therefore imputed the last recorded EHR entry date as the reference date of death for deceased patients. The actual dates of death may have happened at a later point, and our performance estimates are therefore potentially underestimating actual predictive performance, since (i) correct predictions that happened later would mean \themethod{} predicted sooner than we thought for that patient (which is generally harder, see \Cref{fig:baselines}), and (ii)  incorrect predictions of \themethod{} may actually have been outside of the prediction time horizon. We believe this approximation of the exact date of death is therefore an acceptable trade-off, since underestimation of performance is not as much a concern as overestimation would be.

\begin{table}
\caption{International Statistical Classification of Diseases and Related Health Problems (ICD) codes corresponding to the disease classifications and symptoms used in our analysis. HIV = Human Immunodeficiency Virus, COPD = Chronic Obstructive Pulmonary Disease.}
\label{tbl:icd_codes}
\centering
\begin{small}
\begin{tabular}{l@{\hskip 4ex}p{35ex}p{15ex}}
\toprule
Disease classification & ICD-9 & ICD-10 \\
\midrule
Kidney disease & 585 - 586 & N18 - N19 \\
Ischemic heart disease & 410 - 414 & I20 - I25 \\
Other heart disease & 390 - 398, 401 - 405, 416 - 417, 420 - 429, 115.03, 115.04, 115.13, 115.14, 115.93, 115.94 & I27 - I52 \\
Cerebrovascular disease & 430 - 434 & I60 - I69\\
Hypertension & 401 - 405 & I10 - I15 \\
Diabetes & 249 - 250, 357.2, 366.41& E10 - E14\\
Hyperlipidemia & 272 & E78 \\
Cancer & 140 - 239 & C \\
Dyspnea & 786.1, 786.2, 786.8, 786.9 & R06 \\
COPD & 496, 491.21, 491.22 & J44 \\
Asthma & 493 & J45 \\
Pulmonary embolism & 415 & I26 \\
Connective tissue disease & 446, 710 - 711, 713, 725, 136.1, 279.8, 517.2, 728.5 & I30 - I36 \\
Inflammatory bowel disease & 555 - 556 & K50 - K51 \\
Osteoarthritis & 715 & M15 - M19 \\
Rheumatoid arthritis & 274, 712 - 714, 716, 719 & M05 - M14 \\
HIV & 42 & B20 - B24 \\
\bottomrule
\end{tabular}
\end{small}
\end{table}

\subsection{Data Normalisation}
\label{sec:data_normalisation}
The EHR data across both data sources used two different, but compatible, underlying data models consisting of recorded diagnoses, demographics, lab tests, procedures, medications and clinical observations. For our risk factors of interest, we converted records from both datasets into a unified data representation. We used ICD-9 and ICD-10 to extract diagnoses (\Cref{tbl:icd_codes}), Logical Observation Identifiers Names and Codes (LOINC) to extract lab tests, Current Procedural Terminology (CPT), and ICD-9 Clinical Modification (ICD-9-CM) and ICD-10 Procedural Coding System (ICD-10-PCS) to extract intubation events from the EHR records. For lab tests, we additionally normalised the unit of each category of lab tests to be the same for each measured record of that category.

\subsection{Stratification}
\label{sec:stratification}

We split the Optum cohort used for model development into training (50\%), validation (20\%) and held-out test folds (30\% of all patients) at random stratified by patient age, gender, presence of mortality events, presence of intubation events, presence of ICU admission and presence of a human immunodeficiency virus (HIV) diagnosis. We added HIV to the set of stratification covariates since its low prevalence could otherwise have led to imbalances in this risk factor across the folds. Stratification produced balanced cohorts across the three folds (\Cref{tbl:cohort_description}). The Optum training fold was used to train \themethod{}, the validation fold was used to select the optimal hyperparameter configuration for \themethod{}, and the held-out test fold was used in addition to the external TriNetX test cohort to evaluate the out-of-sample generalisation performance of \themethod{}.

\begin{table}[t!] 
\setlength{\tabcolsep}{0.91em}
\caption{Comparison of missingness (in \%), i.e. the fraction of patients that did not have any record for a specific covariate in their EHR, in several important clinical covariates between the Optum test set and the non-hospitalised subgroup of the Optum test set. SBP = Systolic Blood Pressure, RR = Respiratory rate, WBC = White blood cells, CRP = C-reactive protein.}
\label{tbl:missingness_hospitalised}
\centering
\begin{small}
\begin{tabular}{l@{\hskip 2.85ex}cccccc}
\toprule
Cohort & SBP & RR & WBC & CRP & Albumin & Platelets\\
\midrule
Optum Test Set & 38.56 & 42.21 & 48.40 & 67.81 & 53.48 & 48.45 \\
Non-hospitalised &  54.00 & 59.57 & 64.67 & 81.10 & 69.02 & 64.72 \\
\bottomrule
\end{tabular}
\end{small}
\end{table}

\subsection{Preprocessing}
\label{sec:preprocessing}

Discrete covariates with $p$ different values were transformed into their one-hot encoded representation with one out of $p$ indicator variables set to 1 to indicate the discrete value for this patient. All continuous features were standardised to have zero mean and unit standard deviation using observed covariate distribution on the Optum training fold. Missing values of continuous covariates were imputed in an iterative fashion using multiple imputation by chained equations (MICE) \cite{white2011multiple}. After the preprocessing stage, continuous input features were standardised and fully imputed, and discrete input covariates were one-hot encoded. All preprocessing operations were derived only from the training fold, and na\"ively applied without adjustment to other folds and datasets in order to avoid information leakage. 

\subsection{Method}
\label{sec:method}

We adopt a variation of the widely used Cox proportional hazard model that is adapted to accommodate nonlinear and time-varying effects of covariates on the log-hazard function. In the following, the basics of time-to-event analysis that is the main subject of this paper is briefly presented. Then we touch upon the Cox proportional model for continuous-time covariates that is followed by the modifications we applied to this model to prepare it for the current work.

\subsubsection{Survival Analysis}
\label{sec:survival_analysis}
Survival analysis which is also known as Time-To-Event (TTE) analysis included a large body of work consisting of mathematical tools to give a statistical analysis of the time duration until a specified event occurs. In the current work, the event is defined to be the time when a patient dies.

An important tool in time-to-event analysis is~\emph{hazard function}. In discrete-time setting, (e.g. if times are given in specified periods) the hazard function is a conditional probability defined in discrete-time as
\begin{equation}
    \label{eq:hazard_function_discrete}
    h(t|\xb) = P(T=t|T\geq t; \xb),\; t=1, 2,\ldots
\end{equation}
that represents the risk of dying at time $t$ if the patient has survived until that time. The relevant covariates of the patients up to time $t$ are encapsulated in the vector $\xb\in\RR^d$. Age, sex, and lab tests are examples of such covariates that can take either binary or standardised real values after preprocessing. Intuitively, the hazard function captures the underlying dynamics of the transition of the condition of the patient from alive to dead. The larger $h$ is at time $t$, the more likely it is for the patient to die at time $t$.

Another useful function is called~\emph{survival function} that is denoted by $S(t)$ and in discrete-time defined as
\begin{equation}
    \label{eq:survival_function_discrete}
    S(t) = P(T>t) = \prod_{s=1}^t (1-h_s).
\end{equation}

\noindent{}Similar functions can be defined in the continuous-time regime. Let $T_c$ be the continuous survival time with the probability density function $f_c$ and cumulative distribution function $F_c$. Similar to~\Cref{eq:survival_function_discrete}, the continuous survival function represents the probability of surviving until time $t$ that is defined as

\begin{equation}
    \label{eq:survival_function_continuous}
    S_c(t|\xb) = P(T_c>t|\xb)=1-F_c(t|\xb).
\end{equation}

\noindent{}Likewise, the continuous hazard function is defined as
\begin{equation}
    \label{eq:hazard_function_continuous}
    h_c(t|\xb) = \lim_{\Delta t\to 0}\frac{P(t\leq T_c\leq t+\Delta t | T_c\geq t;\xb)}{\Delta t}.
\end{equation}
Notice that unlike discrete hazard function~\labelcref{eq:hazard_function_discrete}, the continuous hazard function~\labelcref{eq:hazard_function_continuous} is not a probability distribution and can take values larger than one.

The last useful function in continuous survival analysis is the cumulative hazard function
\begin{equation}
    \label{eq:cumulative_hazard_function}
    H_c (t) = \int_0^t h_c(u) du.
\end{equation}

\noindent{}The connection between these quantities can be simply derived:

\begin{align}
    h_c(t)&=\frac{f_c(t)}{S_c(t)}\label{eq:hazard_density_survival},\\
    S_c(t)&=\exp(-\int_0^t h_c(u) du) = \exp(-H_c(t))\label{survival_cumulative_hazard},\\
    f_c(t)&=h_c(t)\exp(-\int_0^th_c(u) du)=h_c(t) \exp(-H_c(t))\label{density_hazard_cumulative_hazard}.
\end{align}

Before introducing the simple yet flexible Cox model, we discuss a few important issues that must be taken into account in survival analysis.

\paragraph{Censoring.} What makes the survival analysis different from a simple regression from the covariates $\xb$ to $T$ -- observed duration up to the occurrence of the event -- is the concept of~\emph{censoring}. An observation is called censored -- or more precisely~\emph{right-censored} -- if its survival time has not been fully observed. There are several causes for a censored observation. For example, if a patient is not under observation when the event occurs or if the information of the patient is lost for some reason, only a lower bound to the time-to-event $T$ is observed that is the last time the condition of the patient is recorded. 

\paragraph{Discrete vs. Continuous.} Although time is a continuous physical quantity, in practice, it is measured at discrete points. Especially, in medical sciences, the condition of the patient is measured on a regular daily or bi-daily basis. This implies that even though the change of the covariates of a patient occurs at certain points of time, the exact time is not known. The transition point is only known up to the resolution of the measurement. We assume the time at which an event of interest occurs is denoted by $T$. As the resolution of the measurement is hours in the datasets used in the current work, $T=t$ refers to an event that occurs within the $t^{\rm th}$ hour after the patient is admitted to the hospital and its health condition is recorded.

\paragraph{Ties.} In the limited resolution measurement of time, some observations may have the same survival times, e.g., two patients die on the same day even though it is extremely unlikely that both die at the same moment. However, even in continuous time data, ties may occur which is a hint of underlying discrete sampling in time.

A major difference between continuous and discrete-time survival analysis is that the hazard function is a probability distribution in discrete settings while it can take any positive value in continuous settings. However, the traditional continuous-time approach can still be used for discrete event times especially when the measurements are equally spaced.

\subsubsection{Cox Hazards Model}
\label{sec:cox_hazard}
The most widely known model in the analysis of continuous survival time is {Cox's proportional hazard model}~\cite{cox1972regression} that parameterizes the hazard function as
\begin{equation}
    \label{eq:cox_proportional_hazard}
    h_c(t|\xb)=h_0(t)\exp(\betab\tran\xb),
\end{equation}
where $h_0$ is called the {baseline hazard} that is modulated by the effect of covariates via $\exp(\betab\tran\xb)$. Notice that in the traditional Cox model~\labelcref{eq:cox_proportional_hazard} the covariates $\xb$ are assumed constant over time. Consequently, the temporal variation of the hazard function is separated from the influence of the covariates.

\subsubsection{Time-varying Covariates}
\label{sec:cox_proportional_hazard_with_time_varying_covariates}
In many experimental settings, the assumption of time-invariant covariates in~\labelcref{eq:cox_proportional_hazard} does not hold. For example, many entries in the electronic health records such as heart rate, temperature, and blood measurements do not remain constant over the course of the hospitalisation of a patient. Therefore, the traditional Cox model~\labelcref{eq:cox_proportional_hazard} is extended to a time-varying setting by replacing $\xb$ in~\labelcref{eq:cox_proportional_hazard} with $\xb_t$ that is the measured covariates at time $t$. Assume a dataset consists of $N$ patients indexed by $n=1,2,\ldots,N$. As a notational convention, $\xb^{(n)}_t$ denotes the vector of the corresponding covariates to the patient $n$ at time $t$. 

If the Cox model holds and continuous events are observed, the following function called~\emph{partial likelihood} is maximized to estimate $\betab$:
\begin{equation}
    \label{eq:linear_time_varying_cox_partial_likeliood}
    \Lcal(\betab):=\prod_{i=1}^k\frac{\exp (\betab\tran \xb^{(i)}_{t_i})}{\sum_{j\in\Rcal(t_i)}\exp(\betab\tran \xb^{(j)}_{t=t_i})},
\end{equation}
where $t_1<t_2<\ldots<t_k$ are the ordered times at which the events occur and $\xb^{(1)}_{t_1}, \xb^{(2)}_{t_2}, \ldots, \xb^{(k)}_{t_k}$ are the corresponding set of covariates at those times. Notice that the equality of the superscript of the covariate vector $\xb^{(i)}_{t_i}$(patient's index) and the subscript of time $t_i$ emphasizes the continuous event times and the fact that at most one patient experiences the event at each time. For the moment, we assume time is continuous that results in distinct event times. The set $\Rcal(t_i)$ is the set of the patient's indices that are at risk at time $t_i$. Being at risk means they are alive and can potentially experience the event.

\subsubsection{Nonlinear Time-varying Covariates}
\label{sec:nonlinear_time_varying}
One clear limitation of~\labelcref{eq:linear_time_varying_cox_partial_likeliood} that is caused by the definition of the hazard function~\labelcref{eq:cox_proportional_hazard}, is the fact that the exponent of the modulating function $\exp(\betab\tran \xb)$ is a linear function of $\xb$. Hence higher order interactions among different dimensions of the covariate vector cannot be captured by this method. To improve the expressiveness of the model, we replace the linear function $\betab\tran\xb$ with a nonlinear function realised by a neural network. Let $\phi(\cdot;\thetab):\RR^d\to\RR$ be the function implemented by the neural network and parameterised by $\thetab$. Therefore, the hazard model is represented as

\begin{equation}
    h(t|\xb)=h_0(t)\exp(\phi(\xb_t;\thetab)),
    \label{eq:cox_hazard_nonlinear_body}
\end{equation}
where $h_0(t)$ is the baseline population-level hazard that is independent of the associated covariates to each patient. Time-varying covariates are transformed by the function $\phi(\cdot;\thetab)$ to log-hazard. The parameters $\thetab$ are learned via maximising the partial log-likelihood~\cite{cox1972regression}. Despite traditional Cox proportional hazard model where the gradient and Hessian can be computed analytically, here, we use automatic differentiation to compute gradients with respect to $\thetab$. The nonlinear function $\phi(\cdot;\thetab)$ is implemented as a $2-$layer multilayer perceptron -- see \Cref{sec:method} for a detailed description. The hazard function~\labelcref{eq:cox_hazard_nonlinear_body} estimates the instantaneous risk of death at each time for each patient. Integrating with respect to time and exponentiating the result gives the survival function defined as

\begin{equation}
    S(t|\xb_{0:t})=P(T>t|\xb_{0:t})=\exp\left(-\int_0^t h(u|\xb_u) du\right).\label{eq:survival_function_main}
\end{equation}
Notice that $\xb_{0:t}$ denotes the set of covariates until time $t$, meaning that, the probability of survival up to time $t$ depends on the history of the covariates. 

\noindent{}The partial likelihood~\labelcref{eq:linear_time_varying_cox_partial_likeliood} is re-written as
\begin{equation}
    \label{eq:nonlinear_time_varying_cox_partial_likeliood}
    \Lcal(\thetab):=\prod_{i=1}^k\frac{\exp (\phi(\xb^{(i)}_{t_i};\thetab))}{\sum_{j\in\Rcal(t_i)}\exp (\phi(\xb^{(j)}_{t=t_i};\thetab))}.
\end{equation}

\noindent{}To give an intuition of~\labelcref{eq:nonlinear_time_varying_cox_partial_likeliood}, observe that the partial log-likelihood that is computed by taking logarithm of the right-hand side of~\labelcref{eq:nonlinear_time_varying_cox_partial_likeliood} will consist of $k$ terms corresponding to $k$ observed events. The parameter vector $\thetab$ is perturbed such that the hazard increases for the covariates of a patient who dies at time $t_i$ while it decreases for the covariates of the patients who remain alive at $t_i$.

\subsubsection{Resolving Ties}
\label{sec:resolving_ties}
Even though we adopt a continuous-time approach due to the non-normalised parametric form of the hazard function~\labelcref{eq:cox_proportional_hazard} and the resultant partial likelihood~\labelcref{eq:nonlinear_time_varying_cox_partial_likeliood}, the ties can still occur as we work in hourly resolution. Hence, it is possible that two patients die at the same time. When an event occurs for two patients at the same time, the partial likelihood~\labelcref{eq:nonlinear_time_varying_cox_partial_likeliood} is not valid anymore. Several methods exist in the literature to break the ties and remove the ambiguity such as {average partial likelihood} \cite{cox1972regression} and~\emph{Berslow's method}~\cite{breslow1975analysis} that lives on two ends of a spectrum. The former takes average among all possible orders of the events that can break the tie. Hence, it is the most accurate method but computationally prohibitive. The latter gives a partial likelihood almost exactly like the original Cox likelihood by assuming that every ordering of tied events results in the same partial likelihood. This method gives a crude estimate but is easy to implement. A midway approach that we adopted in this work is called Efron's tie-breaker \cite{efron1977efficiency}. In this method, a weighted average likelihood of tied cases is subtracted from the denominator of \Cref{eq:nonlinear_time_varying_cox_partial_likeliood}. Efron's method gives good accuracy and is moderately easy to work with -- see \cite{efron1977efficiency} for details. 

\begin{figure}[t!]
    \centering
  \includegraphics[width=0.50\textwidth]{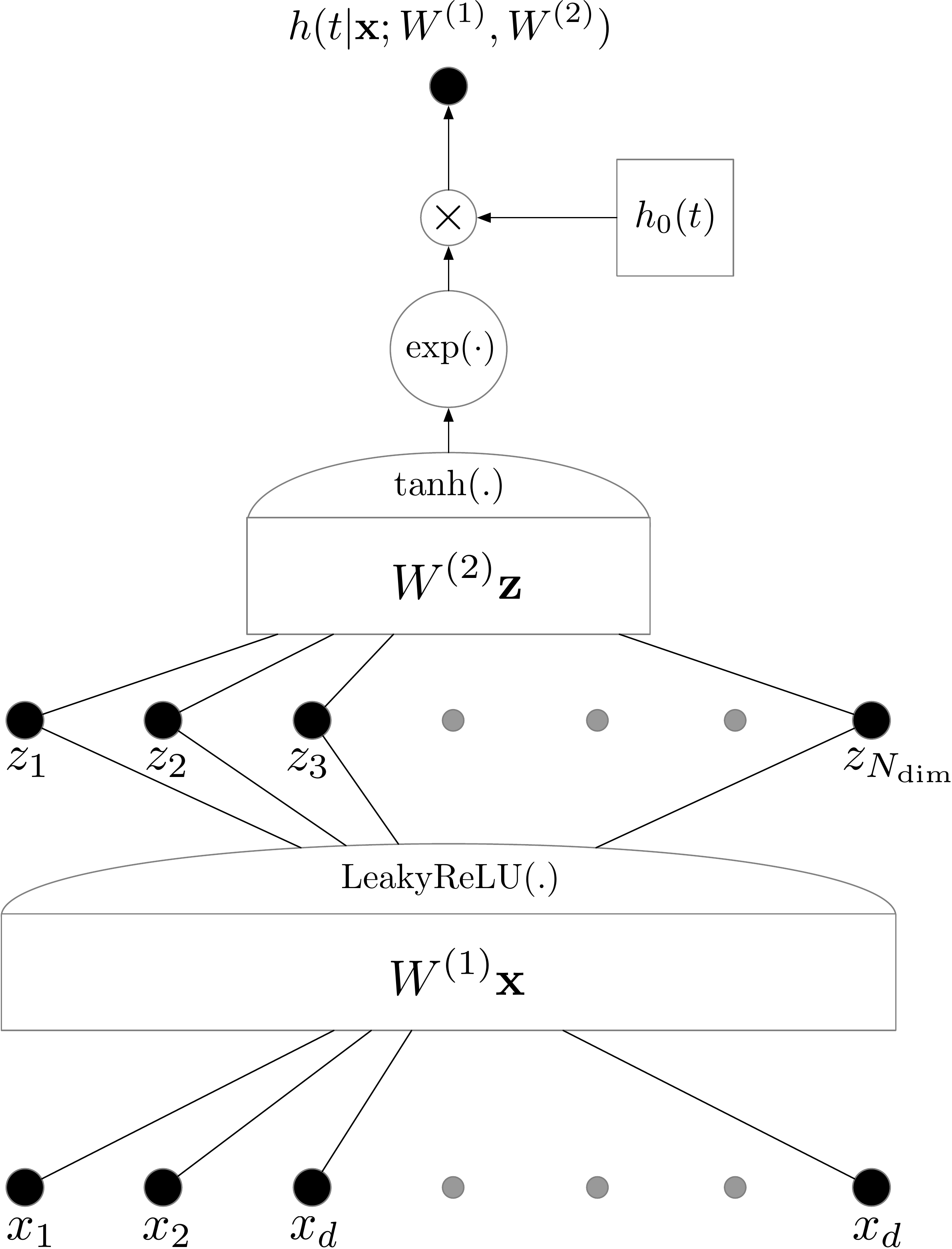}
    \caption{Schematic illustration of a neural network realisation of the nonlinear exponent of the time-varying Cox hazard function with $L=1$ hidden layers with $N_\text{dim}$ hidden units and one output layer.}
    \label{fig:neuralnet_schematic}
\end{figure}

\subsection{Algorithm Details}
\label{sec:algorithm_details}
Survival analysis by the Cox model is done via  maximum likelihood estimation, where the aim is to maximise the logarithm of~\labelcref{eq:linear_time_varying_cox_partial_likeliood} in the original Cox's proportional hazard model and~\labelcref{eq:nonlinear_time_varying_cox_partial_likeliood} in the nonlinear extension. In the original method with linear exponent both gradient $\partial \log \Lcal / \partial \betab$ and Hessian $\partial^2 \log \Lcal / \partial^2 \betab$ can be computed analytically. This is not the case for our proposed extension~\labelcref{eq:nonlinear_time_varying_cox_partial_likeliood} with nonlinear exponent. Instead of an analytical gradient, we use automatic differentiation to compute the gradient $\partial \log \Lcal/\partial \thetab$. Once the gradient is derived, an appropriate gradient-based method is used to perturb $\thetab$ in the direction that increases the partial likelihood.

As mentioned in~\cref{sec:nonlinear_time_varying}, the linear exponent $\betab\tran \xb$ in~\labelcref{eq:linear_time_varying_cox_partial_likeliood} is replaced with a nonlinear function $\phi(\cdot;\thetab)$. We use a neural network with $L$ hidden layers to realise this function. The employed network linearly transforms the input features to a $N_\text{dim}$-dimensional hidden layer. The transformed features are passed through a leaky rectified linear unit (LeakyReLU) \cite{xu2015empirical} nonlinear activation function. The hidden activations are then transformed by a linear transformation to a single node and finally passes through a tangent hyperbolic (${\rm tanh}$) activation function. In summary the network function can be represented as

\begin{equation}
    \phi(\xb;\thetab)=\rm{tanh}(W_2 (\rm{LeakyReLU}(W_1 \xb))),
\end{equation}
 where $\thetab=\{W_1, W_2\}$ and $W_i,\;i=1,2$ are the trainable weight matrices of the network (\Cref{fig:neuralnet_schematic}). We used Xavier's method~\cite{glorot2010understanding} to initialise the weights $\thetab$ of the model. To prevent overfitting, we additionally applied dropout with a dropout probability of $p_\text{dropout}$. In our PyTorch \cite{paszke2017automatic} implementation of \themethod{}, we observed stable  convergence of our model using the Adam \cite{kingma2014adam} optimiser with a learning rate of $0.001$ for up to 100 epochs.

\begin{table}[t!] 
\caption{Hyperparameter ranges used for hyperparameter optimisation of \themethod{} and \themethod{} (linear). Comma-delimited lists indicate discrete choices with equal selection probability. Hyperparameters selected after hyperparameter optimisation (\Cref{sec:hyperparameter_optimisation}) are highlighted in bold.}
\label{tbl:hyperparameters}
\centering
\begin{small}
\begin{tabular}{l@{\hskip 5.25ex}l@{\hskip 30.5ex}r}
\toprule
& Hyperparameter & Range / Choices\\
\midrule
\parbox[t]{2mm}{\multirow{3}{*}{\rotatebox[origin=c]{90}{Linear}}} & & \\
&Regularisation strength $\lambda$ & 0.01, \textbf{0.1}, 1.0\\
& & \\
\midrule
\parbox[t]{2mm}{\multirow{5}{*}{\rotatebox[origin=c]{90}{\themethod{}}}} & & \\
& Number of layers $L$ & \textbf{1}, 2\\
& Number of hidden units $N_\text{dim}$ & 64, \textbf{128}, 256\\
& Dropout percentage $p_\text{dropout}$ & 10\%, \textbf{20\%}\\
& & \\
\bottomrule
\end{tabular}
\end{small}
\end{table}

\subsection{Hyperparameter Optimisation}
\label{sec:hyperparameter_optimisation}

For the methods trained on the Optum training fold (\themethod{} and \themethod{} [linear]), we used a systematic approach to hyperparameter optimisation where each prediction algorithm was given a maximum of 15 hyperparameter optimisation runs with different hyperparameter configurations chosen at random without duplicates from predefined ranges (see \Cref{tbl:hyperparameters}). Out of the models generated in the hyperparameter optimisation runs, we then selected the model that achieved the highest specificity at greater than 95\% sensitivity on the validation set of the Optum cohort.

\subsection{Postprocessing and Calibration}
\label{sec:postprocessing}

\begin{figure}[t!]
{\centering\textbf{\textsf{\themethod{} Calibration}}\par\medskip}
\begin{subfigure}{.49\textwidth}
\includegraphics[width=\linewidth,page=1]{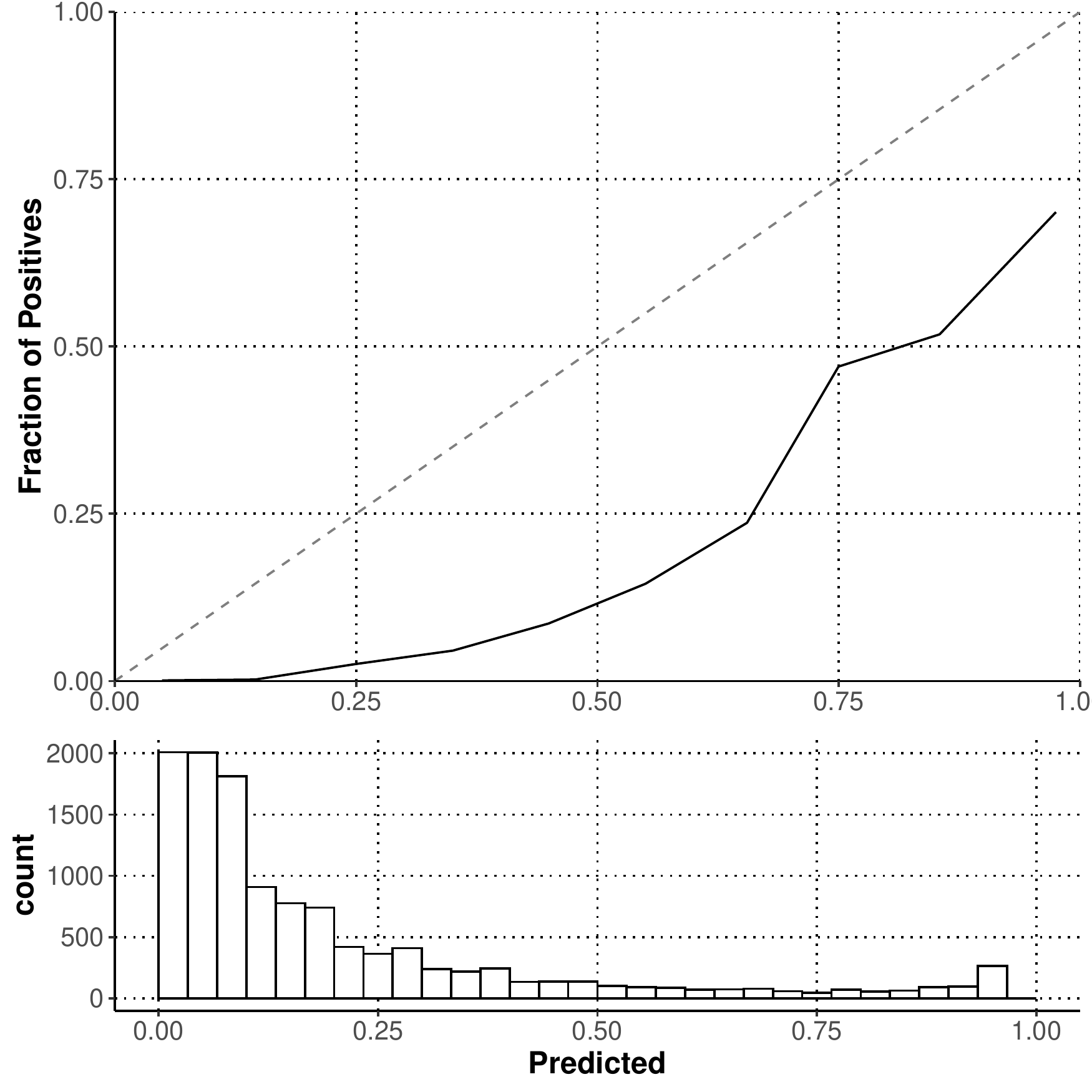}
\caption{Held-out Optum Test Set} 
\end{subfigure}\quad
\begin{subfigure}{.49\textwidth}
\includegraphics[width=\linewidth,page=1]{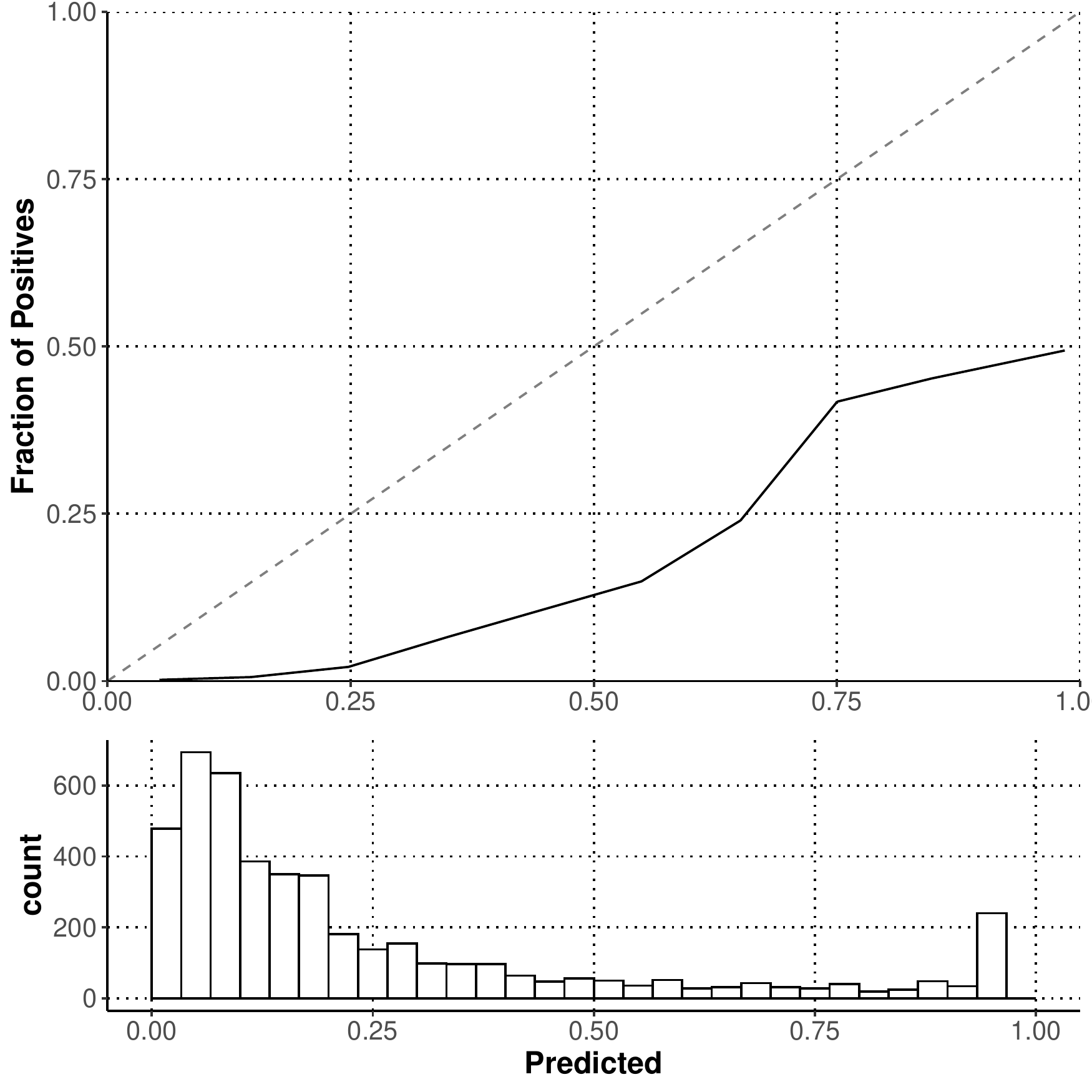}
\caption{External TriNetX Test Set} 
\end{subfigure}
\caption{Calibration plots \cite{niculescu2005predicting} and counts of observed predictions of \themethod{} on the held-out Optum test set (left) and the external TriNetX test set (right) when interpreting the predicted risk percentile of \themethod{} as the probability of a mortality event being observed within the next 24 hours. The reference time point for those patients that did not have an observed mortality event is the date of their respective last observed EHR entry. When interpreted as a patient's mortality probability, \themethod{} overestimates the risk since it can not account a priori for clinical interventions and potential future patient trajectory changes that may occur rapidly and frequently. Direct interpretation of \themethod{} scores as a mortality probability is discouraged, and \themethod{} scores should instead be seen as a relative risk score to stratify patients into risk groups (\Cref{fig:survival_curves}).}
\label{fig:calibration_curves}
\end{figure}

\begin{figure}[t!]
{\centering\textbf{\textsf{\themethod{} (linear) Calibration}}\par\medskip}
\begin{subfigure}{.49\textwidth}
\includegraphics[width=\linewidth,page=1]{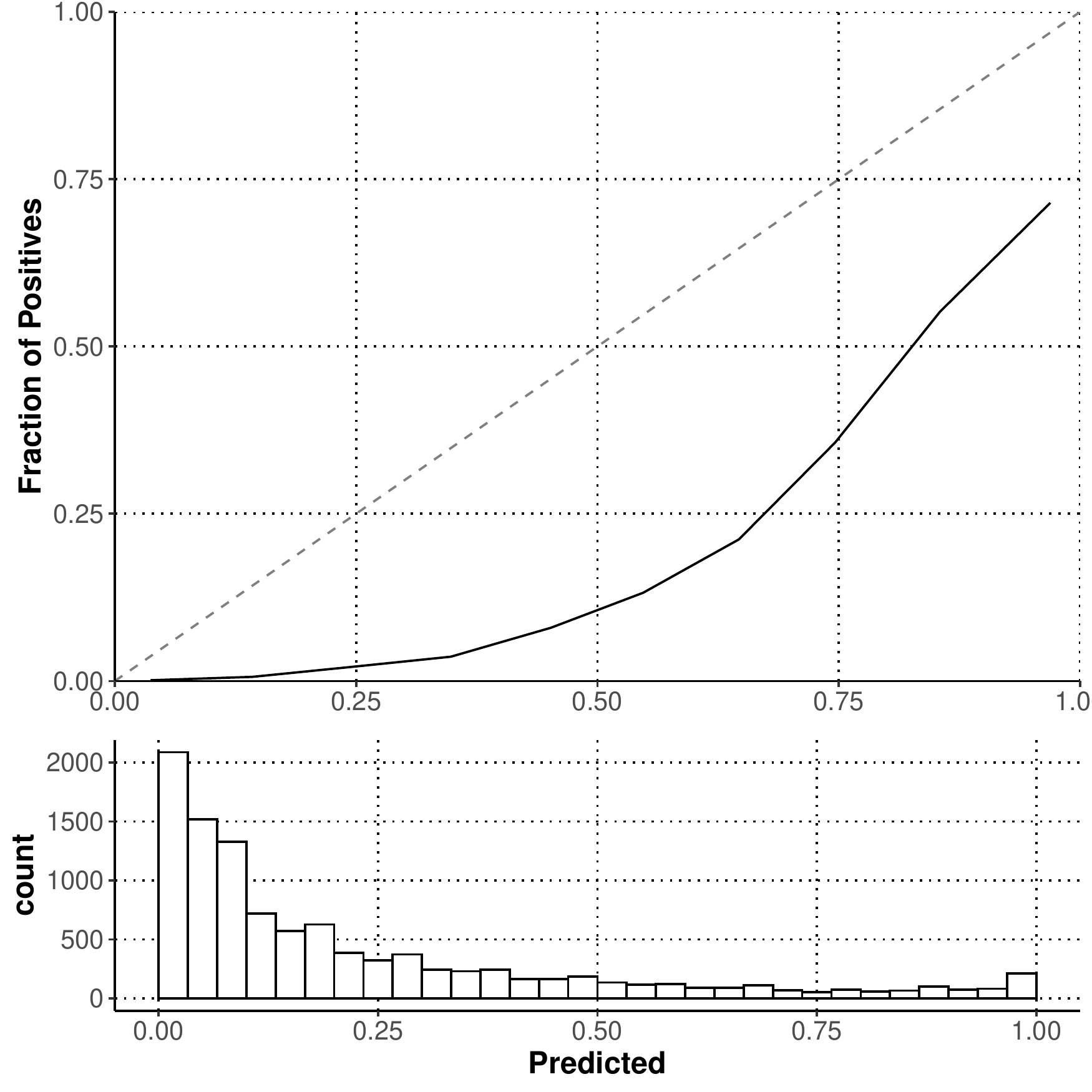}
\caption{Held-out Optum Test Set} 
\end{subfigure}\quad
\begin{subfigure}{.49\textwidth}
\includegraphics[width=\linewidth,page=1]{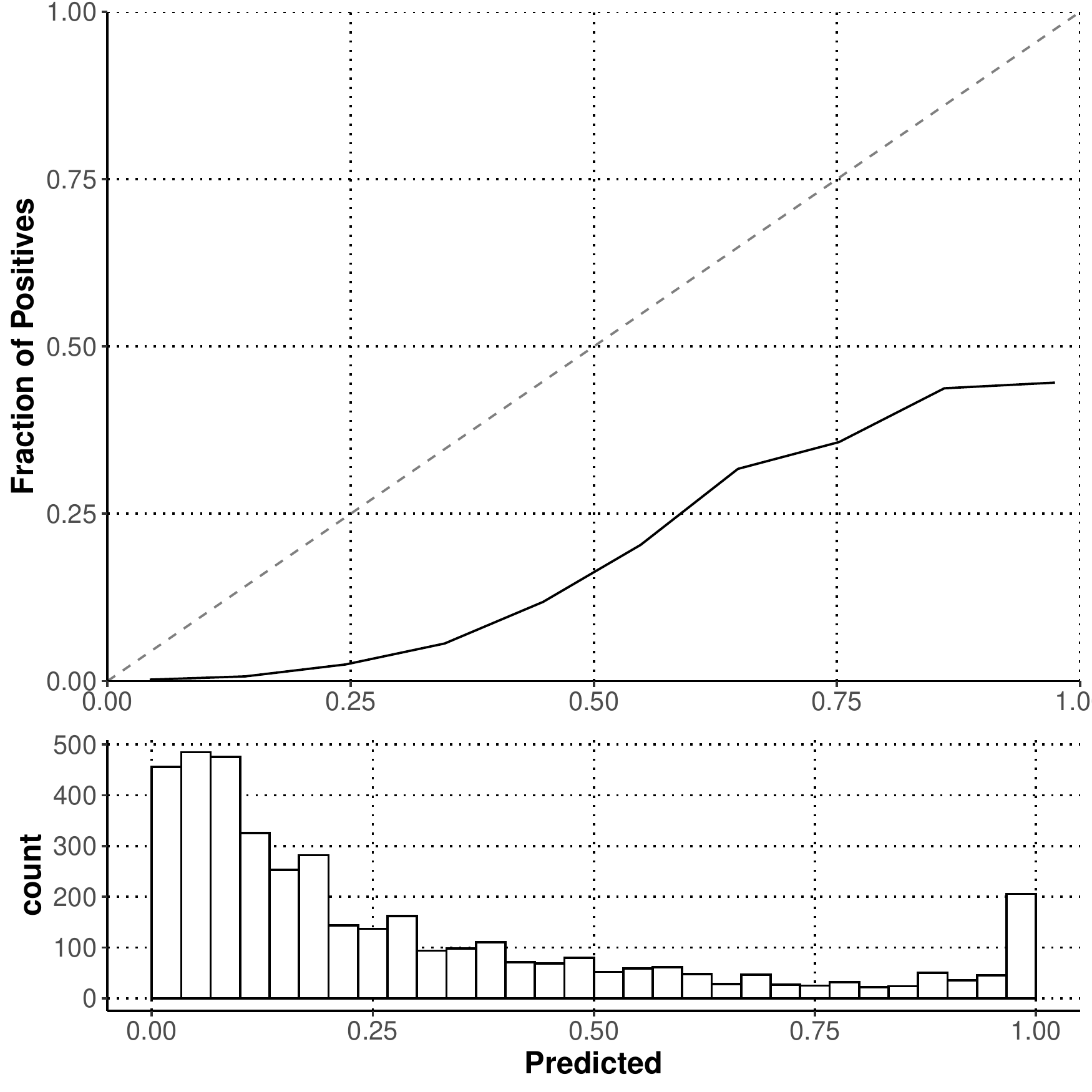}
\caption{External TriNetX Test Set} 
\end{subfigure}
\caption{Calibration plots \cite{niculescu2005predicting} and counts of observed predictions of \themethod{} (linear) on the held-out Optum test set (left) and the external TriNetX test set (right) when interpreting the predicted risk percentile of \themethod{} (linear) as the probability of a mortality event being observed within the next 24 hours. The reference time point for those patients that did not have an observed mortality event is the date of their respective last observed EHR entry. Like the version of \themethod{} that models non-linear interactions, \themethod{} (linear) overestimates mortality risk when interpreted as a patient's mortality probability.}
\label{fig:calibration_curves_linear}
\end{figure}

After training \themethod{} using the Optum training cohort, the predicted hazard for a patient with state $\xb$ is the hazard function $h(t|\xb)$ \cref{eq:cox_proportional_hazard} evaluated at $t=128 \text{ hours}$ ($\approx 5.33$ days) given the current patient covariates $\xb$ and under the assumption that patient covariates stay constant. To produce \themethod{} scores, we additionally apply post-processing using a percentile transformation that converts $h(t|\xb)$ into the percentile of patient states in the Optum validation set that are assigned a lower $h(t|\xb)$ than the evaluated patient state $\xb$. We chose to output \themethod{} scores in form of percentiles to aid in the clinical interpretation of \themethod{} as a risk score relative to a representative set of reference states, and to discourage interpretation as a mortality probability. Interpretation of \themethod{} scores as a mortality probability is difficult since the mortality risk of a patient depends on their uncertain future trajectory and the prediction horizon, and is influenced by clinical interventions that may be initiated in the future. As shown in \Cref{fig:timeline_1}, patients' states may change rapidly and frequently, and clinical interventions can significantly and positively alter the trajectory of patients. We also verified experimentally that, when interpreted as a probability of mortality, \themethod{} scores overestimate the actually observed probability of death on the Optum and TriNetX test sets since patients' states may improve, due to intervention or otherwise, between the prediction time and the end of the prediction horizon (\Cref{fig:calibration_curves}; similar results with \themethod{} [linear] \Cref{fig:calibration_curves_linear}). We, therefore, decided to instead output \themethod{} scores as relative risk percentiles between 0 and 100 that discourages interpretation as a probability of mortality. To aid in the use of \themethod{}, the following \Cref{sec:thresholds} outlines calibrated thresholds that can be used to maximise specificity at the desired target level of sensitivity for different prediction horizons.

\subsection{Thresholds}
\label{sec:thresholds}

\begin{table}[t!] 
\setlength{\tabcolsep}{0.162em}
\caption{Optimal thresholds of \themethod{} scores to maximise specificity at greater than 85\%, 90\% and 95\% sensitivity (Sens.) for each prediction horizon as selected on their respective receiver operator characteristic (ROC) curves on the held-out Optum test set.}
\label{tbl:thresholds}
\centering
\begin{small}
\begin{tabular}{l@{\hskip 0.35ex}|ccccccccc}
\toprule
Sens. & 1 hour & 2 hours & 4 hours & 8 hours & 16 hours & 24 hours & 48 hours & 96 hours & 192 hours \\
\midrule
85\% & 61 & 62 & 56 & 54 & 51 & 48 & 45 & 39 & 36 \\
90\% & 44 & 44 & 42 & 41 & 39 & 38 & 38 & 32 & 27 \\
95\% & 34 & 34 & 33 & 31 & 28 & 27 & 27 & 22 & 19 \\
\bottomrule
\end{tabular}
\end{small}
\end{table}

A key question for clinical decision making is which threshold should be used for \themethod{} scores to indicate severe risk, and potentially trigger an automated warning. To provide guidance in choosing the appropriate \themethod{} score depending on the desired trade-off between sensitivity and specificity, we evaluated the optimal observed thresholds of \themethod{} scores for various target sensitivity levels using their respective receiver operator characteristic (ROC) curves for each prediction horizon (\Cref{tbl:thresholds}). Optimal score thresholds to maximise specificity at high levels of sensitivity were between 61 and 36, 44 and 27, and 34 and 19 depending on the prediction horizon for target sensitivity levels greater than 85\%, 90\% and 95\%, respectively. We note that lower thresholds are necessary to achieve high sensitivity for prediction horizons farther in the future as patients' deterioration has to be identified earlier in its progression.

\subsection{Feature Importance}
\label{sec:feature_importance}

Highlighting the clinical risk factors that positively or negatively influenced \themethod{} to output a certain score is of high utility as it enables clinical users to contextualise \themethod{} scores, and, in some cases, these highlights could potentially even point towards opportunities for timely intervention. We utilised the differentiability of our prediction model as outlined in \Cref{sec:algorithm_details} to provide a real-time visualisation of the clinical covariates that are most important for \themethod{} at any given time point (see \Cref{fig:timeline_1} for an example). To compute the importance scores at each time point, we used the Integrated Gradients (IG) \cite{sundararajan2017axiomatic} method that calculates relative importance scores $a_i \in (-100\%,100\%)$ for each input feature $x_{t,i}$ in the feature vector $\xb_t$ with $i\in[0 \isep d-1]$ where $d$ is the number of input features. We used IG with the mean feature vector $\bar{\xb}_t$ across the Optum training set as a reference, calculated $50$ intermediate steps for each explained $\xb_t$, and normalised $a_i$ to the range of $(-100\%,100\%)$ by dividing each $a_i$ by the sum $\Sigma_{i=0}^{d-1}|a_i|$ of all feature attributions for $\xb_t$. To obtain a timeline of attributions as shown in \Cref{fig:timeline_1}, we calculate attributions $a_i$ whenever a change in patient status was recorded in the patient's EHR.

\subsection{Baselines}
\label{sec:baselines}
In our analysis, we compared the performance of \themethod{} to the following existing generic and COVID-19 specific clinical risk scores, and  baselines:

\paragraph{\themethod{} (linear).} A linear time-varying survival Cox model as described in \Cref{sec:cox_proportional_hazard_with_time_varying_covariates} trained using the same Optum training set and using the same pipeline as the non-linear \themethod{}. We used the implementation provided in version 0.24.8 of the lifelines \citep{cameron2020lifelines} Python package.

\paragraph{COVID-19 Estimated Risk for Fatality (COVER\_F).} The COVER\_F scoring system for COVID-19 as described in \citep{williams2020seek}. Since COVER\_F uses a single flag for any heart disease diagnosis, we aggregated all diagnoses in the diagnosis categories ischemic heart disease, pulmonary embolism, other heart diseases in our dataset into one single joint diagnosis code if any diagnosis in those three categories was present. All other input features used by COVER\_F were direct matches with the input covariates of the same name also used by \themethod{} (\Cref{tbl:cohort_description}).

\paragraph{Sequential Organ Failure Assessment (SOFA).} SOFA scoring is commonly used in clinical contexts to indicate the risk of organ failure in critical patients. We, therefore, used SOFA as a generic risk scoring baseline that was not specifically designed for COVID-19 to demonstrate the comparative benefits in the predictive performance of a COVID-19 specific risk scoring system. Since we did not have FiO$_2$ values available in our EHR datasets, we assumed a default FiO$_2$ value of $21\%$ (equal to the fraction of oxygen in inhaled air) for patients that were not intubated, and an average of 71\% for patients that are intubated (FiO$_2$ is often set to 100\% initially and then progressively lowered as the patient stabilises, see e.g. \cite{rachmale2012practice} for an example). In addition, we did not have access to Glasgow coma scale (GCS) scores in the EHRs, and potential additional points from a high GCS score (a maximum of +4) were therefore not reflected in our calculated SOFA scores.

\paragraph{Yan et al. 2020.} \citet{yan2020interpretable} derived a simple and interpretable decision rule using three features (Figure 2 in \cite{yan2020interpretable}) for mortality prediction in COVID-19 patients from data collected from \numprint{485} COVID-19 positive patients seen in Wuhan, China. In their validation cohort, the decision rule showed a respectable cross-validated prediction performance of $96.1 \pm 0.03$ (mean $\pm$ standard deviation, 5-fold cross validation) \cite{yan2020interpretable}. All input features used by \citet{yan2020interpretable} were direct matches with the input covariates of the same name also used by \themethod{} (\Cref{tbl:cohort_description}).

\paragraph{Liang et al. 2020.} \citet{liang2020early} developed a prediction model for critical COVID-19 related illness using data from \numprint{1590} patients seen at \numprint{575} medical centers in China using deep learning and 10 input covariates, including observed X-ray abnormalities. On three external cohorts from different Chinese provinces, they reported a predictive performance in terms of concordance index (c-index) of 0.890, 0.852 and 0.967 for predicting critical illness under the missingness of input covariates, respectively. Since we did not have access to radiologic assessments in our EHR datasets, we evaluated their model with the X-ray abnormality covariate missing for all evaluated patients (i.e. set to zero). To the best of our knowledge, \citet{liang2020early} did not specify which co-morbidities were included in their collected dataset. However, their study reports a maximum of 6 co-morbidities diagnosed in one patient. In our evaluation, we counted existing patient diagnoses of pulmonary embolism, kidney disease, inflammatory bowel disease, asthma, rheumatoid arthritis, and diabetes towards these 6 co-morbidities.

\subsection{Software}
\label{sec:software}

The source code used for developing \themethod{} and for conducting the presented experiments and analyses was implemented using \texttt{Python} (version 3.7), \texttt{scikit-learn} (version 0.22.2), \texttt{numpy} (version 1.19.1), \texttt{scipy} (version 1.4.1), \texttt{pandas} (version 1.5.0), \texttt{PyTorch} (version 1.5.1), and \texttt{lifelines} (version 0.24.8). All plots shown were generated using the \texttt{ggpot2} R package \cite{wickham2016ggplot2}.

\subsection{Hardware}
\label{sec:hardware}

We used the high-performance computing (HPC) infrastructure provided by the Personalised Healthcare Informatics group at F. Hoffmann-La Roche Ltd to run the presented experiments. The compute nodes used 1st and 2nd generation Intel Xeon Platinum 8000 series processors and had access to  72 GB random access memory (RAM) each.

\subsection{Performance Evaluation}
\label{sec:performance}

\begin{figure}[h]
{\centering\textbf{\textsf{\themethod{} Receiver Operating Characteristic (Optum Test Set)}}\par\medskip}
\begin{subfigure}{.31\textwidth}
\includegraphics[width=\linewidth,page=1]{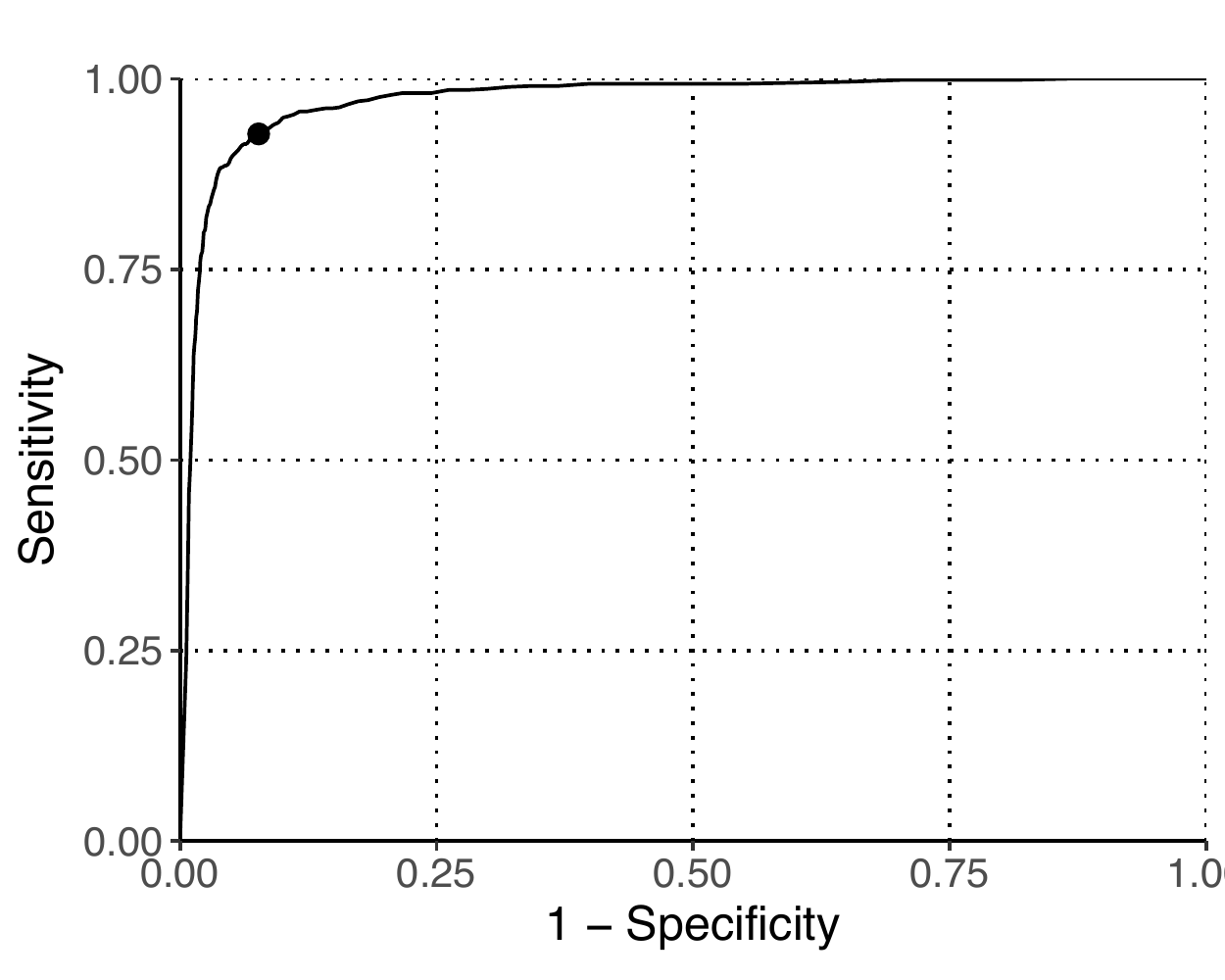}
\caption{1 hour} 
\end{subfigure}\quad
\begin{subfigure}{.31\textwidth}
\includegraphics[width=\linewidth,page=1]{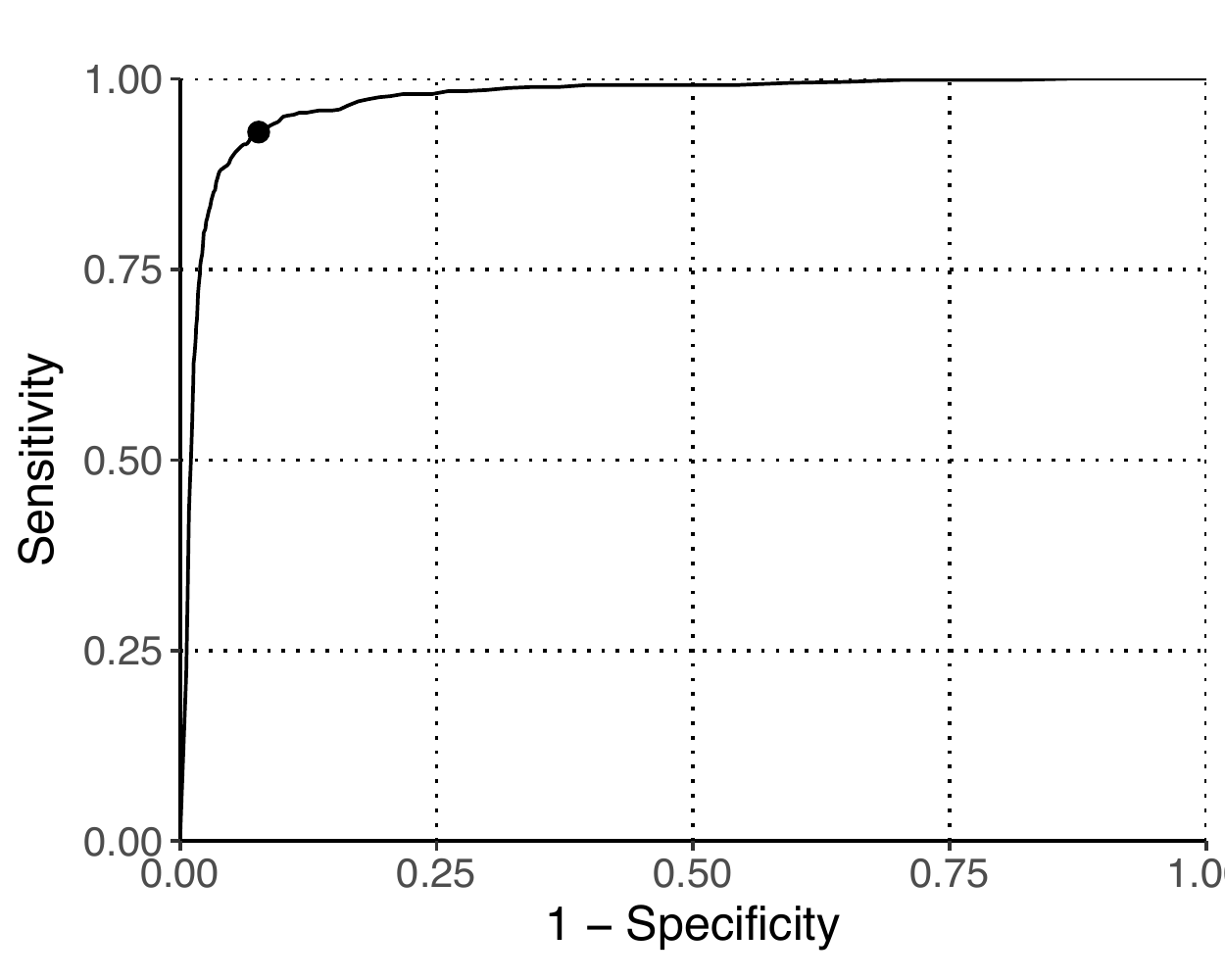}
\caption{2 hours} 
\end{subfigure}\quad
\begin{subfigure}{.31\textwidth}
\includegraphics[width=\linewidth,page=1]{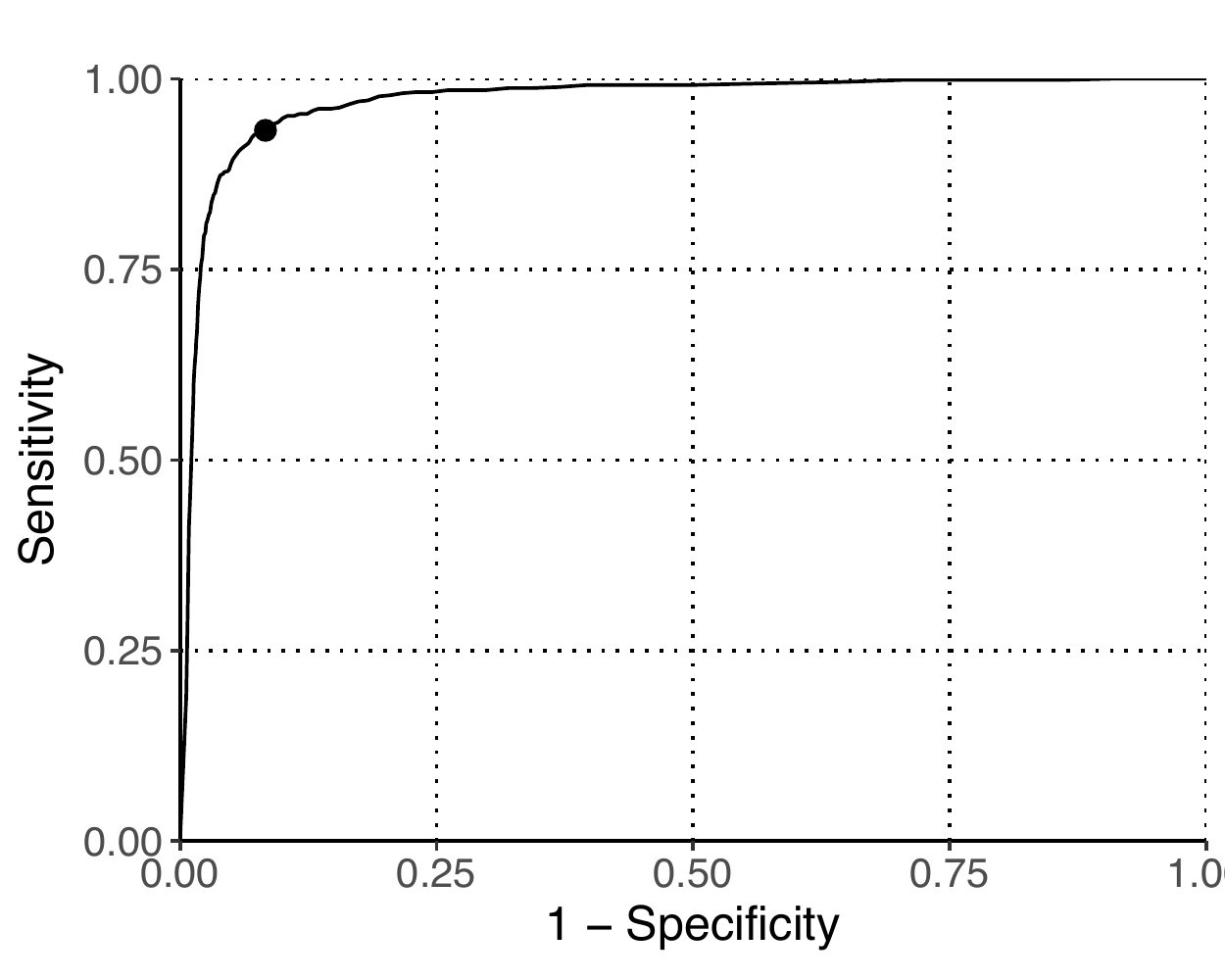}
\caption{4 hours} 
\end{subfigure}\quad
\begin{subfigure}{.31\textwidth}
\includegraphics[width=\linewidth,page=1]{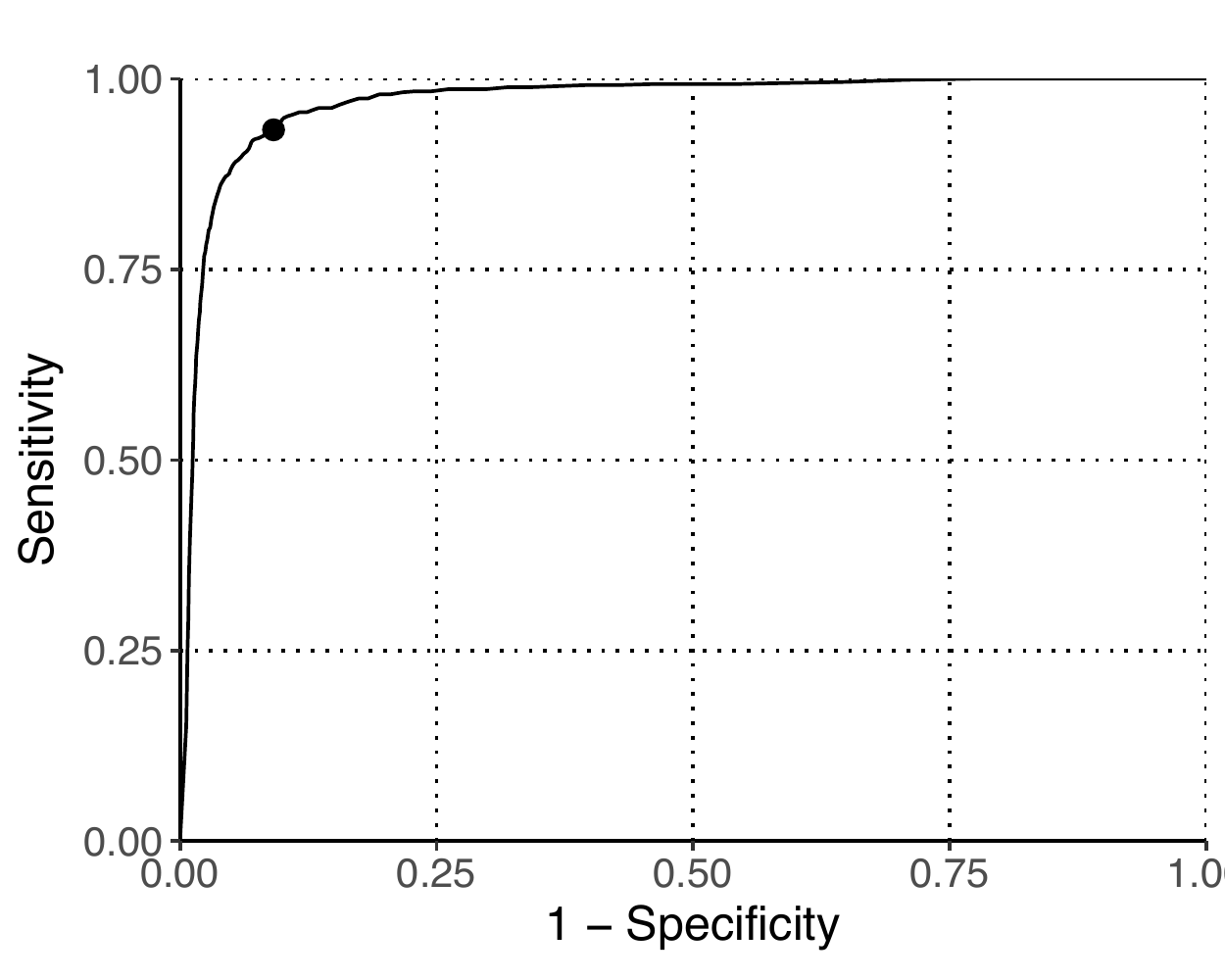}
\caption{8 hours} 
\end{subfigure}\quad
\begin{subfigure}{.31\textwidth}
\includegraphics[width=\linewidth,page=1]{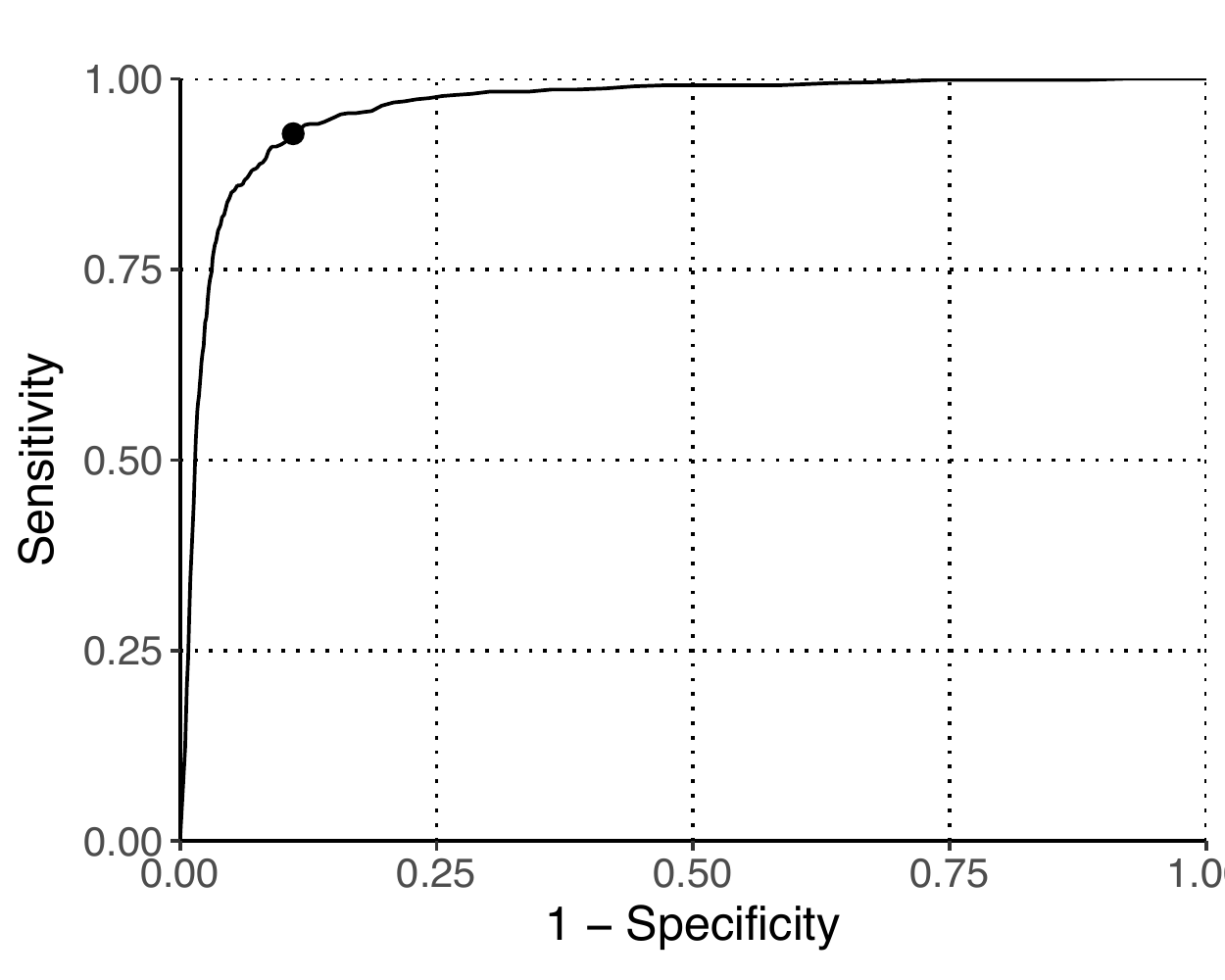}
\caption{16 hours} 
\end{subfigure}\quad
\begin{subfigure}{.31\textwidth}
\includegraphics[width=\linewidth,page=1]{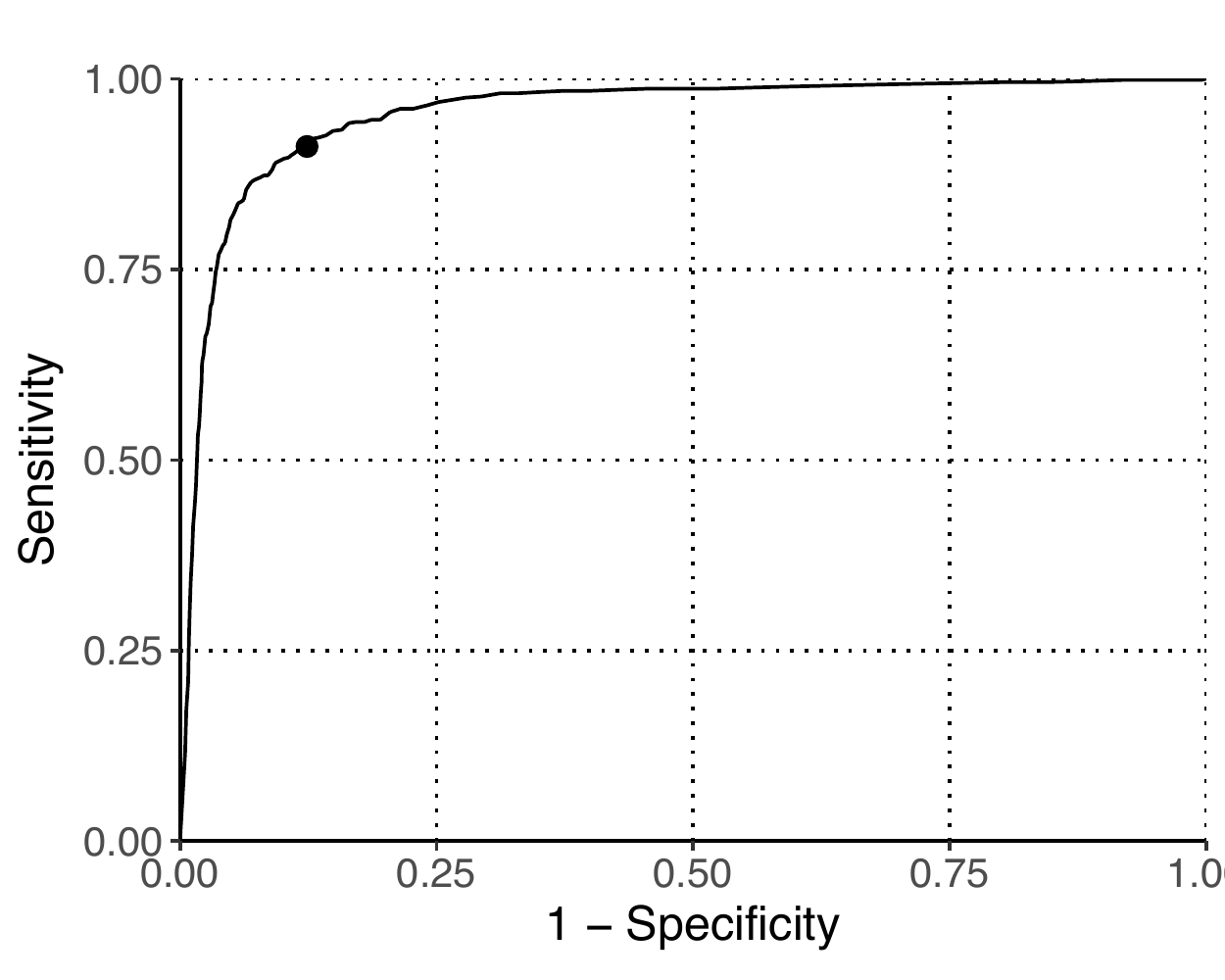}
\caption{24 hours} 
\end{subfigure} \quad
\begin{subfigure}{.31\textwidth}
\includegraphics[width=\linewidth,page=1]{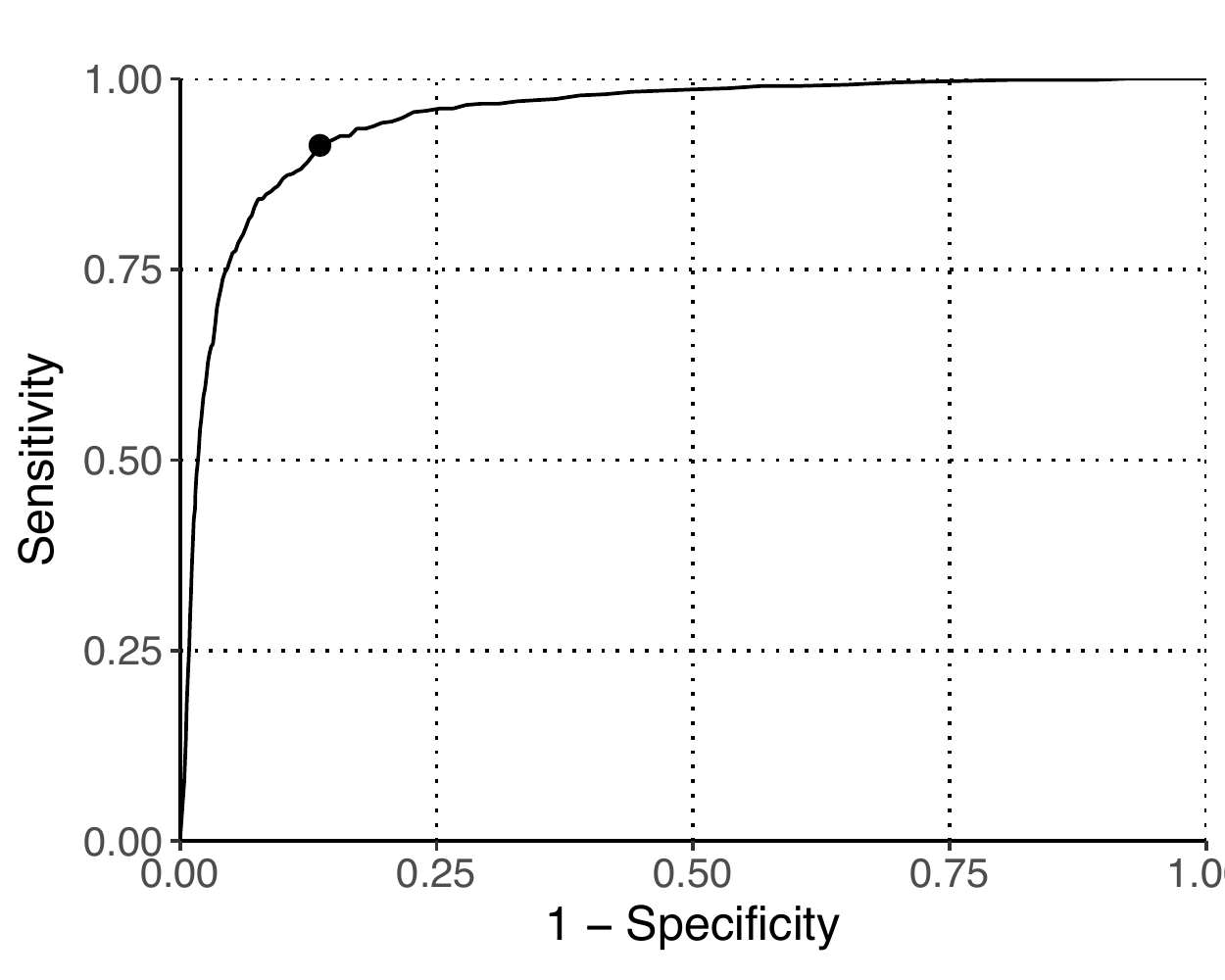}
\caption{48 hours} 
\end{subfigure}\quad
\begin{subfigure}{.31\textwidth}
\includegraphics[width=\linewidth,page=1]{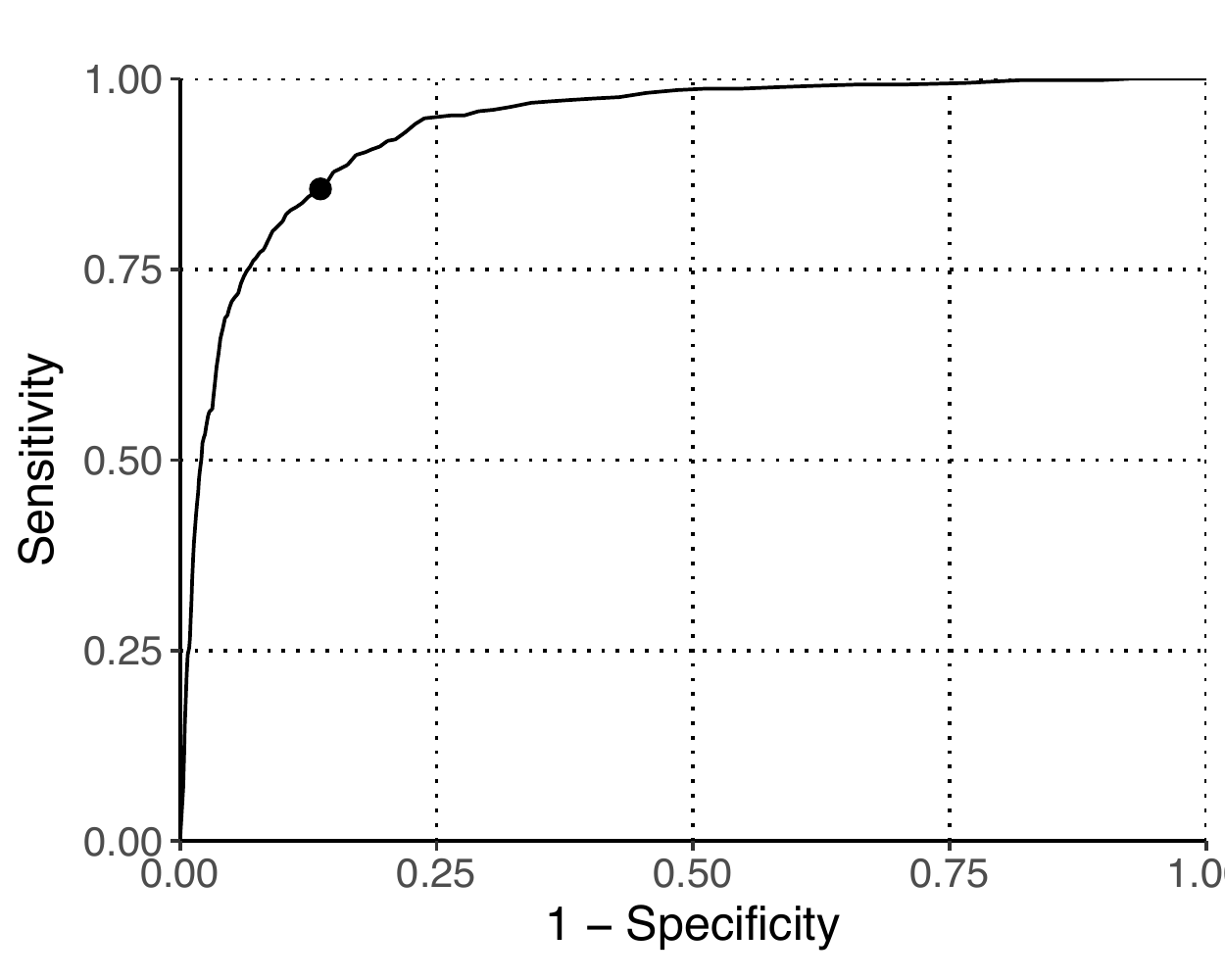}
\caption{96 hours} 
\end{subfigure}\quad
\begin{subfigure}{.31\textwidth}
\includegraphics[width=\linewidth,page=1]{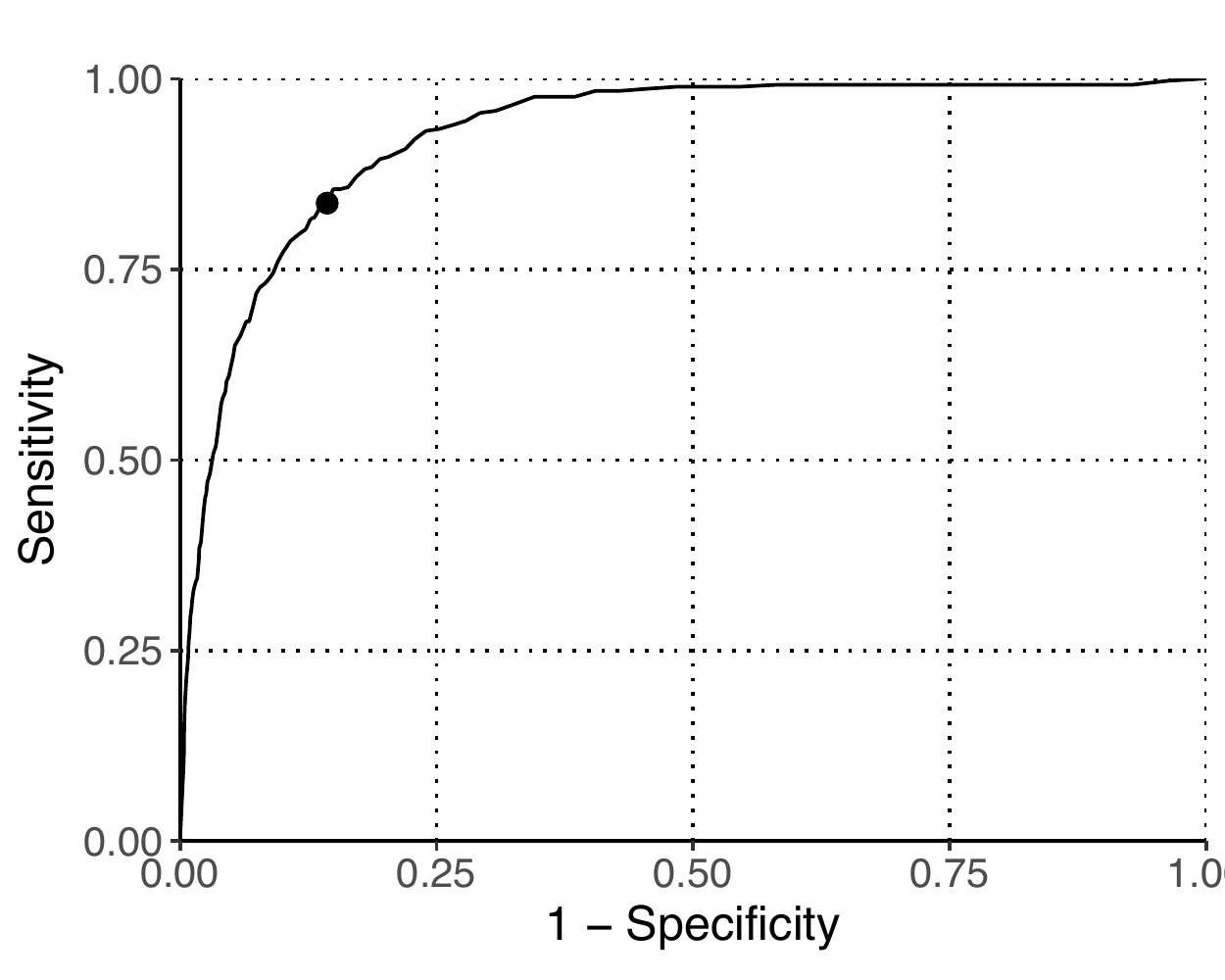}
\caption{192 hours} 
\end{subfigure}\quad
\caption{Receiver operating characteristic (ROC) curves for \themethod{} for various prediction horizons between 1 and 192 hours evaluated on the held-out Optum test set. The black dot indicates the optimal decision threshold for each prediction horizon selected on the Optum validation set as the closest point on the ROC curve to the top left coordinate (closest-to-top-left heuristic).}
\label{fig:roc_optum}
\end{figure}

\begin{figure}[h]
{\centering\textbf{\textsf{\themethod{} Receiver Operating Characteristic (TriNetX)}}\par\medskip}
\begin{subfigure}{.31\textwidth}
\includegraphics[width=\linewidth,page=1]{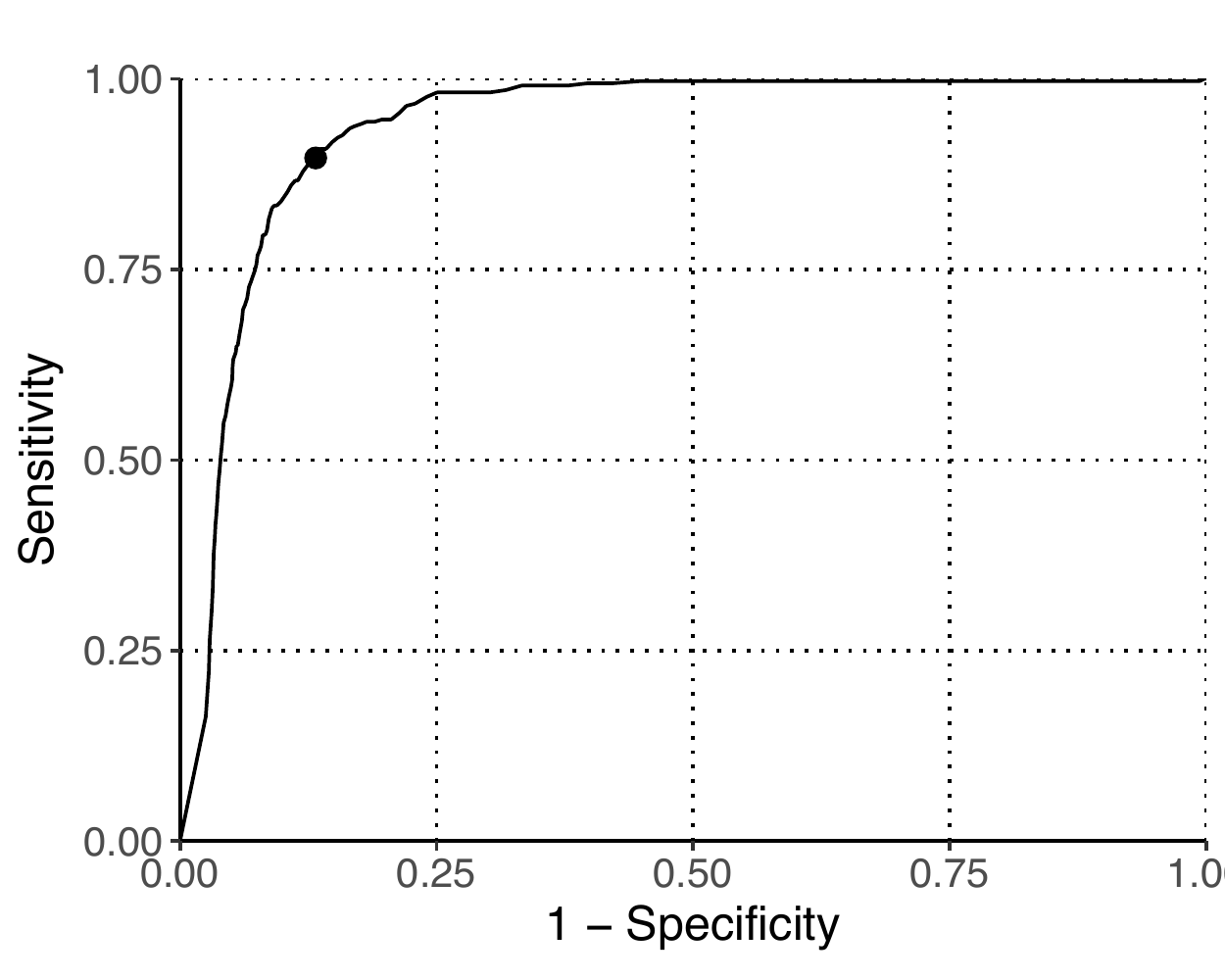}
\caption{1 hour} 
\end{subfigure}\quad
\begin{subfigure}{.31\textwidth}
\includegraphics[width=\linewidth,page=1]{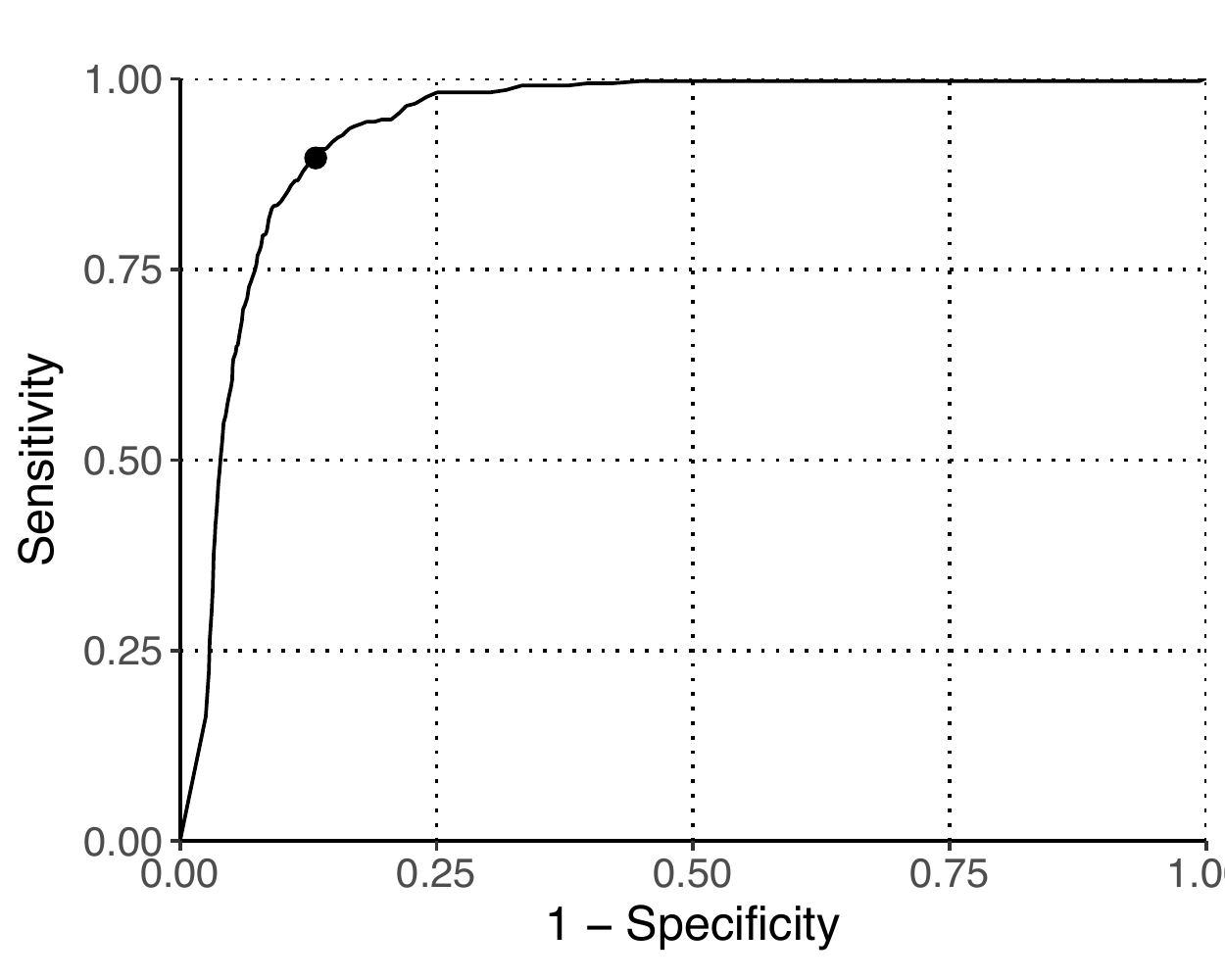}
\caption{2 hours} 
\end{subfigure}\quad
\begin{subfigure}{.31\textwidth}
\includegraphics[width=\linewidth,page=1]{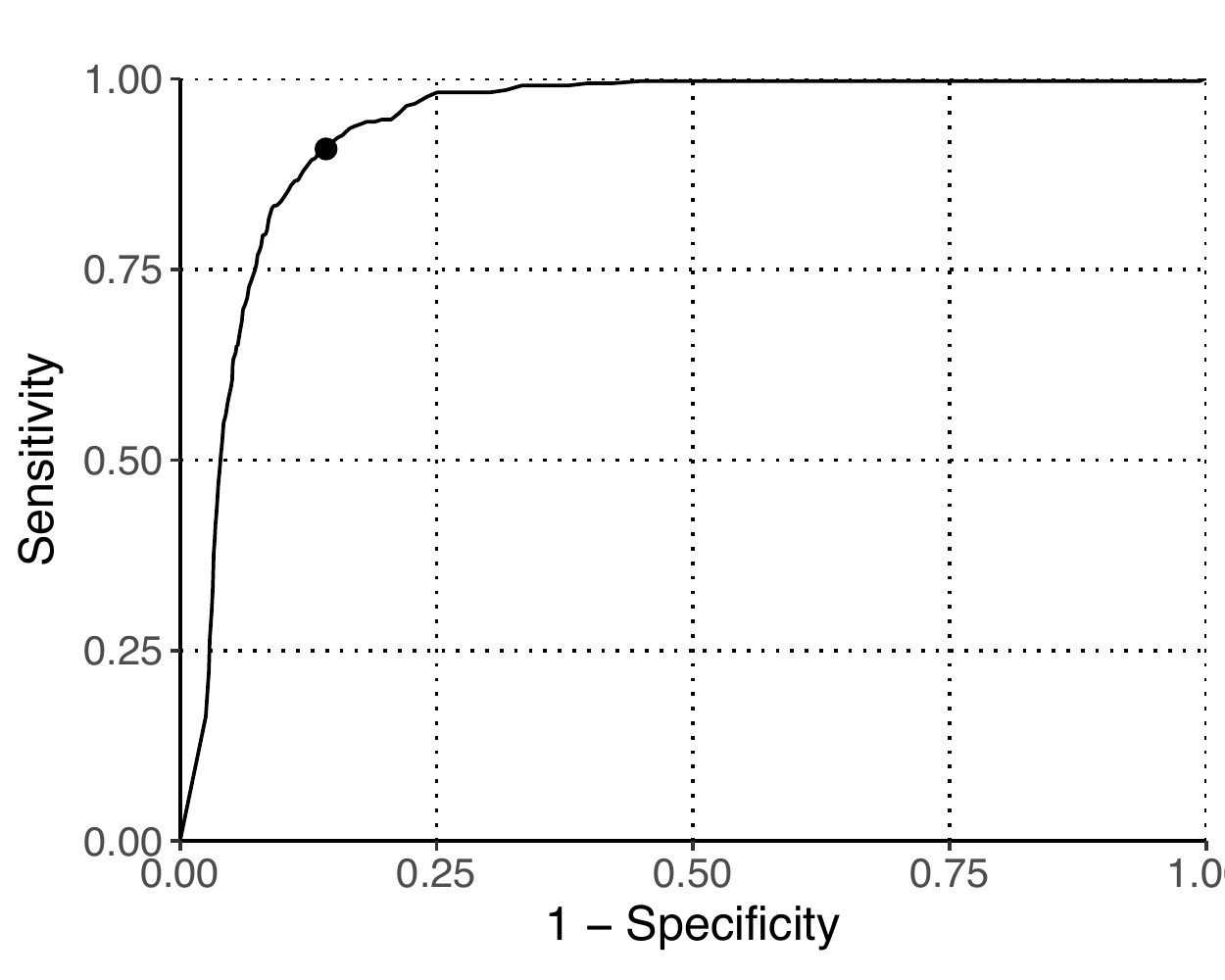}
\caption{4 hours} 
\end{subfigure}\quad
\begin{subfigure}{.31\textwidth}
\includegraphics[width=\linewidth,page=1]{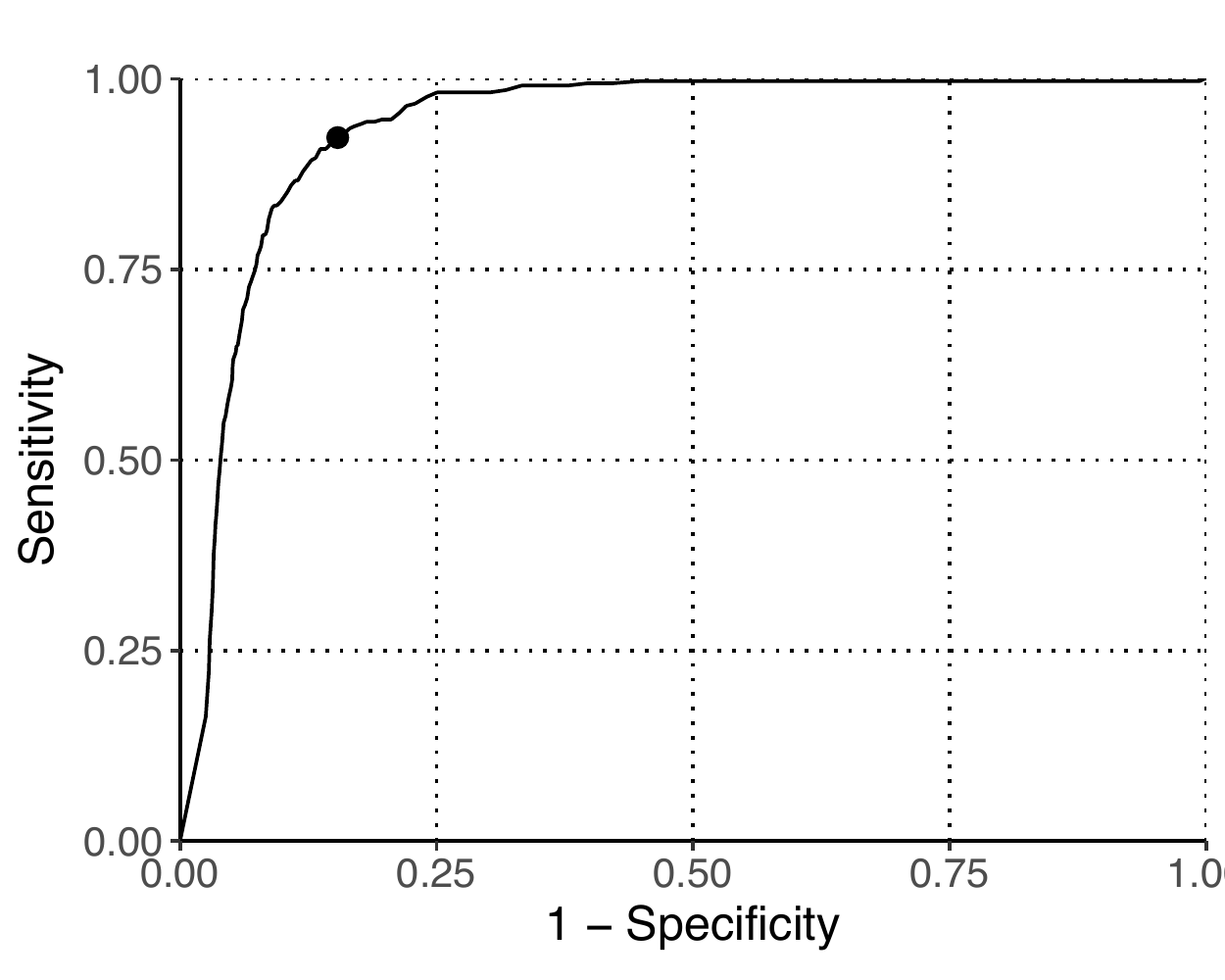}
\caption{8 hours} 
\end{subfigure}\quad
\begin{subfigure}{.31\textwidth}
\includegraphics[width=\linewidth,page=1]{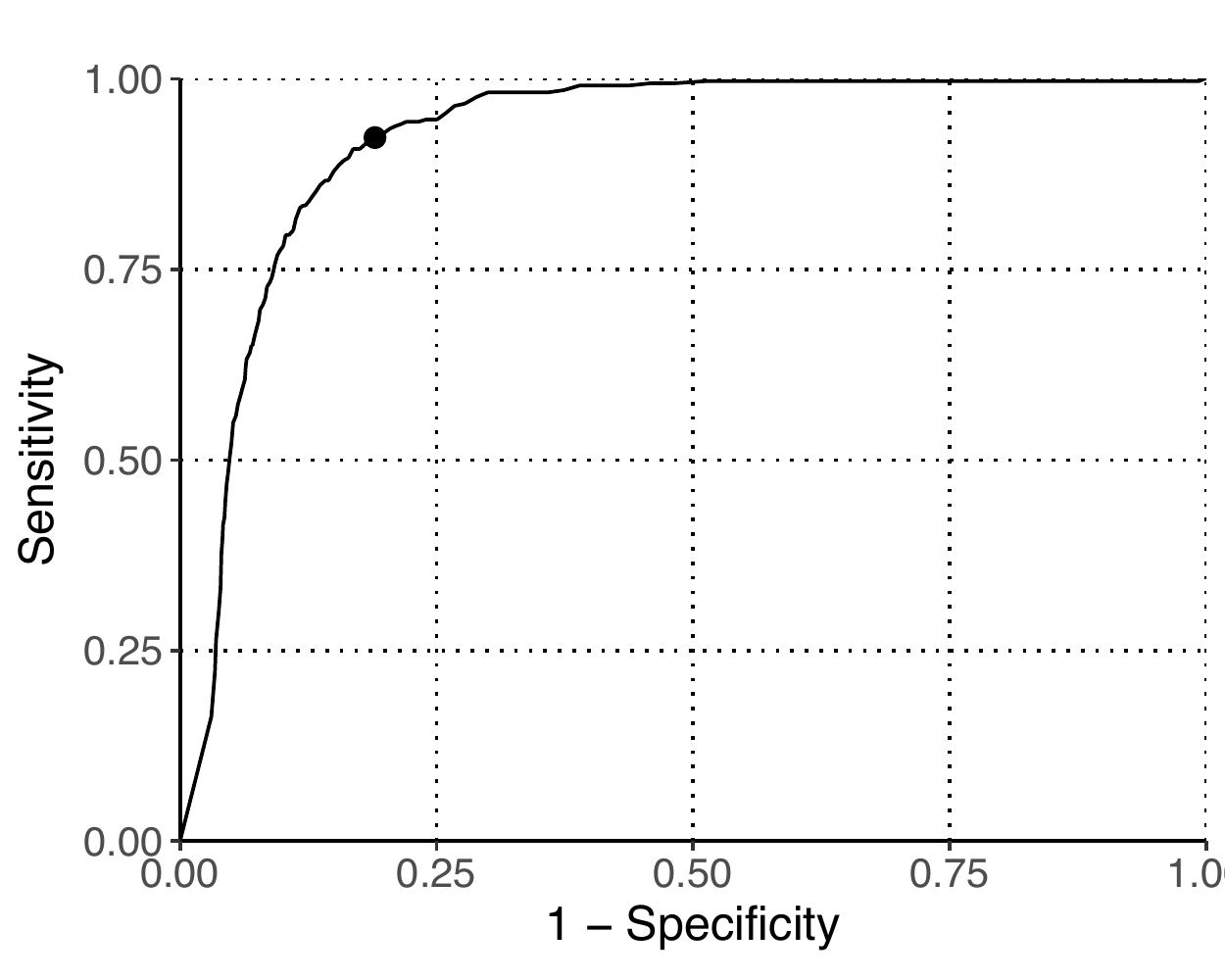}
\caption{16 hours} 
\end{subfigure}\quad
\begin{subfigure}{.31\textwidth}
\includegraphics[width=\linewidth,page=1]{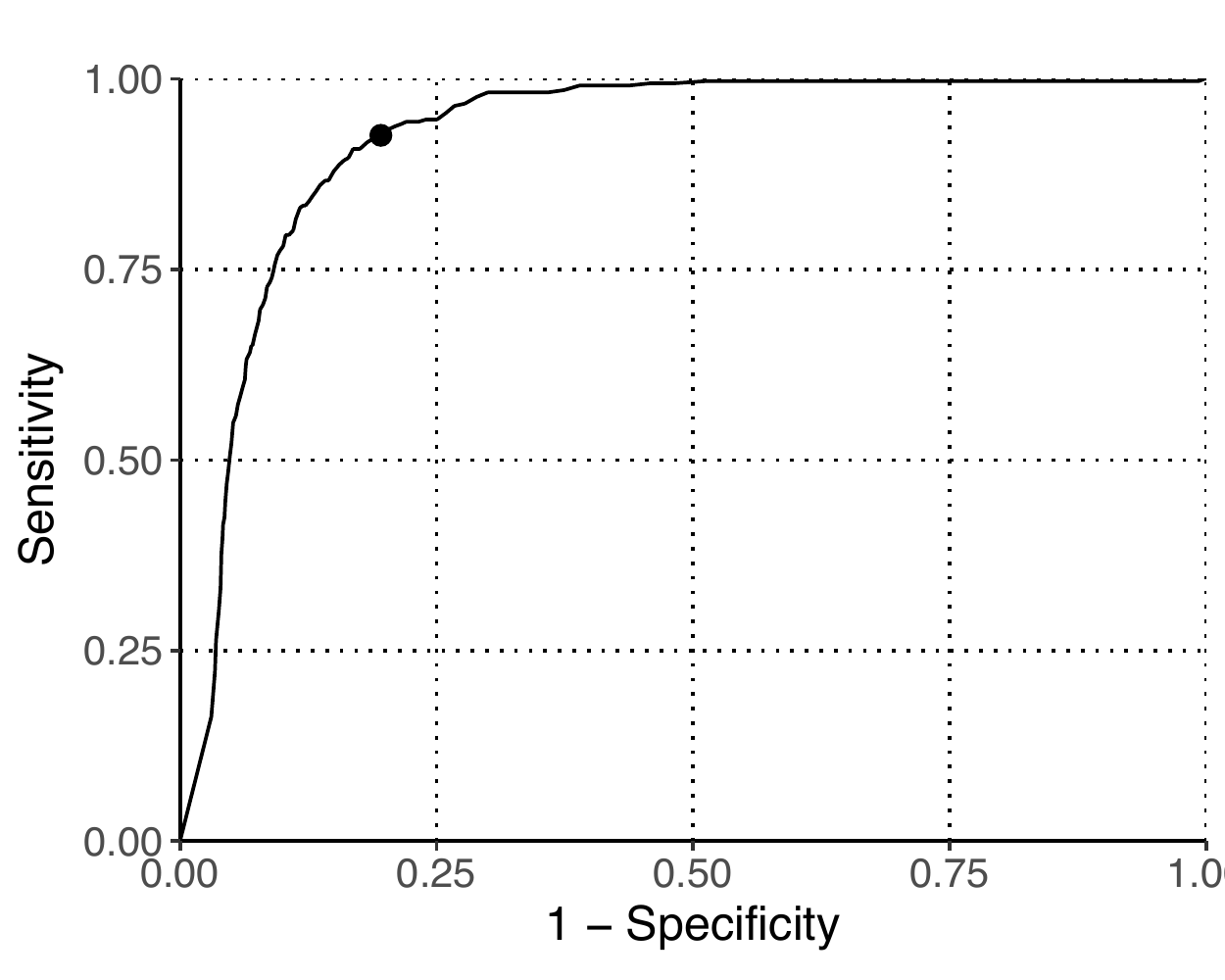}
\caption{24 hours} 
\end{subfigure} \quad
\begin{subfigure}{.31\textwidth}
\includegraphics[width=\linewidth,page=1]{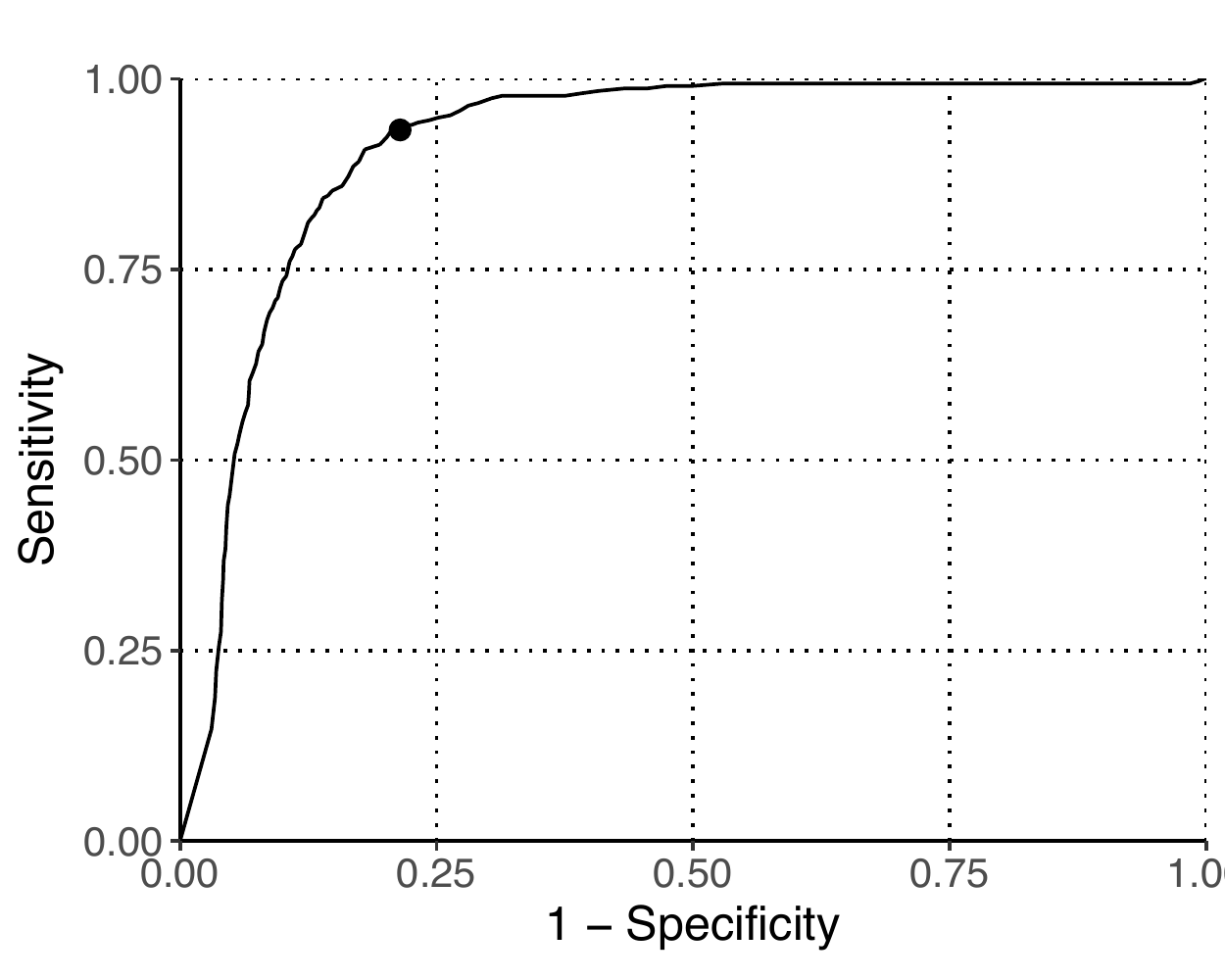}
\caption{48 hours} 
\end{subfigure}\quad
\begin{subfigure}{.31\textwidth}
\includegraphics[width=\linewidth,page=1]{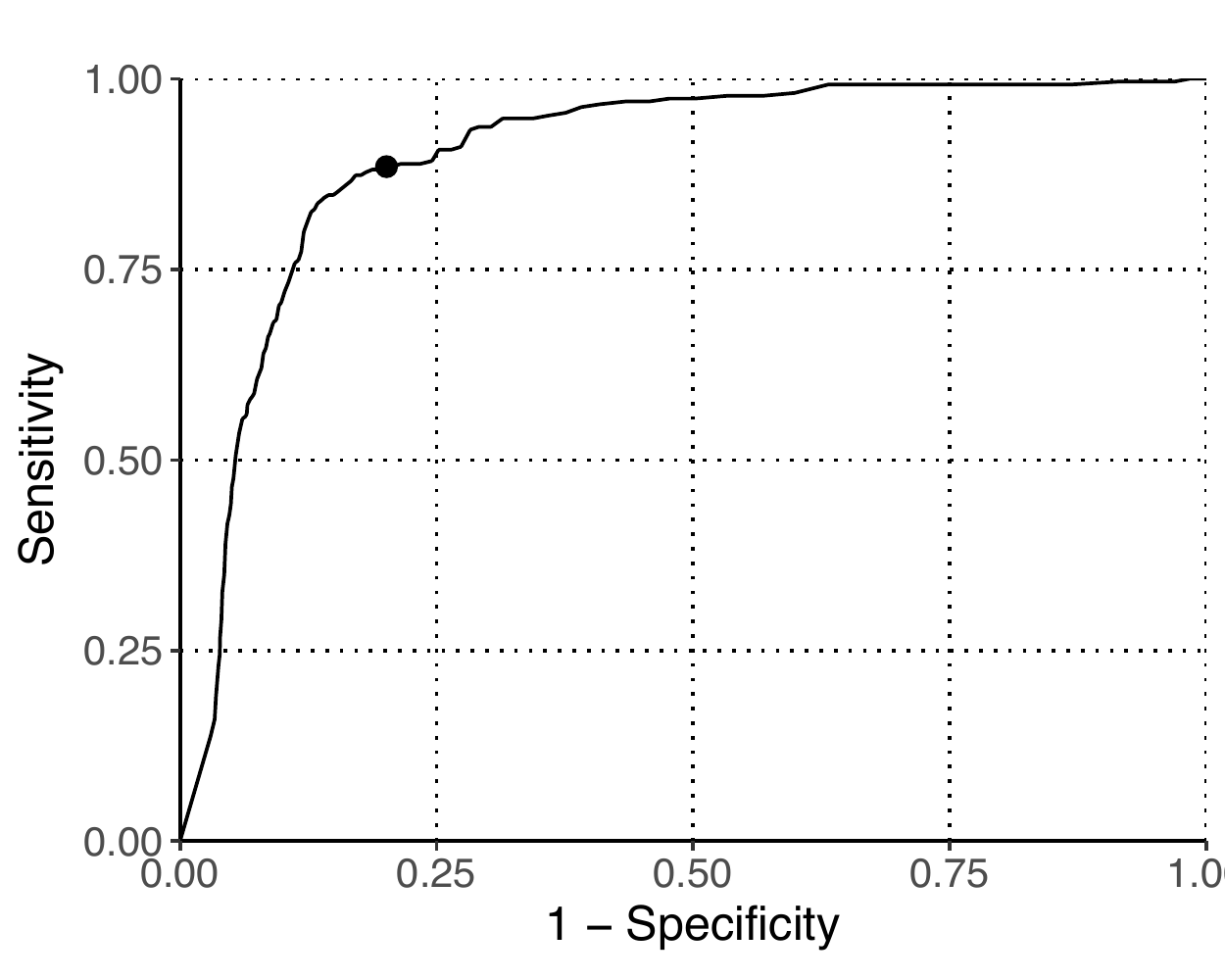}
\caption{96 hours} 
\end{subfigure}\quad
\begin{subfigure}{.31\textwidth}
\includegraphics[width=\linewidth,page=1]{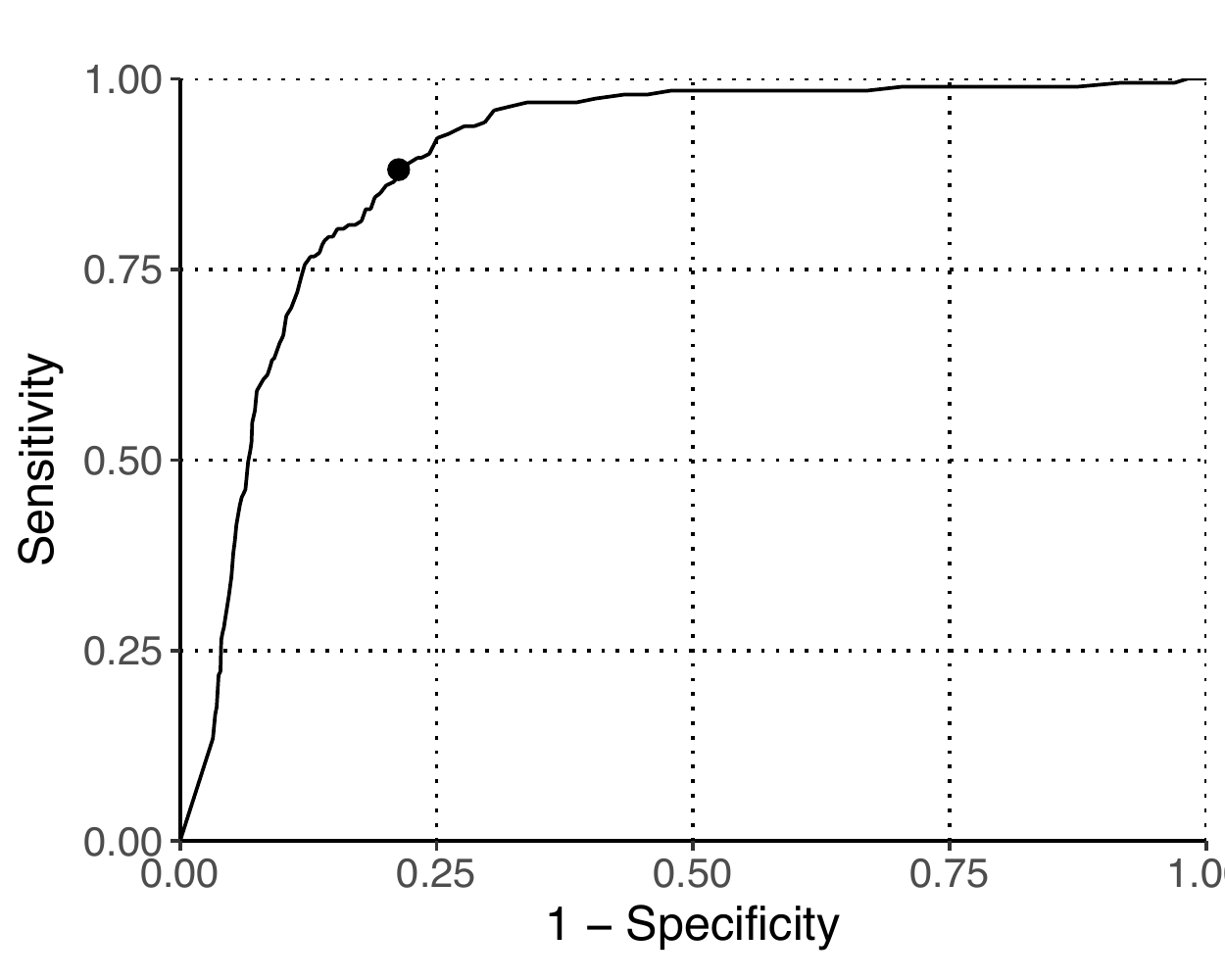}
\caption{192 hours} 
\end{subfigure}\quad
\caption{Receiver operating characteristic (ROC) curves for \themethod{} for various prediction horizons between 1 and 192 hours evaluated on the external TriNetX test set. The black dot indicates the optimal decision threshold for each prediction horizon selected on the Optum validation set as the closest point on the ROC curve to the top left coordinate (closest-to-top-left heuristic).}
\label{fig:roc_trinetx}
\end{figure}

\begin{figure}[h]
{\centering\textbf{\textsf{\themethod{} (linear) Receiver Operating Characteristic (Optum Test Set)}}\par\medskip}
\begin{subfigure}{.31\textwidth}
\includegraphics[width=\linewidth,page=1]{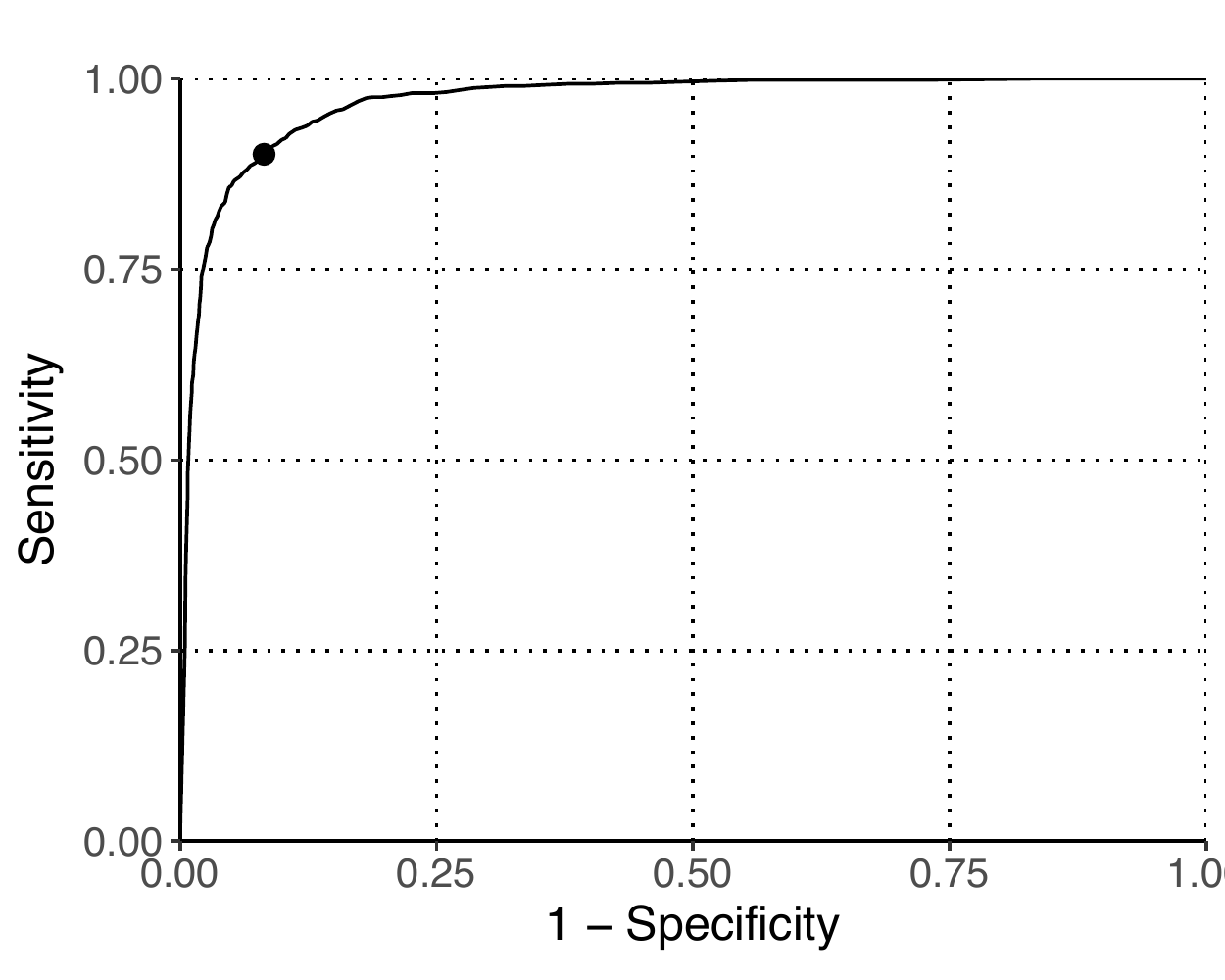}
\caption{1 hour} 
\end{subfigure}\quad
\begin{subfigure}{.31\textwidth}
\includegraphics[width=\linewidth,page=1]{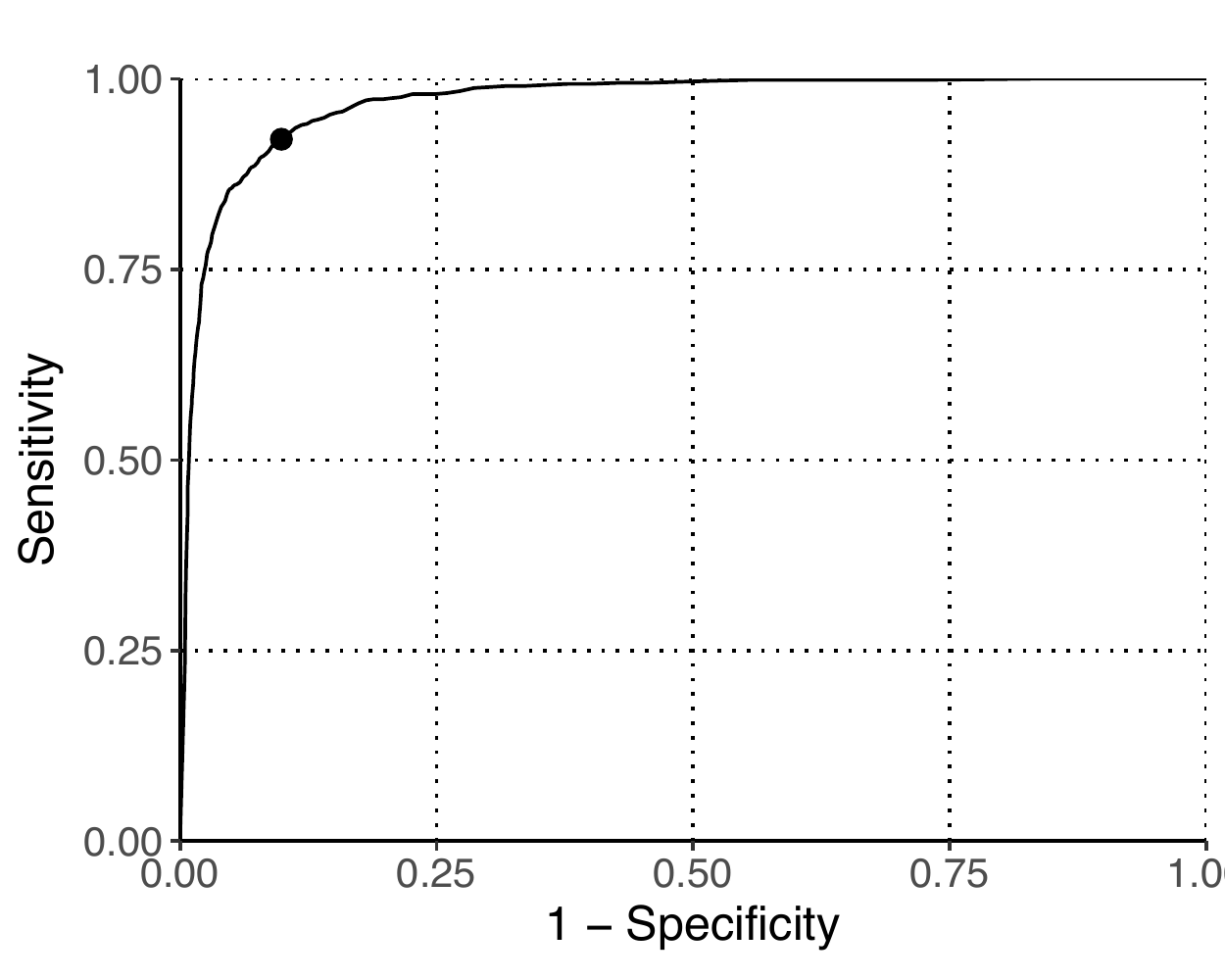}
\caption{2 hours} 
\end{subfigure}\quad
\begin{subfigure}{.31\textwidth}
\includegraphics[width=\linewidth,page=1]{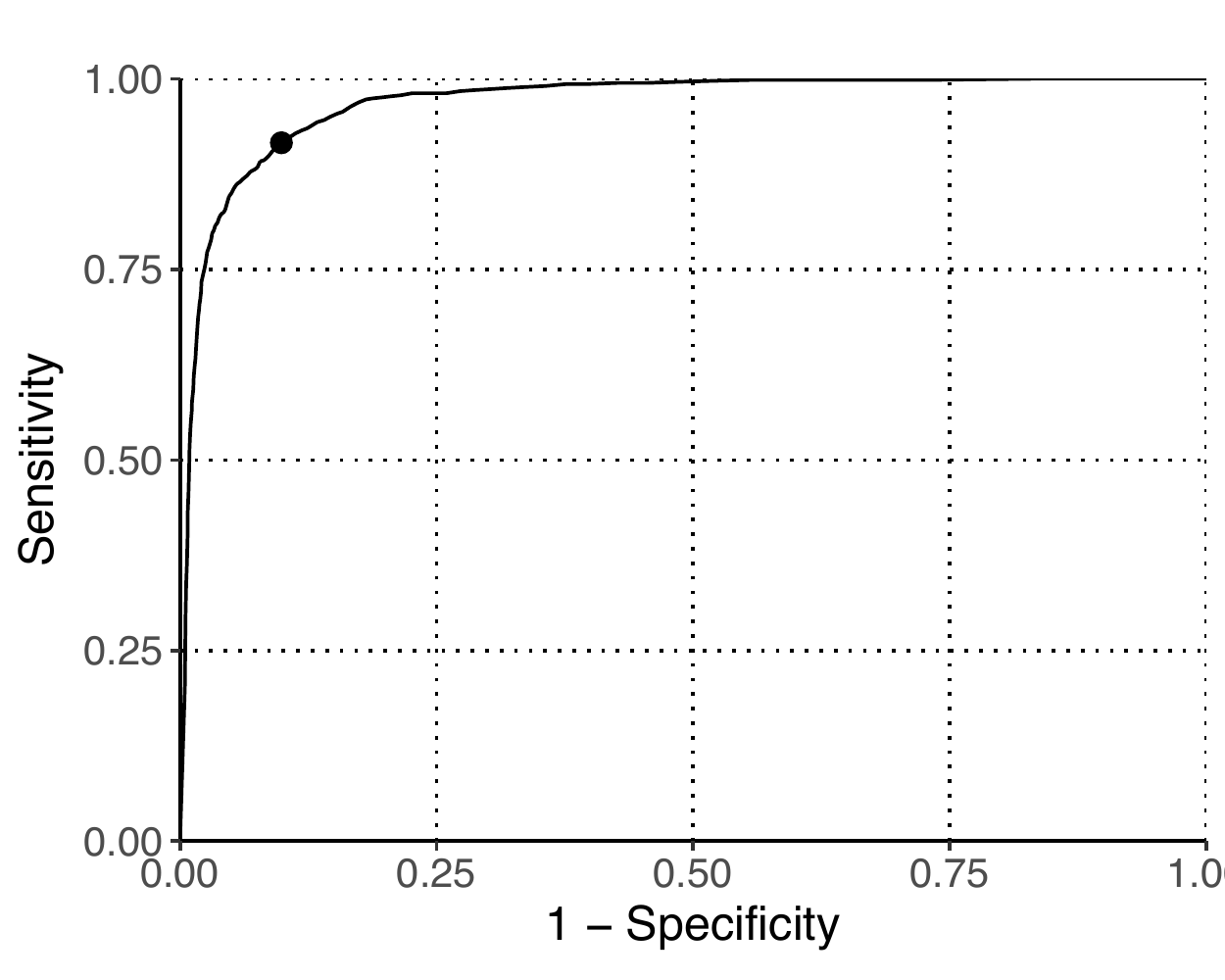}
\caption{4 hours} 
\end{subfigure}\quad
\begin{subfigure}{.31\textwidth}
\includegraphics[width=\linewidth,page=1]{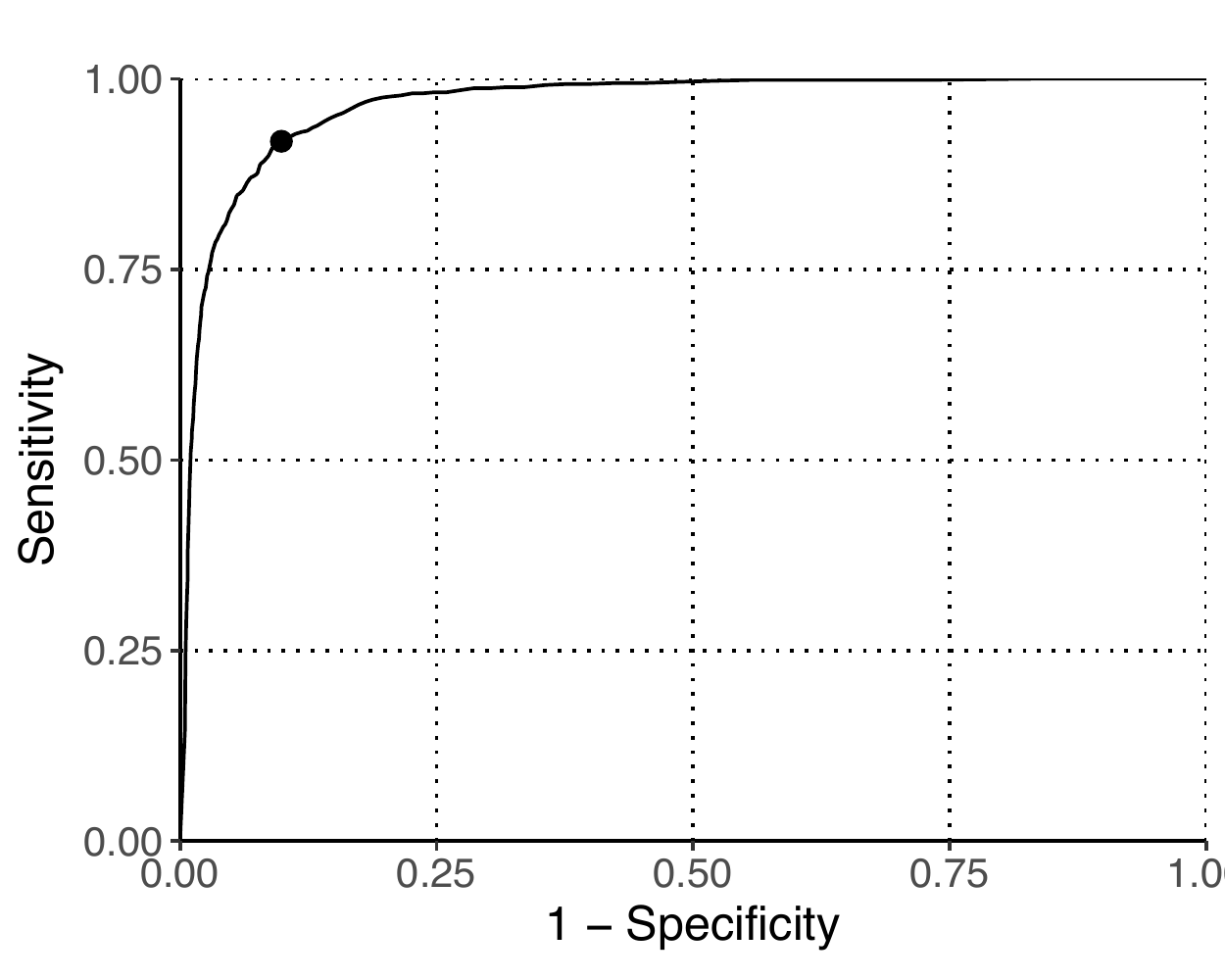}
\caption{8 hours} 
\end{subfigure}\quad
\begin{subfigure}{.31\textwidth}
\includegraphics[width=\linewidth,page=1]{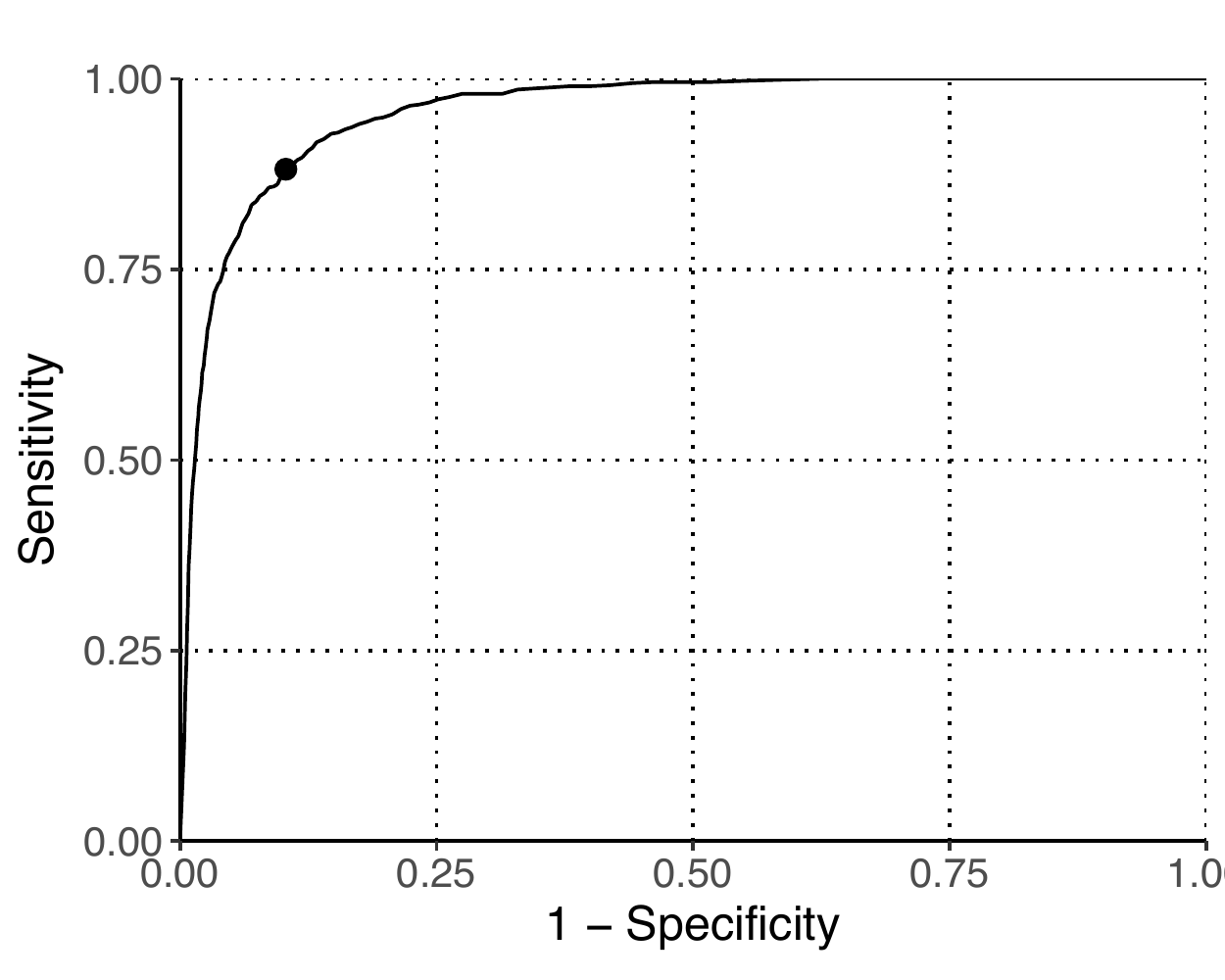}
\caption{16 hours} 
\end{subfigure}\quad
\begin{subfigure}{.31\textwidth}
\includegraphics[width=\linewidth,page=1]{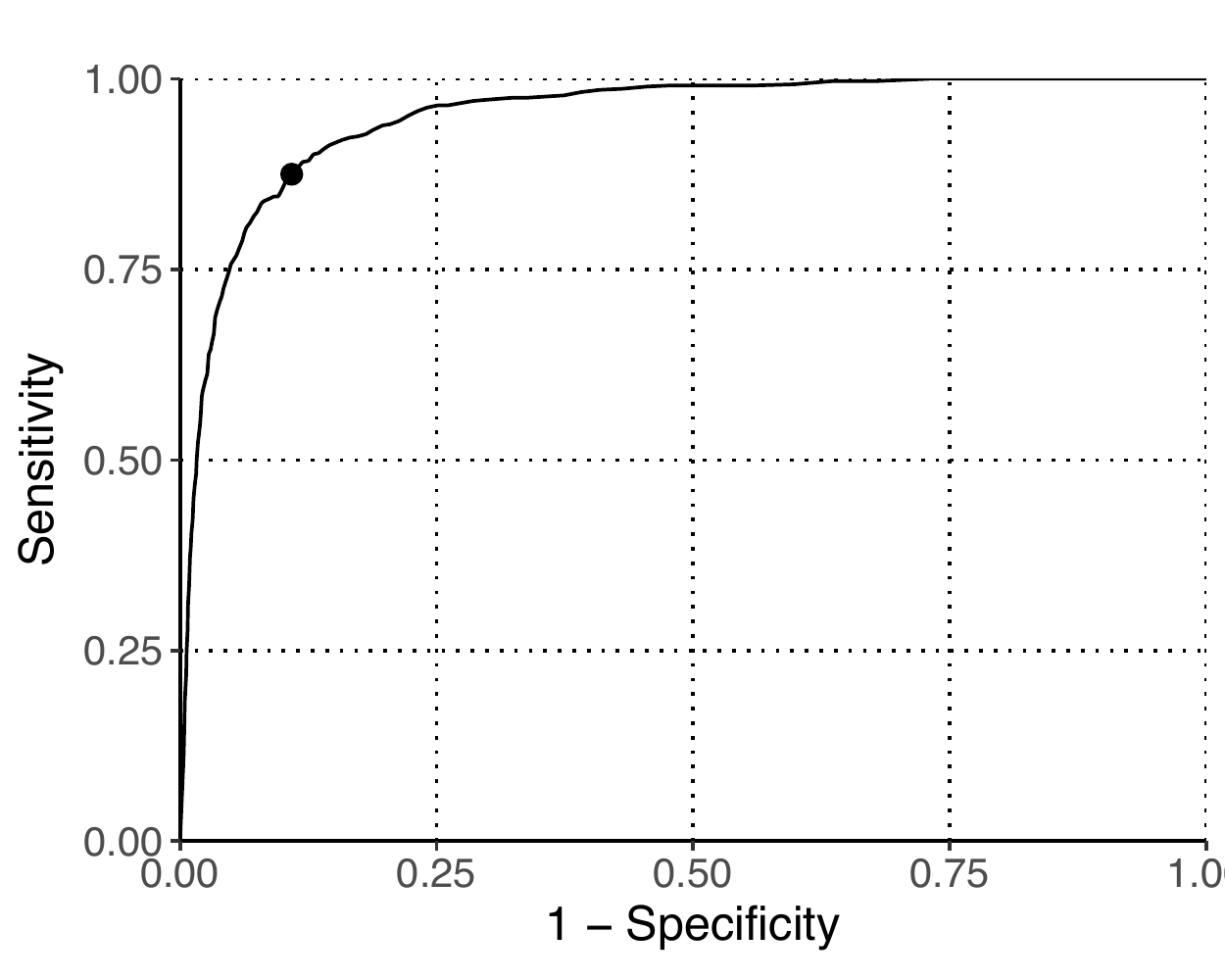}
\caption{24 hours} 
\end{subfigure} \quad
\begin{subfigure}{.31\textwidth}
\includegraphics[width=\linewidth,page=1]{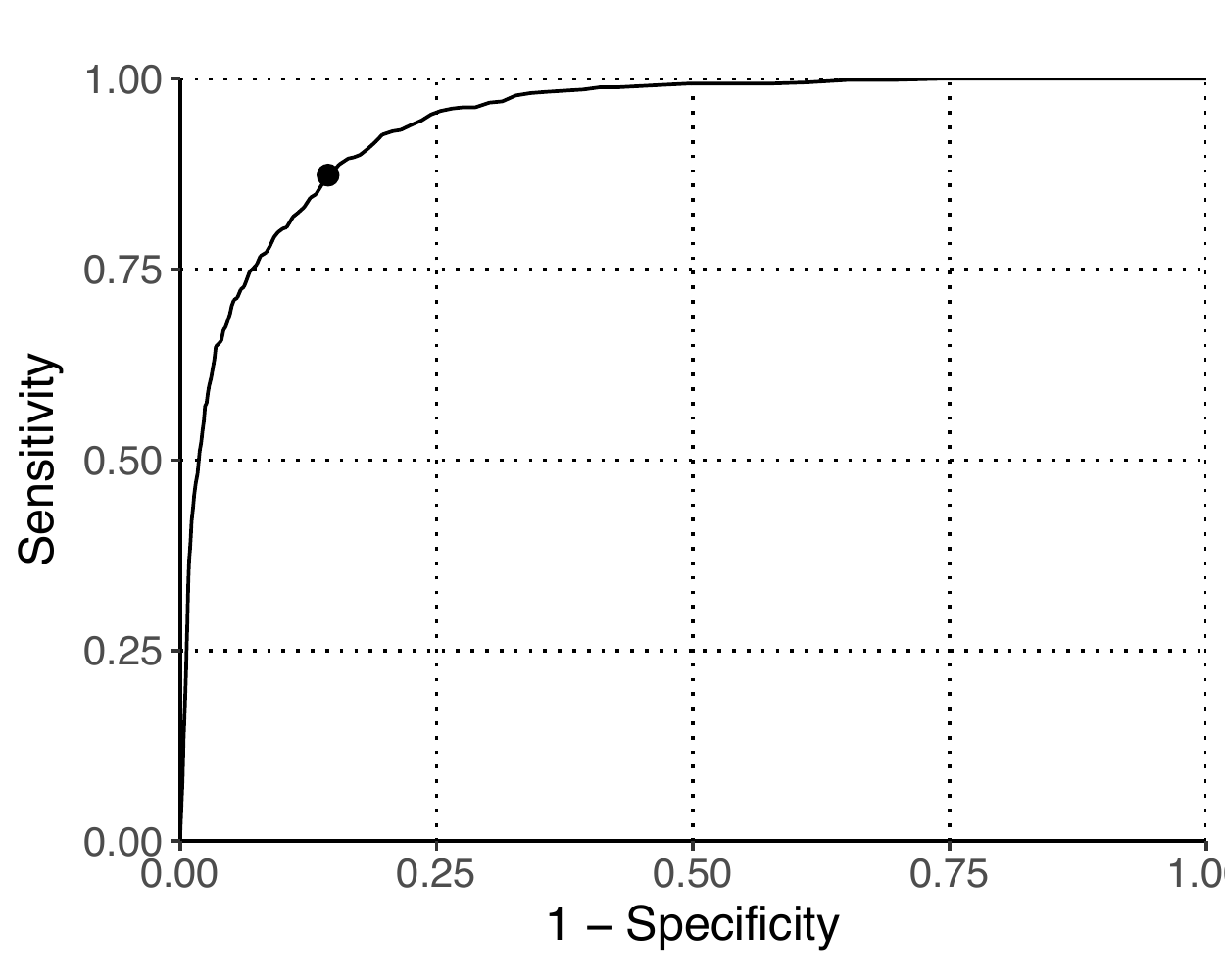}
\caption{48 hours} 
\end{subfigure}\quad
\begin{subfigure}{.31\textwidth}
\includegraphics[width=\linewidth,page=1]{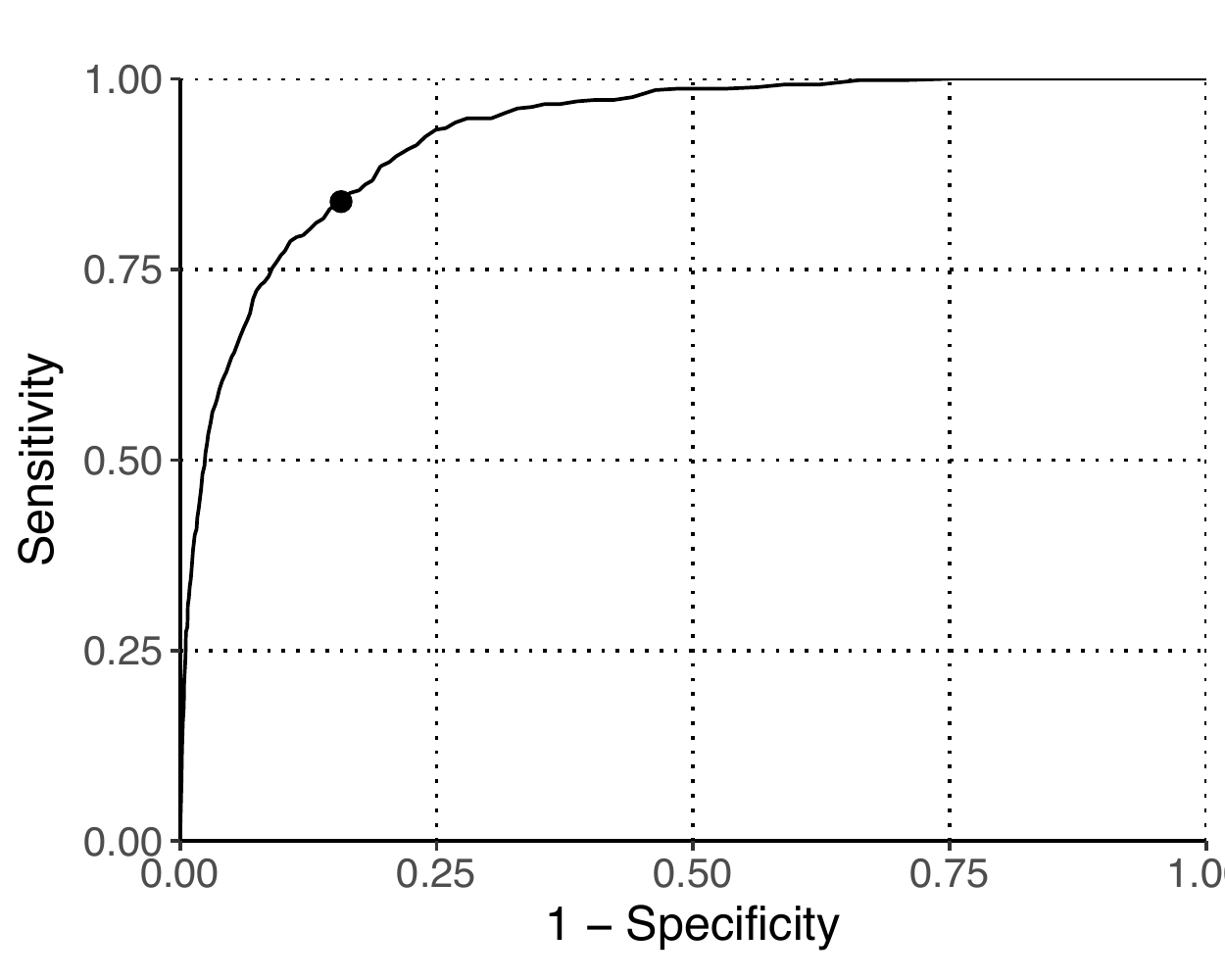}
\caption{96 hours} 
\end{subfigure}\quad
\begin{subfigure}{.31\textwidth}
\includegraphics[width=\linewidth,page=1]{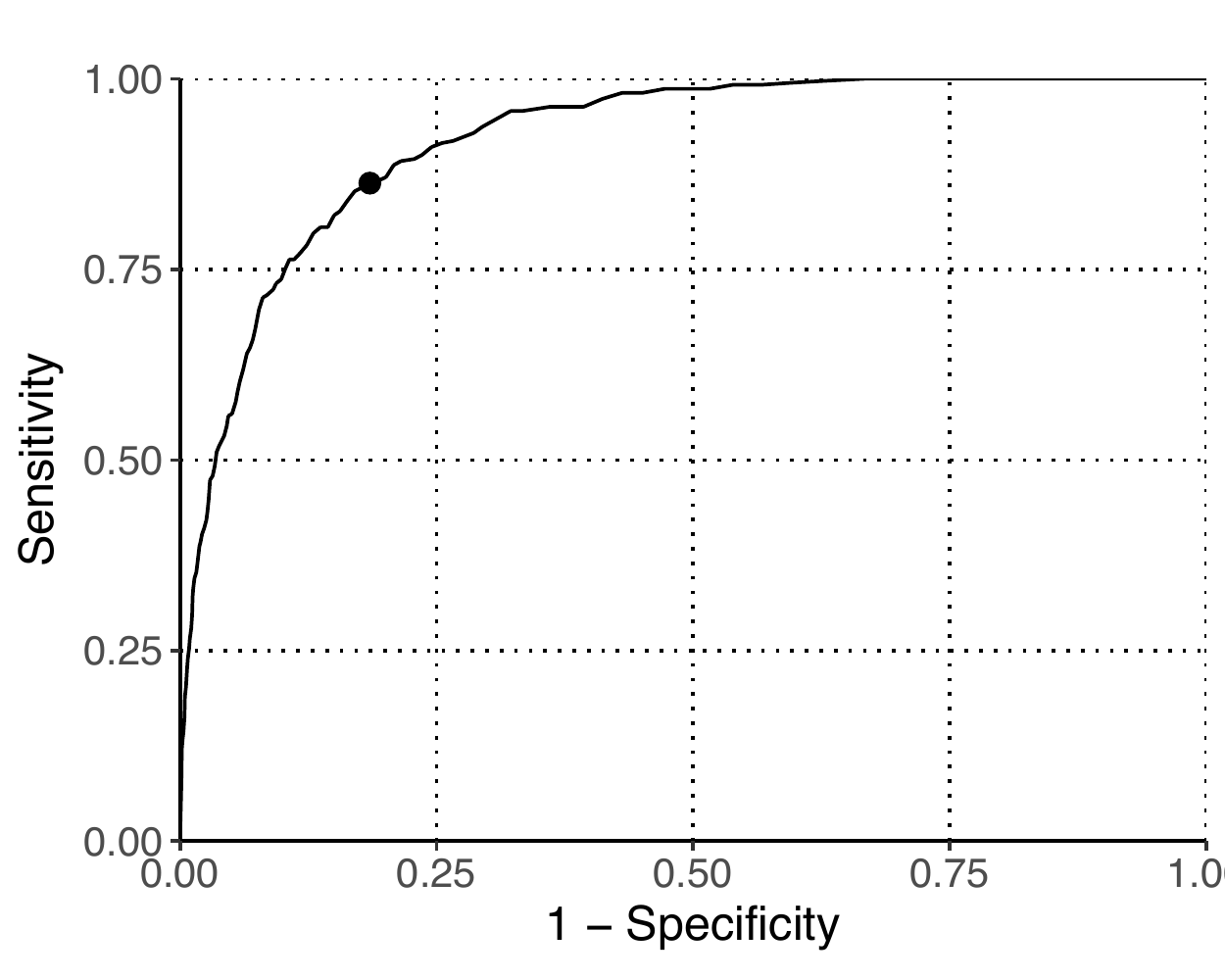}
\caption{192 hours} 
\end{subfigure}\quad
\caption{Receiver operating characteristic (ROC) curves for \themethod{} (linear) for various prediction horizons between 1 and 192 hours evaluated on the held-out Optum test set. The black dot indicates the optimal decision threshold for each prediction horizon selected on the Optum validation set as the closest point on the ROC curve to the top left coordinate (closest-to-top-left heuristic).}
\label{fig:roc_optum_linear}
\end{figure}

\begin{figure}[h]
{\centering\textbf{\textsf{\themethod{} (linear) Receiver Operating Characteristic (TriNetX)}}\par\medskip}
\begin{subfigure}{.31\textwidth}
\includegraphics[width=\linewidth,page=1]{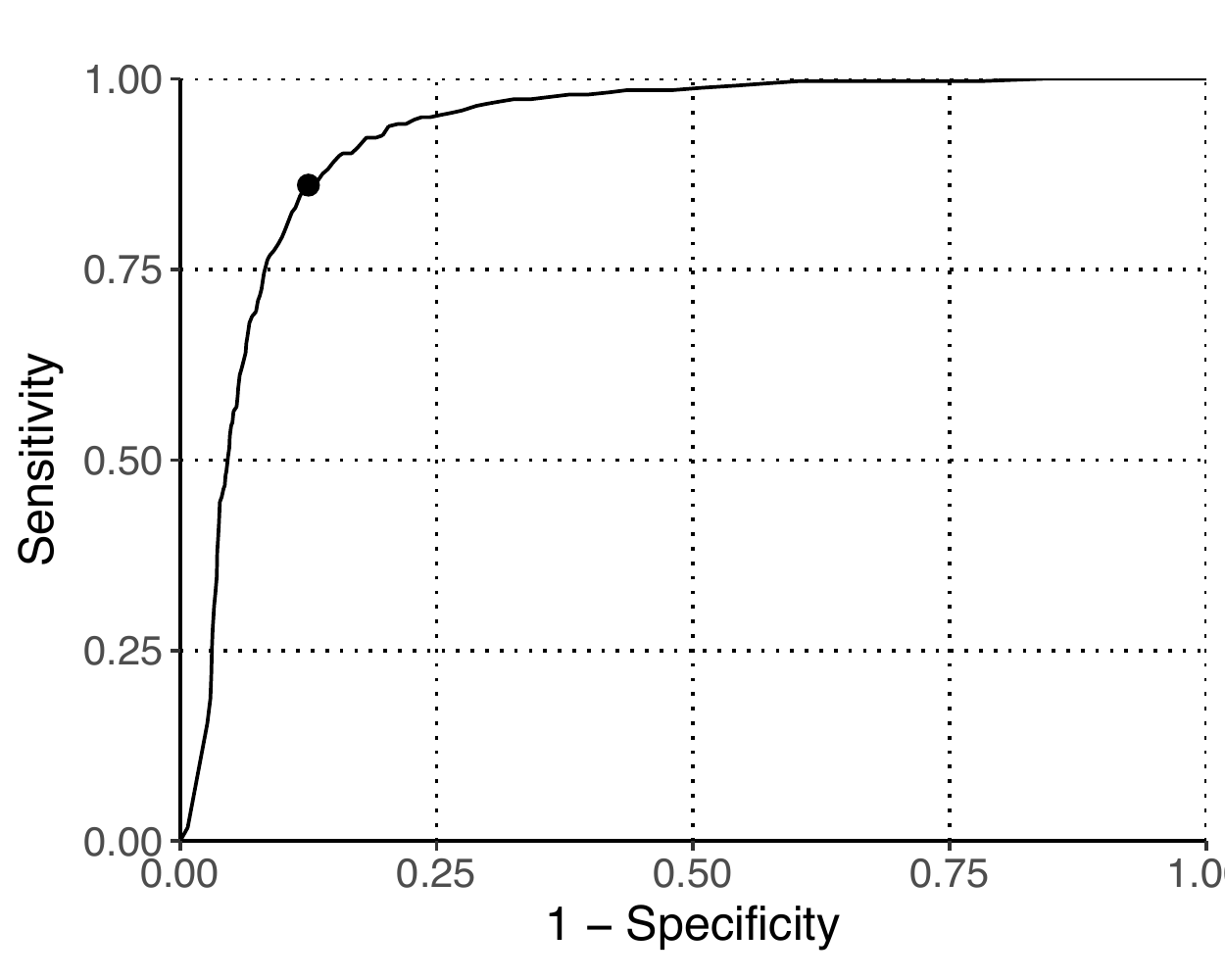}
\caption{1 hour} 
\end{subfigure}\quad
\begin{subfigure}{.31\textwidth}
\includegraphics[width=\linewidth,page=1]{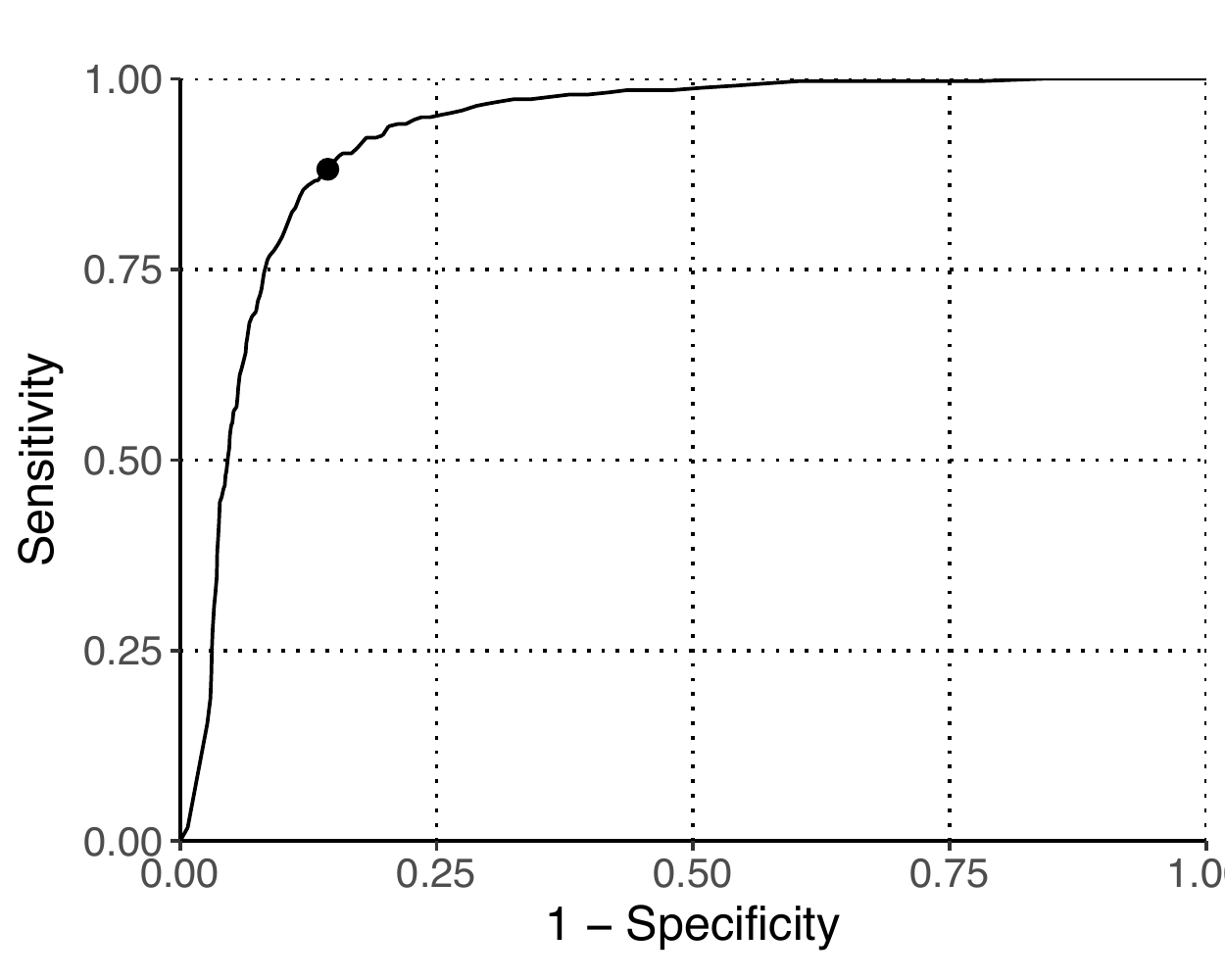}
\caption{2 hours} 
\end{subfigure}\quad
\begin{subfigure}{.31\textwidth}
\includegraphics[width=\linewidth,page=1]{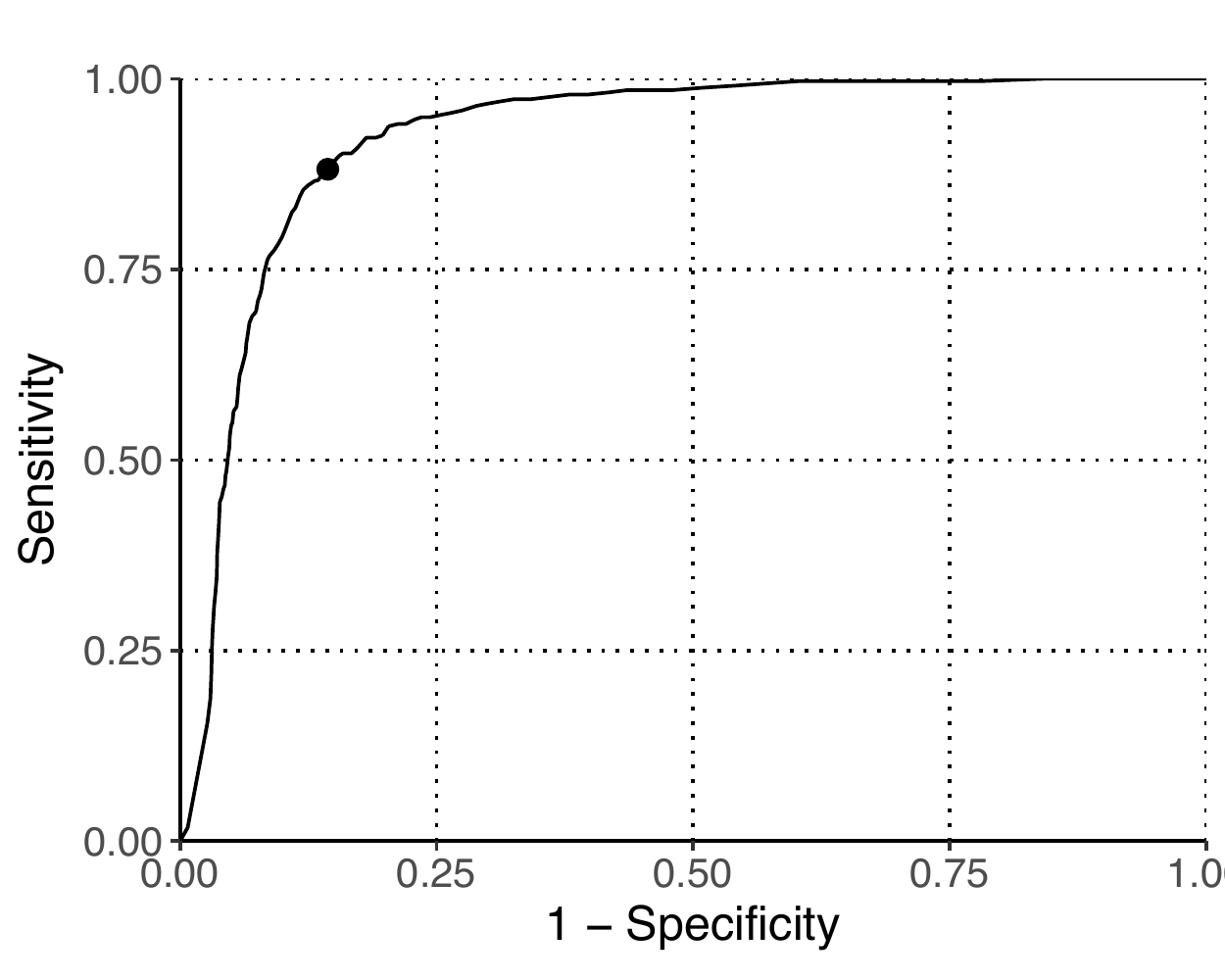}
\caption{4 hours} 
\end{subfigure}\quad
\begin{subfigure}{.31\textwidth}
\includegraphics[width=\linewidth,page=1]{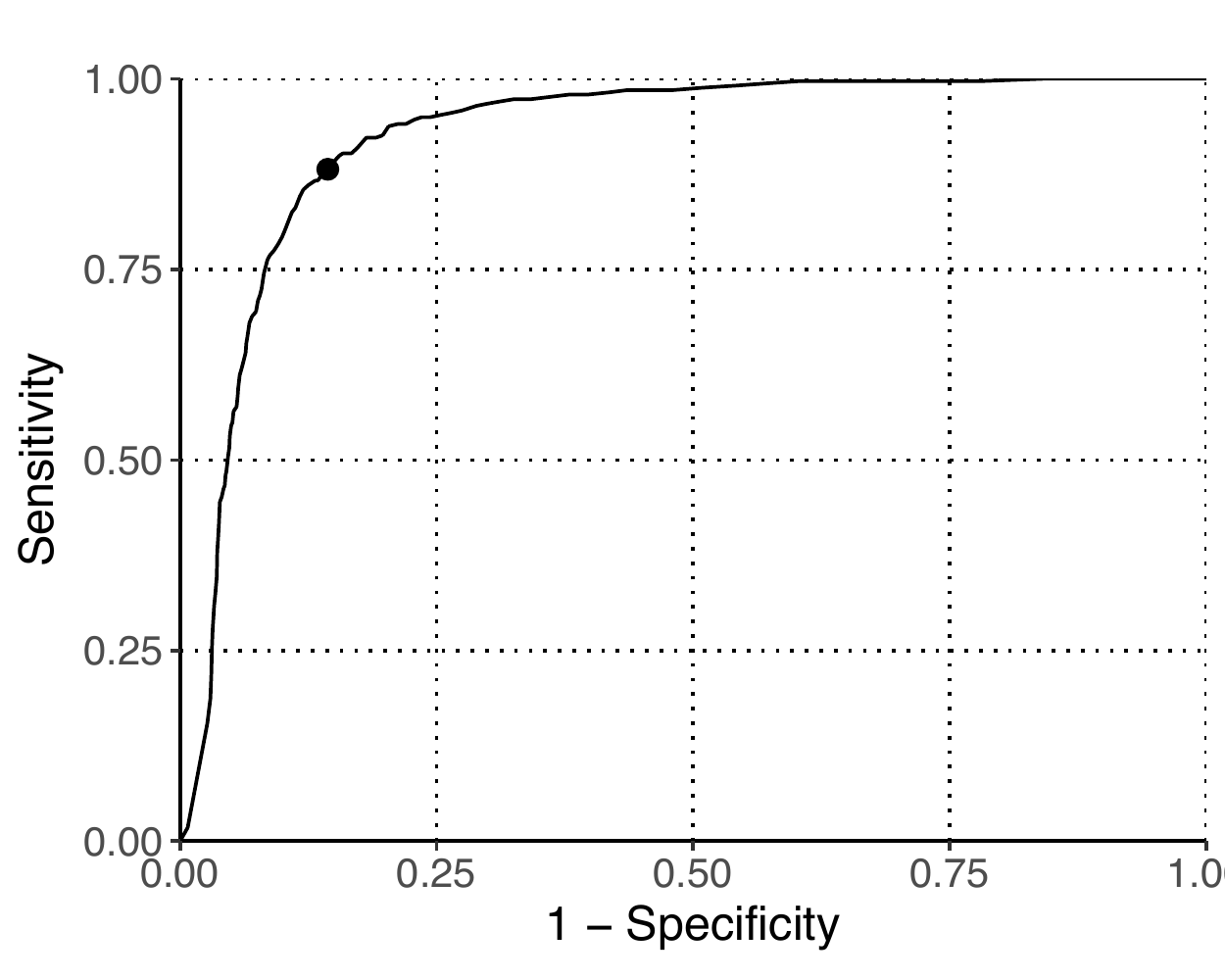}
\caption{8 hours} 
\end{subfigure}\quad
\begin{subfigure}{.31\textwidth}
\includegraphics[width=\linewidth,page=1]{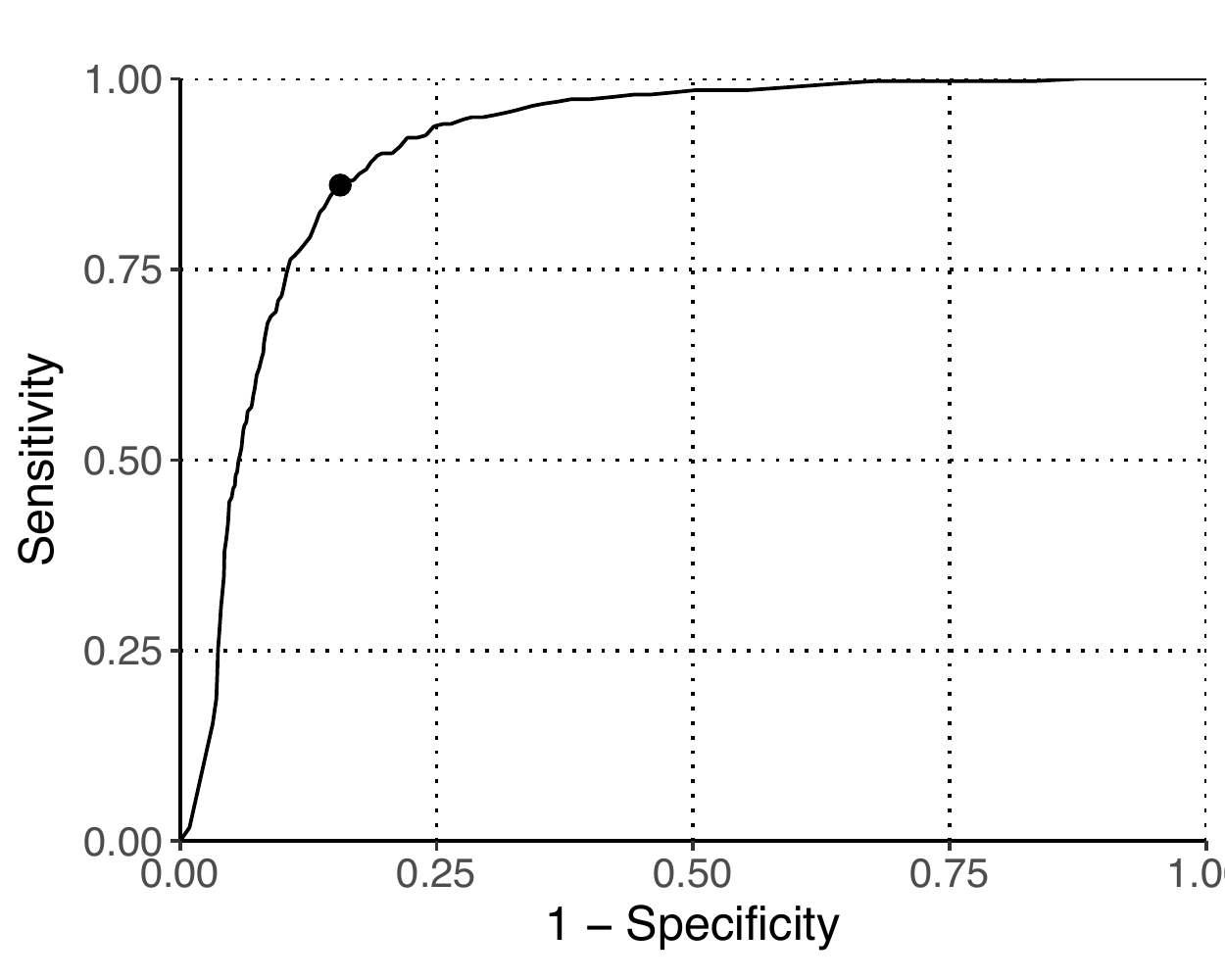}
\caption{16 hours} 
\end{subfigure}\quad
\begin{subfigure}{.31\textwidth}
\includegraphics[width=\linewidth,page=1]{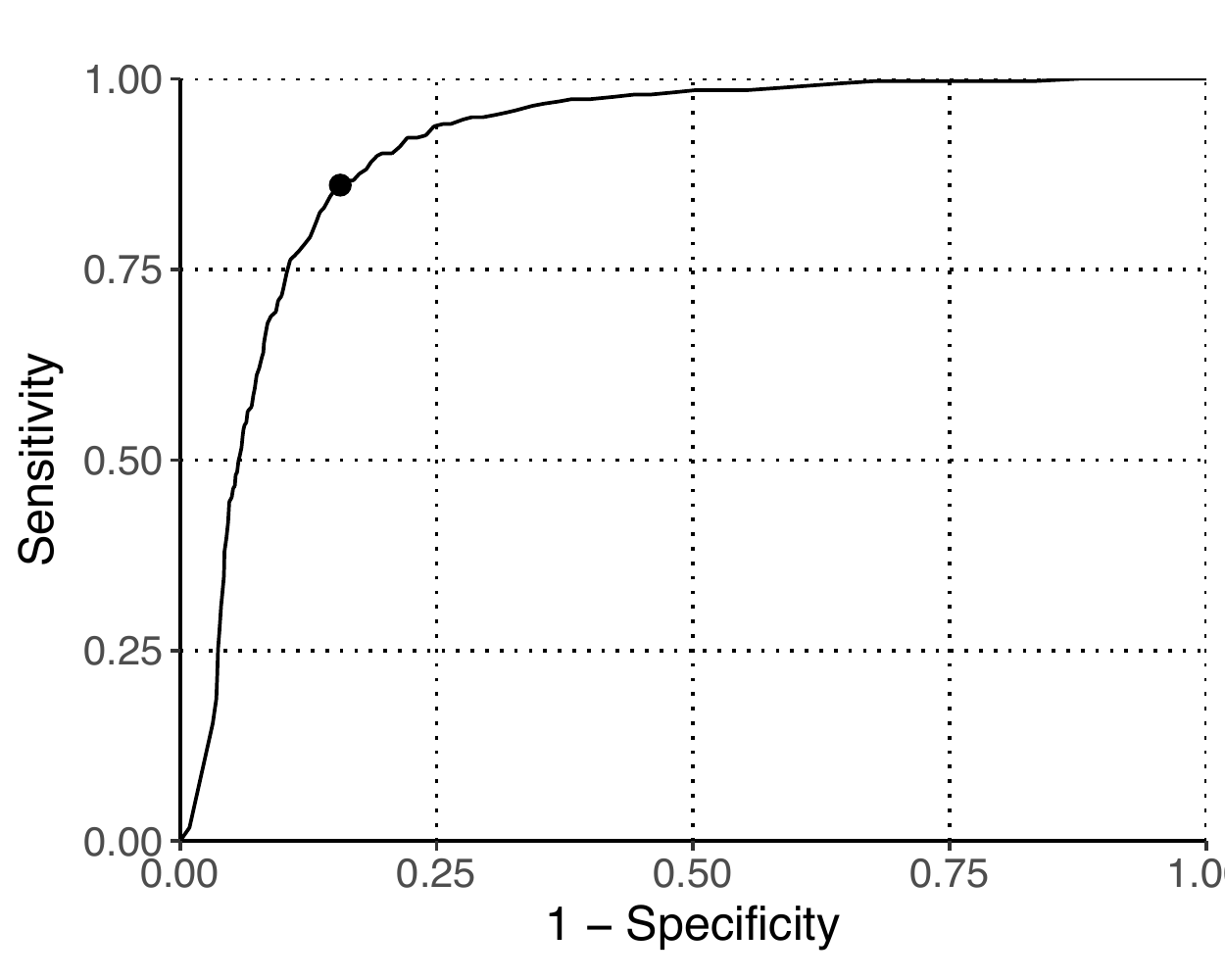}
\caption{24 hours} 
\end{subfigure} \quad
\begin{subfigure}{.31\textwidth}
\includegraphics[width=\linewidth,page=1]{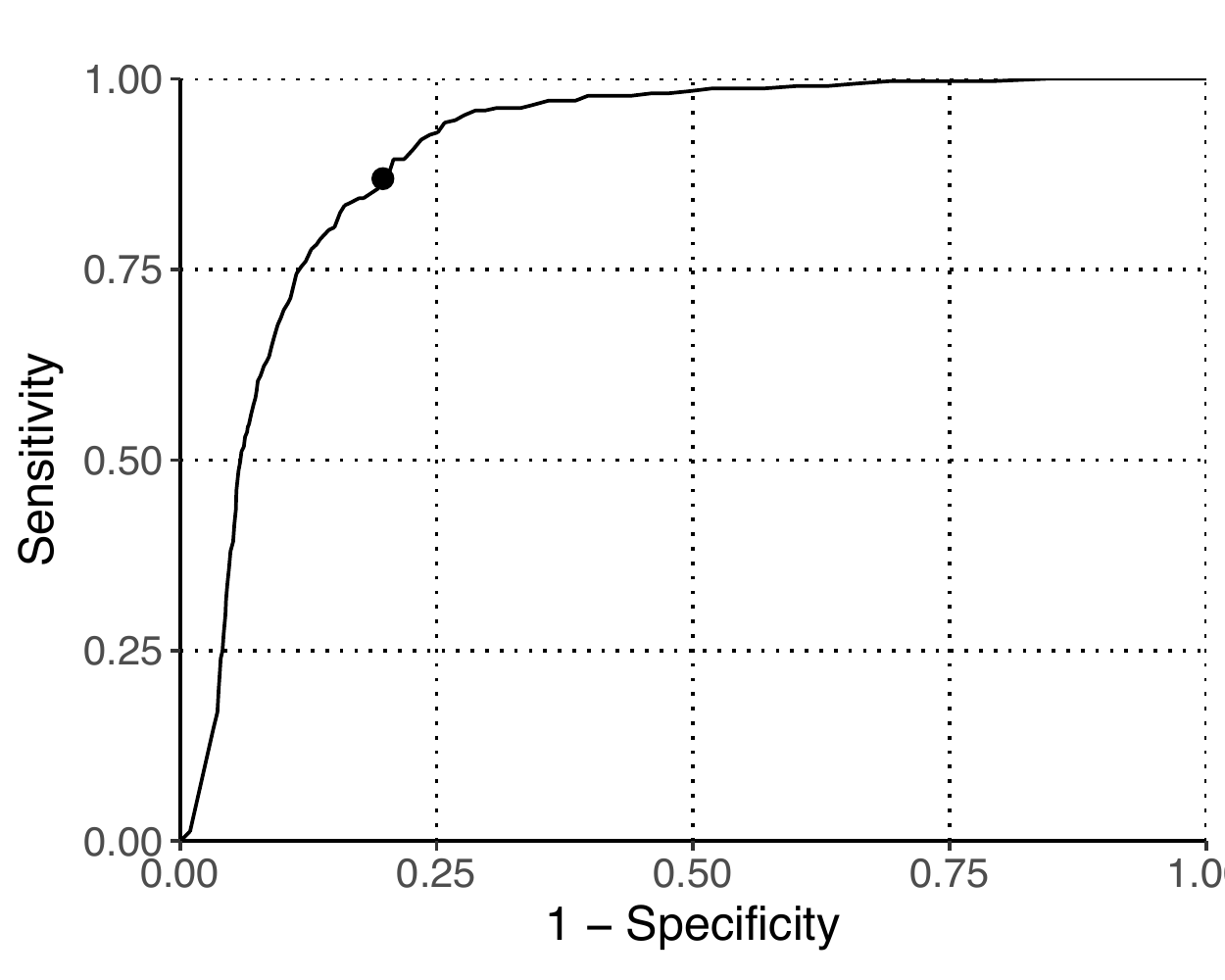}
\caption{48 hours} 
\end{subfigure}\quad
\begin{subfigure}{.31\textwidth}
\includegraphics[width=\linewidth,page=1]{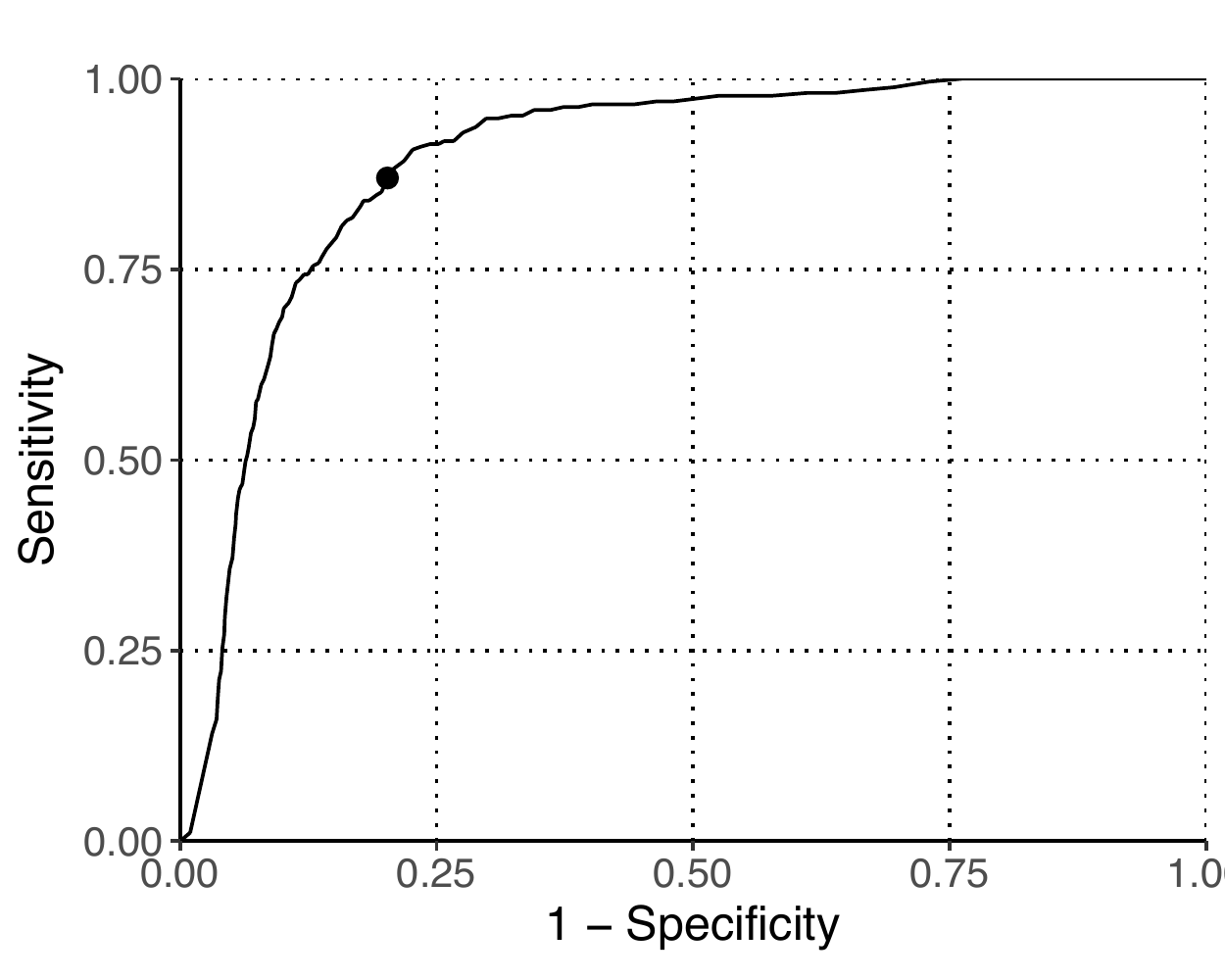}
\caption{96 hours} 
\end{subfigure}\quad
\begin{subfigure}{.31\textwidth}
\includegraphics[width=\linewidth,page=1]{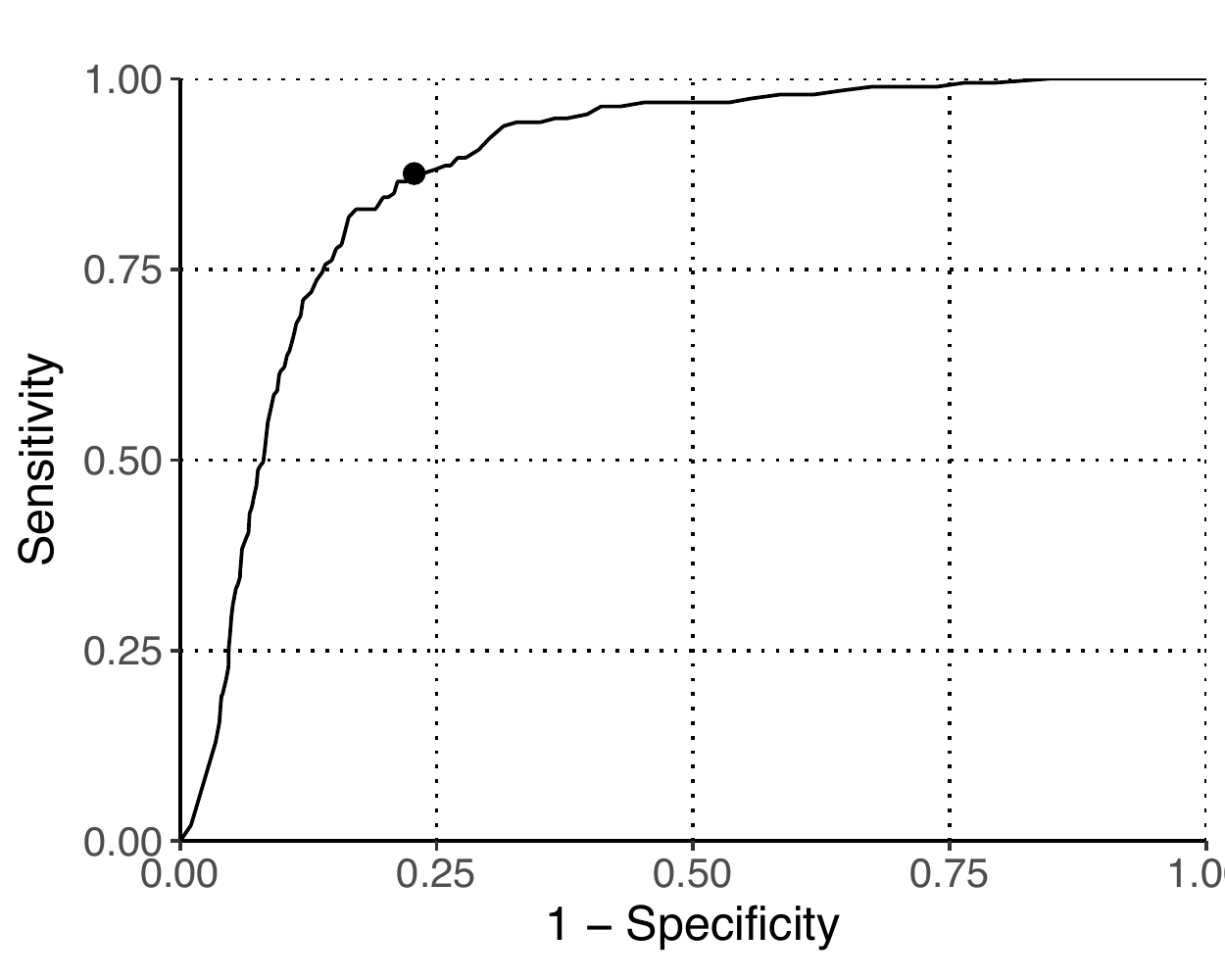}
\caption{192 hours} 
\end{subfigure}\quad
\caption{Receiver operating characteristic (ROC) curves for \themethod{} (linear) for various prediction horizons between 1 and 192 hours evaluated on the external TriNetX test set. The black dot indicates the optimal decision threshold for each prediction horizon selected on the Optum validation set as the closest point on the ROC curve to the top left coordinate (closest-to-top-left heuristic).}
\label{fig:roc_trinetx_linear}
\end{figure}

In addition to the results presented in the main body of this work, we also present Receiver operator characteristic (ROC) curves for \themethod{} for various prediction horizons between 1 and 192 hours evaluated on the held-out Optum test set (\Cref{fig:roc_optum}) and the external TriNetX test set (\Cref{fig:roc_trinetx}), the same ROC curves for \themethod{} (linear) (\Cref{fig:roc_optum_linear} an \Cref{fig:roc_trinetx_linear}) a comparison of \themethod{}, Time Varying Cox \citep{cameron2020lifelines}, COVER\_F \citep{williams2020seek}, SOFA \citep{vincent1996sofa}, \citet{yan2020interpretable}, and \citet{liang2020early} at various prediction horizons in terms of AUC, AUPR, F$_1$, sensitivity, specificity and specificity at greater than 95\% sensitivity (Spec.@95\%Sens.) for predicting COVID-19 related mortality on the held-out Optum test set (\Cref{tbl:results_all_optum}), on the external TriNetX test set (\Cref{tbl:results_all_trinetx}), on the Optum Future cohort (\Cref{tbl:results_all_optumfuture}), and on the Black or African American (\Cref{tbl:results_all_black}), Hispanic (\Cref{tbl:results_all_hispanic}), Asian (\Cref{tbl:results_all_asian}), Caucasian (\Cref{tbl:results_all_caucasian}) and non-hospitalised (\Cref{tbl:results_all_optumhospitalised}) subgroups of the Optum test set.

%% file: 7_performance_tables.tex
\input{7_1_optum.tex}

\input{7_2_trinetx.tex}

\input{7_3_optumblack.tex}

\input{7_4_optumhispanic.tex}

\input{7_5_optumasian.tex}

\input{7_6_optumcaucasian.tex}

\input{7_7_optumfuture.tex}

\input{7_8_optumhospitalised.tex}

%% file: 7_1_optum.tex
\begin{landscape}
\begin{scriptsize}
\begin{longtable}{ll@{\hskip 9ex}l@{\hskip 4ex}l@{\hskip 4ex}l@{\hskip 4ex}l@{\hskip 4ex}l@{\hskip 4ex}l}
\caption{Comparison of \themethod{}, \themethod{} (linear), COVER\_F \citep{williams2020seek}, SOFA \citep{vincent1996sofa}, \citet{yan2020interpretable}, and \citet{liang2020early} at various prediction horizons in terms of AUC, AUPR, F$_1$, sensitivity, specificity and specificity at greater than 95\% sensitivity (Spec.@95\%Sens.) for predicting COVID-19 related mortality on the held-out Optum test set. Values are the median and the 95\% confidence intervals (CIs, in parentheses) obtained via bootstrap resampling with 200 samples. \sig = significant at $p < 0.05$ to \themethod{} (one-sided Mann-Whitney-Wilcoxon, Bonferroni corrected). Best results for each metric are highlighted in bold.}
\label{tbl:results_all_optum}
\centering
\endfirsthead
\caption*{\Cref{tbl:results_all_optum} cont.}
\endhead
\toprule
\multicolumn{8}{c}{Held-out Optum Test Set} \\
& Model & AUC & AUPR & F$_1$ & Sensitivity & Specificity & Spec.@95\%Sens.\\
\midrule
\parbox[t]{0.5mm}{\multirow{6}{*}{\rotatebox[origin=c]{90}{1 hour}}}& \themethod{} & \hspace{1ex}\textbf{97.2} (96.7, 97.7) & \hspace{1ex}{69.2} (66.1, 72.3) & \hspace{1ex}\textbf{56.1} (54.2, 58.3) & \hspace{1ex}\textbf{92.8} (90.9, 94.7) & \hspace{1ex}\textbf{92.4} (91.9, 92.8) & \hspace{1ex}\textbf{89.3} (83.0, 91.6) \\
& \themethod{} (linear) & \sig{}{97.0} (96.5, 97.4) & \hspace{1ex}\textbf{69.9} (66.9, 72.8) & \sig{}{53.3} (51.4, 55.5) & \sig{}{90.1} (88.2, 92.2) & \sig{}{91.8} (91.4, 92.3) & \sig{}{85.8} (83.1, 87.9) \\
& Liang et al. 2020 & \sig{}{86.7} (85.4, 88.2) & \sig{}{32.4} (29.5, 36.1) & \sig{}{30.2} (28.6, 32.3) & \sig{}{82.0} (79.2, 85.6) & \sig{}{80.1} (79.3, 80.7) & \sig{}{40.0} (32.8, 60.4) \\
& COVER\_F & \sig{}{65.6} (63.8, 67.8) & \sig{}{07.9} (07.1, 08.6) & \sig{}{14.7} (13.4, 15.9) & \sig{}{67.5} (63.9, 71.0) & \sig{}{58.1} (57.4, 58.9) & \sig{}{09.4} (04.7, 16.2) \\
& SOFA & \sig{}{90.7} (89.5, 91.8) & \sig{}{46.4} (42.5, 50.1) & \sig{}{38.8} (36.6, 41.0) & \sig{}{83.8} (81.2, 86.0) & \sig{}{86.2} (85.6, 86.8) & \sig{}{00.0} (00.0, 65.3) \\
& Yan et al. 2020 & \sig{}{82.8} (81.6, 84.0) & \sig{}{53.5} (52.2, 54.9) & \sig{}{29.0} (27.4, 30.6) & \sig{}{89.3} (87.2, 91.5) & \sig{}{76.3} (75.6, 77.0) & \sig{}{00.0} (00.0, 00.0) \\
\bottomrule
\end{longtable}
\end{scriptsize}
\end{landscape}

%% file: 7_2_trinetx.tex
\begin{landscape}
\begin{scriptsize}
\begin{longtable}{ll@{\hskip 9ex}l@{\hskip 4ex}l@{\hskip 4ex}l@{\hskip 4ex}l@{\hskip 4ex}l@{\hskip 4ex}l}
\caption{Comparison of \themethod{}, \themethod{} (linear), COVER\_F \citep{williams2020seek}, SOFA \citep{vincent1996sofa}, \citet{yan2020interpretable}, and \citet{liang2020early} at various prediction horizons in terms of AUC, AUPR, F$_1$, sensitivity, specificity and specificity at greater than 95\% sensitivity (Spec.@95\%Sens.) for predicting COVID-19 related mortality on the external TriNetX test set. Values are the median and the 95\% confidence intervals (CIs, in parentheses) obtained via bootstrap resampling with 200 samples. \sig = significant at $p < 0.05$ to \themethod{} (one-sided Mann-Whitney-Wilcoxon, Bonferroni corrected). Best results for each metric are highlighted in bold.}
\label{tbl:results_all_trinetx}
\centering
\endfirsthead
\caption*{\Cref{tbl:results_all_trinetx} cont.}
\endhead
\toprule
\multicolumn{8}{c}{External TriNetX Test Set} \\
& Model & AUC & AUPR & F$_1$ & Sensitivity & Specificity & Spec.@95\%Sens.\\
\midrule
\parbox[t]{0.5mm}{\multirow{6}{*}{\rotatebox[origin=c]{90}{1 hour}}}& \themethod{} & \hspace{1ex}\textbf{93.6} (92.8, 94.4) & \hspace{1ex}{44.8} (40.9, 49.4) & \hspace{1ex}\textbf{48.1} (45.0, 51.1) & \hspace{1ex}\textbf{89.8} (86.3, 93.0) & \hspace{1ex}{86.8} (86.0, 87.6) & \hspace{1ex}\textbf{78.8} (76.0, 84.7) \\
& \themethod{} (linear) & \sig{}{92.3} (91.3, 93.4) & \sig{}{35.8} (32.3, 39.3) & \hspace{1ex}\textbf{48.1} (44.5, 51.1) & \sig{}{86.0} (82.3, 89.5) & \hspace{1ex}\textbf{87.5} (86.7, 88.3) & \sig{}{75.0} (68.6, 81.9) \\
& Liang et al. 2020 & \sig{}{83.6} (81.3, 85.8) & \sig{}{32.5} (27.9, 38.1) & \sig{}{32.6} (29.8, 35.7) & \sig{}{76.7} (72.1, 81.8) & \sig{}{78.8} (77.7, 80.0) & \sig{}{34.4} (19.8, 44.3) \\
& COVER\_F & \sig{}{73.3} (70.4, 76.1) & \sig{}{15.3} (13.0, 17.7) & \sig{}{22.6} (20.4, 25.0) & \sig{}{70.8} (66.0, 75.7) & \sig{}{67.1} (66.0, 68.4) & \sig{}{13.3} (05.1, 34.0) \\
& SOFA & \sig{}{88.2} (86.1, 90.6) & \hspace{1ex}{46.1} (40.7, 50.9) & \sig{}{43.1} (39.9, 46.5) & \sig{}{81.3} (76.9, 85.7) & \sig{}{85.8} (84.9, 86.7) & \sig{}{00.0} (00.0, 00.0) \\
& Yan et al. 2020 & \sig{}{56.6} (55.0, 58.2) & \hspace{1ex}\textbf{49.0} (47.4, 50.4) & \sig{}{14.4} (12.9, 16.0) & \hspace{1ex}{89.4} (86.5, 92.1) & \sig{}{23.6} (22.4, 24.8) & \sig{}{00.0} (00.0, 00.0) \\
\bottomrule
\end{longtable}
\end{scriptsize}
\end{landscape}

%% file: 7_3_optumblack.tex
\begin{landscape}
\begin{scriptsize}
\begin{longtable}{ll@{\hskip 9ex}l@{\hskip 4ex}l@{\hskip 4ex}l@{\hskip 4ex}l@{\hskip 4ex}l@{\hskip 4ex}l}
\caption{Comparison of \themethod{}, \themethod{} (linear), COVER\_F \citep{williams2020seek}, SOFA \citep{vincent1996sofa}, \citet{yan2020interpretable}, and \citet{liang2020early} at various prediction horizons in terms of AUC, AUPR, F$_1$, sensitivity, specificity and specificity at greater than 95\% sensitivity (Spec.@95\%Sens.) for predicting COVID-19 related mortality on the subgroup of Black or African American patients of the Optum test set. Values are the median and the 95\% confidence intervals (CIs, in parentheses) obtained via bootstrap resampling with 200 samples. \sig = significant at $p < 0.05$ to \themethod{} (one-sided Mann-Whitney-Wilcoxon, Bonferroni corrected). Best results for each metric are highlighted in bold.}
\label{tbl:results_all_black}
\centering
\endfirsthead
\caption*{\Cref{tbl:results_all_black} cont.}
\endhead
\toprule
\multicolumn{8}{c}{Optum Test Set (Black or African American Subgroup)} \\
& Model & AUC & AUPR & F$_1$ & Sensitivity & Specificity & Spec.@95\%Sens.\\
\midrule
\parbox[t]{0.5mm}{\multirow{6}{*}{\rotatebox[origin=c]{90}{1 hour}}}& \themethod{} & \hspace{1ex}\textbf{97.6} (96.3, 98.5) & \hspace{1ex}{73.2} (65.5, 80.9) & \hspace{1ex}\textbf{56.1} (51.0, 60.5) & \hspace{1ex}\textbf{93.7} (89.1, 97.0) & \hspace{1ex}\textbf{92.7} (91.9, 93.6) & \hspace{1ex}\textbf{90.9} (69.3, 94.2) \\
& \themethod{} (linear) & \sig{}{96.8} (95.2, 98.1) & \hspace{1ex}\textbf{74.6} (67.1, 80.5) & \sig{}{51.1} (45.6, 55.5) & \sig{}{89.7} (84.7, 94.4) & \sig{}{91.6} (90.6, 92.5) & \sig{}{82.4} (68.9, 90.8) \\
& Liang et al. 2020 & \sig{}{89.4} (86.9, 91.9) & \sig{}{34.3} (27.2, 42.8) & \sig{}{33.9} (30.2, 38.5) & \sig{}{78.6} (72.3, 84.8) & \sig{}{85.3} (84.0, 86.5) & \sig{}{62.8} (36.8, 69.9) \\
& COVER\_F & \sig{}{83.5} (80.5, 86.1) & \sig{}{19.3} (14.4, 24.9) & \sig{}{24.3} (20.7, 27.9) & \sig{}{78.5} (72.0, 83.6) & \sig{}{76.0} (74.4, 77.6) & \sig{}{48.7} (37.6, 61.1) \\
& SOFA & \sig{}{92.1} (89.7, 93.9) & \sig{}{47.1} (39.1, 55.4) & \sig{}{41.5} (36.4, 47.0) & \sig{}{75.6} (68.9, 81.6) & \sig{}{90.4} (89.4, 91.5) & \sig{}{41.7} (40.1, 79.1) \\
& Yan et al. 2020 & \sig{}{75.5} (72.8, 77.7) & \sig{}{50.2} (47.5, 52.4) & \sig{}{19.3} (16.7, 21.8) & \sig{}{88.9} (83.9, 93.3) & \sig{}{62.1} (60.3, 63.5) & \sig{}{00.0} (00.0, 00.0) \\
\bottomrule
\end{longtable}
\end{scriptsize}
\end{landscape}

%% file: 7_4_optumhispanic.tex
\begin{landscape}
\begin{scriptsize}
\begin{longtable}{ll@{\hskip 9ex}l@{\hskip 4ex}l@{\hskip 4ex}l@{\hskip 4ex}l@{\hskip 4ex}l@{\hskip 4ex}l}
\caption{Comparison of \themethod{}, \themethod{} (linear), COVER\_F \citep{williams2020seek}, SOFA \citep{vincent1996sofa}, \citet{yan2020interpretable}, and \citet{liang2020early} at various prediction horizons in terms of AUC, AUPR, F$_1$, sensitivity, specificity and specificity at greater than 95\% sensitivity (Spec.@95\%Sens.) for predicting COVID-19 related mortality on the subgroup of Hispanic patients of the Optum test set. Values are the median and the 95\% confidence intervals (CIs, in parentheses) obtained via bootstrap resampling with 200 samples. \sig = significant at $p < 0.05$ to \themethod{} (one-sided Mann-Whitney-Wilcoxon, Bonferroni corrected). Best results for each metric are highlighted in bold.}
\label{tbl:results_all_hispanic}
\centering
\endfirsthead
\caption*{\Cref{tbl:results_all_hispanic} cont.}
\endhead
\toprule
\multicolumn{8}{c}{Optum Test Set (Hispanic Subgroup)} \\
& Model & AUC & AUPR & F$_1$ & Sensitivity & Specificity & Spec.@95\%Sens.\\
\midrule
\parbox[t]{0.5mm}{\multirow{6}{*}{\rotatebox[origin=c]{90}{1 hour}}}& \themethod{} (linear) & \hspace{1ex}\textbf{98.7} (97.9, 99.3) & \hspace{1ex}\textbf{75.0} (62.0, 87.0) & \hspace{1ex}\textbf{51.3} (40.6, 59.7) & \hspace{1ex}\textbf{92.7} (86.0, 100.0) & \hspace{1ex}\textbf{94.8} (93.6, 95.8) & \hspace{1ex}\textbf{92.8} (87.2, 97.3) \\
& \themethod{} & \hspace{1ex}{98.3} (97.0, 99.1) & \hspace{1ex}{71.2} (58.9, 82.6) & \hspace{1ex}{49.4} (38.2, 57.7) & \hspace{1ex}\textbf{92.7} (83.7, 100.0) & \hspace{1ex}{94.2} (93.1, 95.3) & \hspace{1ex}{85.9} (83.1, 97.8) \\
& Liang et al. 2020 & \sig{}{95.2} (92.9, 96.8) & \sig{}{43.8} (28.4, 58.4) & \sig{}{29.5} (22.0, 36.2) & \sig{}{85.5} (75.0, 93.8) & \sig{}{87.8} (86.5, 89.4) & \sig{}{79.1} (74.7, 86.5) \\
& COVER\_F & \sig{}{85.4} (81.7, 89.0) & \sig{}{11.6} (07.5, 17.1) & \sig{}{16.7} (11.1, 21.6) & \sig{}{73.1} (60.5, 84.6) & \sig{}{78.3} (75.8, 80.4) & \sig{}{56.4} (43.1, 73.9) \\
& SOFA & \sig{}{91.3} (85.9, 95.9) & \sig{}{46.5} (32.7, 60.5) & \sig{}{29.3} (22.2, 35.2) & \sig{}{89.1} (79.5, 97.6) & \sig{}{87.2} (85.3, 88.7) & \sig{}{00.0} (00.0, 87.4) \\
& Yan et al. 2020 & \sig{}{71.7} (67.0, 75.4) & \sig{}{47.6} (42.6, 51.4) & \sig{}{10.7} (07.2, 13.6) & \sig{}{89.1} (80.0, 96.2) & \sig{}{54.1} (51.5, 56.1) & \sig{}{00.0} (00.0, 54.4) \\
\bottomrule
\end{longtable}
\end{scriptsize}
\end{landscape}

%% file: 7_5_optumasian.tex
\begin{landscape}
\begin{scriptsize}
\begin{longtable}{ll@{\hskip 9ex}l@{\hskip 4ex}l@{\hskip 3ex}l@{\hskip 3ex}l@{\hskip 3ex}l@{\hskip 4ex}l}
\caption{Comparison of \themethod{}, \themethod{} (linear), COVER\_F \citep{williams2020seek}, SOFA \citep{vincent1996sofa}, \citet{yan2020interpretable}, and \citet{liang2020early} at various prediction horizons in terms of AUC, AUPR, F$_1$, sensitivity, specificity and specificity at greater than 95\% sensitivity (Spec.@95\%Sens.) for predicting COVID-19 related mortality on the subgroup of Asian patients of the Optum test set. Values are the median and the 95\% confidence intervals (CIs, in parentheses) obtained via bootstrap resampling with 200 samples. \sig = significant at $p < 0.05$ to \themethod{} (one-sided Mann-Whitney-Wilcoxon, Bonferroni corrected). Best results for each metric are highlighted in bold.}
\label{tbl:results_all_asian}
\centering
\endfirsthead
\caption*{\Cref{tbl:results_all_asian} cont.}
\endhead
\toprule
\multicolumn{8}{c}{Optum Test Set (Asian Subgroup)} \\
& Model & AUC & AUPR & F$_1$ & Sensitivity & Specificity & Spec.@95\%Sens.\\
\midrule
\parbox[t]{0.5mm}{\multirow{6}{*}{\rotatebox[origin=c]{90}{1 hour}}}& \themethod{} (linear) & \hspace{1ex}\textbf{99.2} (98.3, 99.7) & \hspace{1ex}{78.7} (58.8, 94.1) & \hspace{1ex}\textbf{58.3} (42.3, 68.1) & \hspace{1ex}\textbf{100.0} (100.0, 100.0) & \hspace{1ex}\textbf{93.5} (90.5, 95.6) & \hspace{1ex}\textbf{97.9} (92.9, 99.4) \\
& \themethod{} & \hspace{1ex}\textbf{99.2} (98.2, 99.7) & \hspace{1ex}\textbf{81.7} (62.3, 93.9) & \hspace{1ex}{51.3} (38.4, 62.7) & \hspace{1ex}\textbf{100.0} (100.0, 100.0) & \hspace{1ex}{91.7} (88.8, 93.9) & \hspace{1ex}{96.9} (93.2, 99.4) \\
& SOFA & \sig{}{96.8} (94.7, 98.3) & \sig{}{51.6} (30.8, 75.3) & \sig{}{42.1} (27.9, 54.2) & \hspace{1ex}\textbf{100.0} (100.0, 100.0) & \sig{}{87.3} (84.1, 90.3) & \sig{}{87.6} (84.7, 92.4) \\
& Yan et al. 2020 & \sig{}{79.2} (73.0, 82.6) & \sig{}{53.6} (46.5, 57.4) & \sig{}{19.3} (13.2, 26.0) & \sig{}{95.6} (84.2, 100.0) & \sig{}{62.3} (58.3, 66.4) & \sig{}{60.3} (00.0, 65.8) \\
& COVER\_F & \sig{}{87.7} (76.4, 95.1) & \sig{}{20.5} (09.9, 37.5) & \sig{}{28.8} (17.5, 39.7) & \sig{}{85.2} (63.6, 100.0) & \sig{}{84.3} (80.6, 87.6) & \sig{}{39.5} (19.8, 86.9) \\
& Liang et al. 2020 & \sig{}{89.0} (77.8, 95.8) & \sig{}{02.2} (01.2, 04.0) & \sig{}{05.9} (03.2, 08.9) & \sig{}{78.9} (60.0, 95.5) & \sig{}{00.4} (00.0, 01.2) & \sig{}{00.0} (00.0, 02.5) \\
\bottomrule
\end{longtable}
\end{scriptsize}
\end{landscape}

%% file: 7_6_optumcaucasian.tex
\begin{landscape}
\begin{scriptsize}
\begin{longtable}{ll@{\hskip 9ex}l@{\hskip 4ex}l@{\hskip 4ex}l@{\hskip 4ex}l@{\hskip 4ex}l@{\hskip 4ex}l}
\caption{Comparison of \themethod{}, \themethod{} (linear), COVER\_F \citep{williams2020seek}, SOFA \citep{vincent1996sofa}, \citet{yan2020interpretable}, and \citet{liang2020early} at various prediction horizons in terms of AUC, AUPR, F$_1$, sensitivity, specificity and specificity at greater than 95\% sensitivity (Spec.@95\%Sens.) for predicting COVID-19 related mortality on the subgroup of Caucasian patients of the Optum test set. Values are the median and the 95\% confidence intervals (CIs, in parentheses) obtained via bootstrap resampling with 200 samples. \sig = significant at $p < 0.05$ to \themethod{} (one-sided Mann-Whitney-Wilcoxon, Bonferroni corrected). Best results for each metric are highlighted in bold.}
\label{tbl:results_all_caucasian}
\centering
\endfirsthead
\caption*{\Cref{tbl:results_all_caucasian} cont.}
\endhead
\toprule
\multicolumn{8}{c}{Optum Test Set (Caucasian Subgroup)} \\
& Model & AUC & AUPR & F$_1$ & Sensitivity & Specificity & Spec.@95\%Sens.\\
\midrule
\parbox[t]{0.5mm}{\multirow{6}{*}{\rotatebox[origin=c]{90}{1 hour}}}& \themethod{} & \hspace{1ex}{96.7} (96.2, 97.4) & \hspace{1ex}{67.5} (63.1, 72.2) & \hspace{1ex}\textbf{56.9} (53.6, 60.0) & \hspace{1ex}\textbf{91.3} (88.6, 94.1) & \hspace{1ex}\textbf{90.8} (90.0, 91.5) & \hspace{1ex}\textbf{87.3} (82.7, 89.4) \\
& \themethod{} (linear) & \hspace{1ex}\textbf{96.8} (96.1, 97.3) & \hspace{1ex}\textbf{71.7} (66.8, 75.3) & \sig{}{54.6} (51.6, 57.2) & \sig{}{89.8} (87.1, 92.3) & \sig{}{90.4} (89.8, 91.1) & \sig{}{85.3} (81.9, 87.8) \\
& COVER\_F & \sig{}{65.2} (62.7, 67.7) & \sig{}{08.8} (07.8, 09.7) & \sig{}{16.9} (15.5, 18.3) & \sig{}{74.9} (70.9, 78.9) & \sig{}{51.2} (50.0, 52.5) & \sig{}{21.0} (12.8, 32.8) \\
& Liang et al. 2020 & \sig{}{82.1} (80.1, 84.4) & \sig{}{04.1} (03.6, 05.2) & \sig{}{02.3} (01.1, 03.7) & \sig{}{03.0} (01.5, 04.7) & \sig{}{88.7} (88.0, 89.5) & \sig{}{00.4} (00.3, 00.6) \\
& SOFA & \sig{}{90.3} (88.6, 91.6) & \sig{}{48.5} (44.0, 52.8) & \sig{}{49.2} (45.8, 51.8) & \sig{}{77.6} (73.6, 81.1) & \sig{}{90.2} (89.5, 90.8) & \sig{}{00.0} (00.0, 00.0) \\
& Yan et al. 2020 & \sig{}{82.3} (80.2, 84.0) & \sig{}{54.1} (51.5, 56.2) & \sig{}{38.9} (36.3, 41.4) & \sig{}{81.5} (77.5, 84.9) & \sig{}{83.2} (82.2, 84.0) & \sig{}{00.0} (00.0, 00.0) \\
\bottomrule
\end{longtable}
\end{scriptsize}
\end{landscape}

%% file: 7_7_optumfuture.tex
\begin{landscape}
\begin{scriptsize}
\begin{longtable}{ll@{\hskip 9ex}l@{\hskip 4ex}l@{\hskip 4ex}l@{\hskip 4ex}l@{\hskip 4ex}l@{\hskip 4ex}l}
\caption{Comparison of \themethod{}, \themethod{} (linear), COVER\_F \citep{williams2020seek}, SOFA \citep{vincent1996sofa}, \citet{yan2020interpretable}, and \citet{liang2020early} at various prediction horizons in terms of AUC, AUPR, F$_1$, sensitivity, specificity and specificity at greater than 95\% sensitivity (Spec.@95\%Sens.) for predicting COVID-19 related mortality on the Optum future cohort. Values are the median and the 95\% confidence intervals (CIs, in parentheses) obtained via bootstrap resampling with 200 samples. \sig = significant at $p < 0.05$ to \themethod{} (one-sided Mann-Whitney-Wilcoxon, Bonferroni corrected). Best results for each metric are highlighted in bold.}
\label{tbl:results_all_optumfuture}
\centering
\endfirsthead
\caption*{\Cref{tbl:results_all_optumfuture} cont.}
\endhead
\toprule
\multicolumn{8}{c}{Optum Future Cohort (June 6$^\text{th}$ to July 13$^\text{th}$ 2020)} \\
& Model & AUC & AUPR & F$_1$ & Sensitivity & Specificity & Spec.@95\%Sens.\\
\midrule
\parbox[t]{0.5mm}{\multirow{6}{*}{\rotatebox[origin=c]{90}{1 hour}}}& \themethod{} & \hspace{1ex}{97.7} (97.1, 98.3) & \hspace{1ex}{27.3} (21.8, 32.7) & \hspace{1ex}{22.1} (18.5, 25.4) & \hspace{1ex}\textbf{90.2} (84.7, 95.9) & \hspace{1ex}{95.0} (94.6, 95.3) & \hspace{1ex}\textbf{90.8} (87.6, 95.5) \\
& \themethod{} (linear) & \hspace{1ex}\textbf{97.8} (96.6, 98.6) & \hspace{1ex}\textbf{52.5} (42.5, 61.2) & \hspace{1ex}\textbf{25.0} (21.1, 29.3) & \sig{}{88.4} (81.4, 94.5) & \hspace{1ex}\textbf{95.8} (95.5, 96.1) & \sig{}{89.3} (81.6, 95.1) \\
& COVER\_F & \sig{}{88.2} (85.2, 90.8) & \sig{}{06.0} (03.9, 09.7) & \sig{}{05.7} (04.8, 06.6) & \sig{}{88.6} (82.5, 94.5) & \sig{}{76.3} (75.7, 76.9) & \sig{}{54.1} (45.4, 74.0) \\
& Liang et al. 2020 & \sig{}{90.7} (87.0, 93.8) & \sig{}{14.3} (09.9, 20.9) & \sig{}{09.5} (07.9, 11.2) & \sig{}{86.8} (80.2, 92.9) & \sig{}{86.7} (86.1, 87.2) & \sig{}{53.5} (14.5, 70.8) \\
& SOFA & \sig{}{91.5} (88.8, 93.8) & \sig{}{15.9} (10.8, 22.3) & \sig{}{12.3} (10.2, 14.3) & \sig{}{83.0} (76.2, 89.5) & \sig{}{90.6} (90.1, 91.1) & \sig{}{00.0} (00.0, 71.6) \\
& Yan et al. 2020 & \sig{}{88.4} (84.8, 91.8) & \hspace{1ex}{46.9} (42.9, 50.4) & \sig{}{11.6} (09.7, 13.6) & \sig{}{87.5} (80.4, 94.3) & \sig{}{89.3} (88.9, 89.7) & \sig{}{00.0} (00.0, 00.0) \\
\bottomrule
\end{longtable}
\end{scriptsize}
\end{landscape}

%% file: 7_8_optumhospitalised.tex
\begin{landscape}
\begin{scriptsize}
\begin{longtable}{ll@{\hskip 9ex}l@{\hskip 4ex}l@{\hskip 4ex}l@{\hskip 4ex}l@{\hskip 4ex}l@{\hskip 4ex}l}
\caption{Comparison of \themethod{}, \themethod{} (linear), COVER\_F \citep{williams2020seek}, SOFA \citep{vincent1996sofa}, \citet{yan2020interpretable}, and \citet{liang2020early} at various prediction horizons in terms of AUC, AUPR, F$_1$, sensitivity, specificity and specificity at greater than 95\% sensitivity (Spec.@95\%Sens.) for predicting COVID-19 related mortality on the subgroup of non-hospitalised patients of the Optum test set. Values are the median and the 95\% confidence intervals (CIs, in parentheses) obtained via bootstrap resampling with 200 samples. \sig = significant at $p < 0.05$ to \themethod{} (one-sided Mann-Whitney-Wilcoxon, Bonferroni corrected). Best results for each metric are highlighted in bold.}
\label{tbl:results_all_optumhospitalised}
\centering
\endfirsthead
\caption*{\Cref{tbl:results_all_optumhospitalised} cont.}
\endhead
\toprule
\multicolumn{8}{c}{Optum Test Set (Not Hospitalised Subgroup)} \\
& Model & AUC & AUPR & F$_1$ & Sensitivity & Specificity & Spec.@95\%Sens.\\
\midrule
\parbox[t]{0.5mm}{\multirow{6}{*}{\rotatebox[origin=c]{90}{1 hour}}}& \themethod{} (linear) & \hspace{1ex}\textbf{96.7} (95.6, 97.5) & \hspace{1ex}\textbf{50.0} (43.3, 57.3) & \hspace{1ex}{40.5} (36.5, 44.1) & \hspace{1ex}{85.4} (80.6, 89.3) & \hspace{1ex}{93.8} (93.2, 94.3) & \hspace{1ex}\textbf{86.3} (77.7, 89.6) \\
& \themethod{} & \hspace{1ex}{96.6} (95.1, 97.6) & \hspace{1ex}{49.4} (43.3, 55.8) & \hspace{1ex}\textbf{42.2} (38.3, 46.6) & \hspace{1ex}\textbf{86.9} (82.8, 91.0) & \hspace{1ex}\textbf{94.1} (93.6, 94.6) & \hspace{1ex}{85.5} (77.8, 91.2) \\
& Liang et al. 2020 & \sig{}{84.4} (81.6, 86.7) & \sig{}{18.8} (15.2, 23.6) & \sig{}{16.3} (14.2, 18.7) & \sig{}{77.8} (72.3, 82.9) & \sig{}{79.8} (78.8, 80.5) & \sig{}{34.0} (20.9, 45.4) \\
& COVER\_F & \sig{}{69.8} (66.4, 72.3) & \sig{}{04.4} (03.7, 05.2) & \sig{}{08.2} (07.0, 09.2) & \sig{}{77.0} (70.9, 81.6) & \sig{}{55.5} (54.6, 56.5) & \sig{}{17.8} (14.5, 26.3) \\
& SOFA & \sig{}{91.4} (88.9, 93.1) & \sig{}{28.1} (23.6, 34.0) & \sig{}{29.5} (26.7, 32.7) & \sig{}{81.2} (75.9, 86.1) & \sig{}{90.3} (89.7, 91.0) & \sig{}{00.0} (00.0, 74.5) \\
& Yan et al. 2020 & \sig{}{84.0} (81.3, 85.8) & \sig{}{47.6} (44.6, 50.0) & \sig{}{22.5} (20.2, 25.2) & \sig{}{82.0} (76.8, 86.0) & \sig{}{85.9} (85.0, 86.6) & \sig{}{00.0} (00.0, 00.0) \\
\bottomrule
\end{longtable}
\end{scriptsize}
\end{landscape}